\documentclass[letterpaper,12pt]{article}
\usepackage{tabularx} % extra features for tabular environment

\usepackage{amsmath}  % improve math presentation
\usepackage{amssymb}
\usepackage{graphicx} % takes care of graphic including machinery
\usepackage{subfig}
\usepackage{natbib}
\usepackage{dcolumn}
\usepackage[margin=1in,letterpaper]{geometry} % decreases margins
\usepackage[final]{hyperref} % adds hyper links inside the generated pdf file
 \hypersetup{
 	colorlinks=true,       % false: boxed links; true: colored links
 	linkcolor=blue,        % color of internal links
 	citecolor=blue,        % color of links to bibliography
 	filecolor=magenta,     % color of file links
 	urlcolor=blue         
}
\usepackage{float}

\begin{document}
\title{Internal Technical Report: uGMRT Polarization Observations with Walsh Modulation}

\begin{figure}
\centering{
\includegraphics[height=5cm]{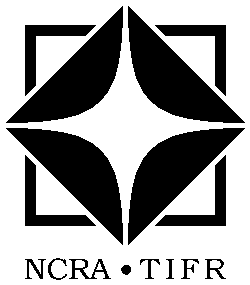}}
\end{figure}

\author{Salmoli Ghosh, Preeti Kharb, Sandeep Chaudhari, Surajit Mondal}
%\date{\today}
\date{October 1, 2026}
\maketitle
\begin{abstract}
In this report, we discuss the performance of the Walsh phase-switching scheme for Band~4 ($550-750$~MHz) polarimetric observations with the upgraded GMRT (uGMRT). We have carried out test observations with the Walsh modulation enabled and disabled, using the polarized calibrator 3C138 as the target and unpolarized calibrator 3C147 for amplitude, phase and leakage calibration. The two Walsh datasets (ON \& OFF) were processed using identical flagging, calibration, and imaging procedures. The Walsh ON data showed smoother leakage solutions and lower residual scatter in the Stokes I, Q, U data. The Walsh scheme also suppressed a fraction of anomalous data at the correlation stage, reducing the amount of subsequent flagging required during data processing. The final self-calibrated images obtained with Walsh ON showed improved image fidelity, lower r.m.s. noise, and reduced error in the fractional polarization of 3C138. However, the possibility that the Walsh scheme itself introduces spurious frequency-dependent structure into the Stokes Q and U signals remains and warrants closer investigation. Overall, these results demonstrate the potential of Walsh phase switching to mitigate instrumental cross-talk and improve the reliability of polarimetric measurements with the uGMRT.

\end{abstract}

\section{Introduction}
The Giant Metrewave Radio Telescope (GMRT) consists of 30 antennas operating with dual polarization feeds at low frequencies ranging from 120 MHz to 1.4 GHz \citep{Swarup1991}. The 60 simultaneous radio-frequency (RF) signals generated by the antennas pass through a long electronic path, which can cause spurious electromagnetic coupling between signal paths from different antennas. This can introduce unwanted correlations, cross-talks, or `leakages'  between signals from different antennas. Since radio telescopes are designed to detect the weakest radio signals from astrophysical sources, even weak cross-coupled signals can significantly degrade calibration accuracy and dynamic range in images. This is even more significant for polarization, where the source signals are a small fraction of their total intensity. In order to mitigate these effects, a Walsh phase modulation and demodulation scheme has been implemented in the upgraded GMRT \citep[uGMRT;][]{Gupta2017}.

Polarization fractions in many astrophysical sources, like `radio-quiet' active galactic nuclei (AGN), relic emission inside galaxy clusters, and diffuse synchrotron emission in general, are typically a few per cent or less \citep[e.g.,][]{Bonafede2010, Govoni2013,
Sebastian2020, Ghosh2025}. Detecting such weak polarization requires well-calibrated cross-correlation products. Suppressing cross-talk in the instrumental path reduces difficulties in finding leakage solutions (D-terms) and maintaining channel independence. Instrumental polarization leakage arises from imperfect isolation between the orthogonal polarization channels due to feed non-orthogonality and cross-coupling within the receiver chain. Additional contributions arise from differential gain and phase errors, frequency-dependent bandpass mismatches, and residual calibration errors in the complex gains and D-terms. While polarization calibration corrects most stable instrumental effects, coherent cross-talk can persist. The Walsh phase-switching scheme suppresses this contribution by de-correlating cross-coupled signals before correlation, thereby improving the fidelity of the recovered polarization products, potentially allowing the study of rotation measures (RM) in weakly polarized sources using the uGMRT.

The Walsh filtering method applies orthonormal Walsh patterns (series of square waves) at the front-end receiver (immediately after the low-noise circuit in the RF) of each antenna and performs a corresponding demodulation in the digital backend before combining the signals, after initially digitizing the signals at the central station. For each antenna, a different Walsh pattern is used, but the modulating and demodulating signal patterns for each antenna precisely match and align. This orthogonal coding ensures that cross-couplings are canceled out after correlation, thereby preserving the coherence of true astronomical signals while suppressing spurious correlations. The sixty RF signals in the uGMRT are being modulated by a set of 64 (128-bit) independent Walsh patterns. More details are available in the NCRA Technical Report by Sandeep Chaudhari et al. (2026): {\small \url{http://192.168.115.167:4000/items/891498f2-f27f-4f8a-b339-2f100fe7afed}}.

In the circular basis (R \& L), the Stokes Q and U parameters that parametrize the linear polarization (linear polarization fraction $f_{pl}=\frac{\sqrt{Q^2+U^2}}{I}$, linear polarization angle $\chi_l=\frac{1}{2} tan^{-1}\left(\frac{U}{Q}\right)$, I being the total intensity) are dependent on cross-hand correlations (RL, LR) as shown in equations 1 and 2 below \citep[also see][]{Hales2017}.

\begin{minipage}{0.47\textwidth}
\centering
\begin{equation}
\begin{aligned}
RR &= I + V, \\
LL &= I - V, \\
RL &= Q + iU, \\
LR &= Q - iU.
\end{aligned}
\end{equation}
\end{minipage}
\hfill
\begin{minipage}{0.47\textwidth}
\centering
\begin{equation}
\begin{aligned}
I &= \frac{RR + LL}{2}, \\
Q &= \frac{RL + LR}{2}, \\
U &= \frac{RL - LR}{2i}, \\
V &= \frac{RR - LL}{2}.
\end{aligned}
\end{equation}
\end{minipage}

\vspace{0.5cm}
Whereas, in the linear basis (X \& Y), the Stokes U and V, affecting both linear and circular polarization (circular polarization fraction $f_{pc}=V/I$), depend on the cross-hand correlation terms (XY, YX; see equations 3 and 4 below). 

\begin{minipage}{0.47\textwidth}
\centering
\begin{equation}
\begin{aligned}
XX &= I + Q, \\
YY &= I - Q, \\
XY &= U + iV, \\
YX &= U - iV.
\end{aligned}
\end{equation}
\end{minipage}
\hfill
\begin{minipage}{0.47\textwidth}
\centering
\begin{equation}
\begin{aligned}
I &= \frac{XX + YY}{2}, \\
Q &= \frac{XX - YY}{2}, \\
U &= \frac{XY + YX}{2}, \\
V &= \frac{XY - YX}{2i}.
\end{aligned}
\end{equation}
\end{minipage}

\vspace{0.5cm}
These cross-hand signals are often orders of magnitude weaker than the parallel-hand (RR, LL, or XX, YY) correlations. Polarization leakages mostly arise due to the imperfections and non-orthogonality of the feeds (i.e., the R/X feed picks up L/Y emission signal, and vice versa). This is primarily taken care of by leakage calibration using a polarized calibrator observed over a large parallactic angle range or an unpolarized calibrator. However, the other comparatively smaller system-induced leakages, can also mimic distort intrinsic source polarization, making it harder to calibrate. Walsh phase switching ensures that only signals that follow the correct modulation-demodulation sequence survive in correlation, thereby preserving true cross-hand coherence while suppressing instrumental artifacts. Walsh modulation is implemented in the front-end electronics using a Walsh generator located in the antenna-based receiver chain. A 1~MHz clock derived from the uGMRT time and frequency standard is divided to produce a fundamental Walsh frequency ($\sim$244 Hz), resulting in a full 128-bit pattern period of $\sim$0.671~s. 

As the introduction of a new scheme could potentially create other problems for the existing working systems, we needed to test the performance of Walsh filtering for polarimetric observations with the uGMRT. For this, we observed a polarized AGN with known polarization characteristics, 3C138, with and without the implementation of the Walsh scheme, to find out how the Walsh scheme affects our polarization observations and results. The results of this exercise are reported below. 

\section{Observations and strategy}
We observed the polarized calibrator 3C138, which has well-established\footnote{It has been and continues to be monitored by the Karl G. Jansky Very Large Array (VLA) for various surveys and projects.} polarization properties, on 12 March 2024 from 13:00--19:00 IST (6 hours for the whole experiment), using the uGMRT band-4 (550--750 MHz; bandwidth of 200 MHz) with 2048 channels. The unpolarized calibrator 3C147 was also observed and served both as a phase calibrator and an instrumental leakage calibrator. The observing strategy was designed such that both sources were observed consecutively with the Walsh phase-switching scheme alternately enabled (Walsh ON) and disabled (Walsh OFF). This approach ensured that the datasets acquired with and without Walsh modulation were subject to nearly identical radio frequency interference (RFI) and instrumental conditions, enabling a controlled comparison of system performance. Fringe solutions were successfully obtained for all antennas except E05 and C11 (Pol-1), W06 (Pol-2), and C04 when the Walsh was enabled, and hence these antennas were not considered further for both modes. Three antennas (C01, C10, and S03) were non-operational during these observations. 24 antennas were finally used in both the Walsh ON and OFF modes.

The long-term archival (LTA) data from the test observations were retrieved and converted to measurement set (MS) format. The dataset was then separated into two subsets corresponding to timestamps with Walsh ON and Walsh OFF. Both subsets were processed independently using an identical calibration and imaging pipeline \citep[see \href{https://github.com/astrouser-salm/radio-imaging}{Salmoli Ghosh's GitHub page;}][{\color{blue}Ghosh et al. 2026, under review}]{Ghosh2025b} using the Common Astronomy Software Applications package \citep[{\tt CASA};][]{McMullin2007}. The processing steps included RFI flagging, gain calibration, polarization calibration, application of calibration solutions, and imaging. In the following section, we present a comparative analysis of the results obtained before and after calibration for both Walsh ON and Walsh OFF datasets.

\section{Results}
\subsection{Raw visibilities}
\begin{figure}
\centering
\includegraphics[width=8cm]{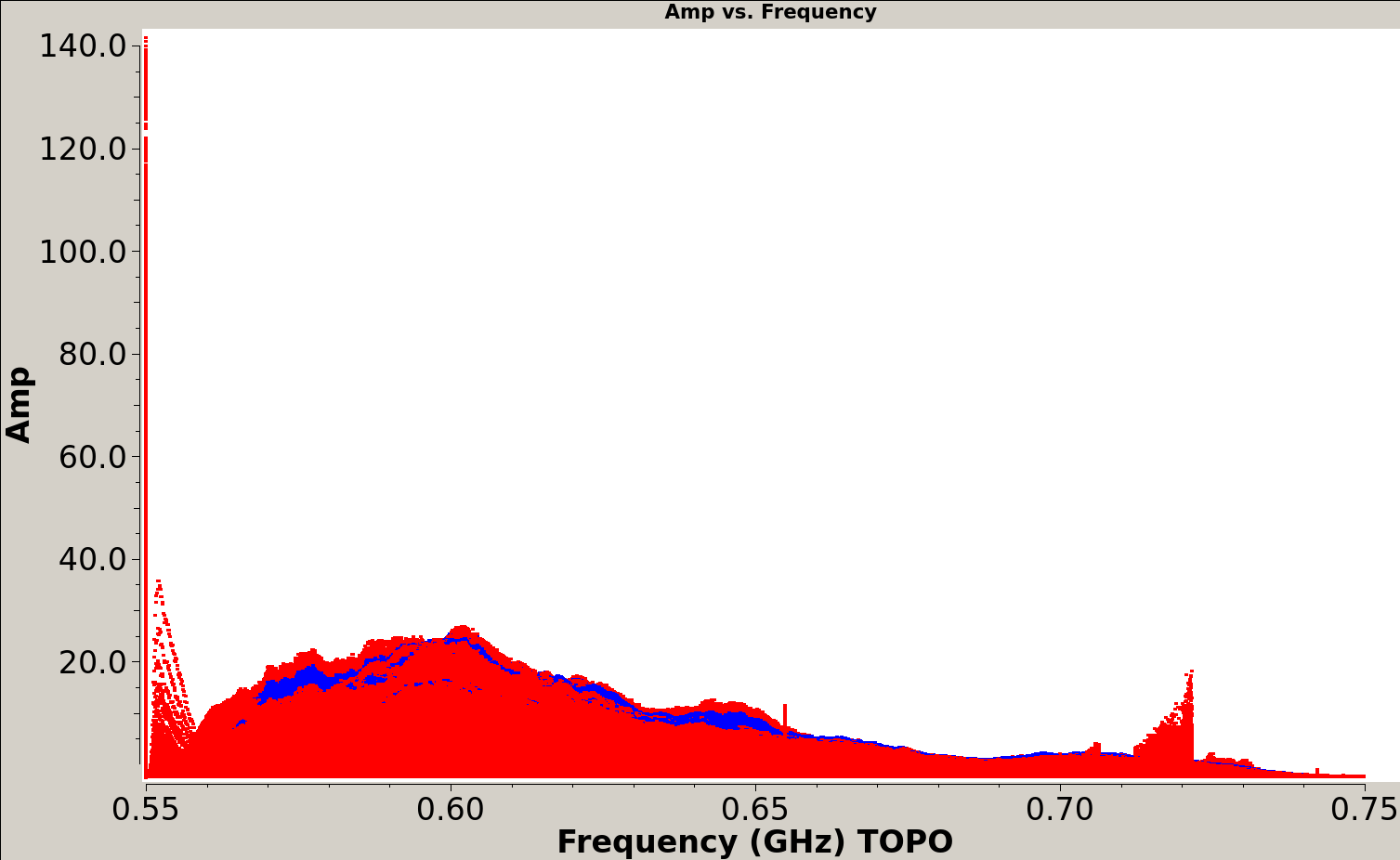}
\includegraphics[width=8cm]{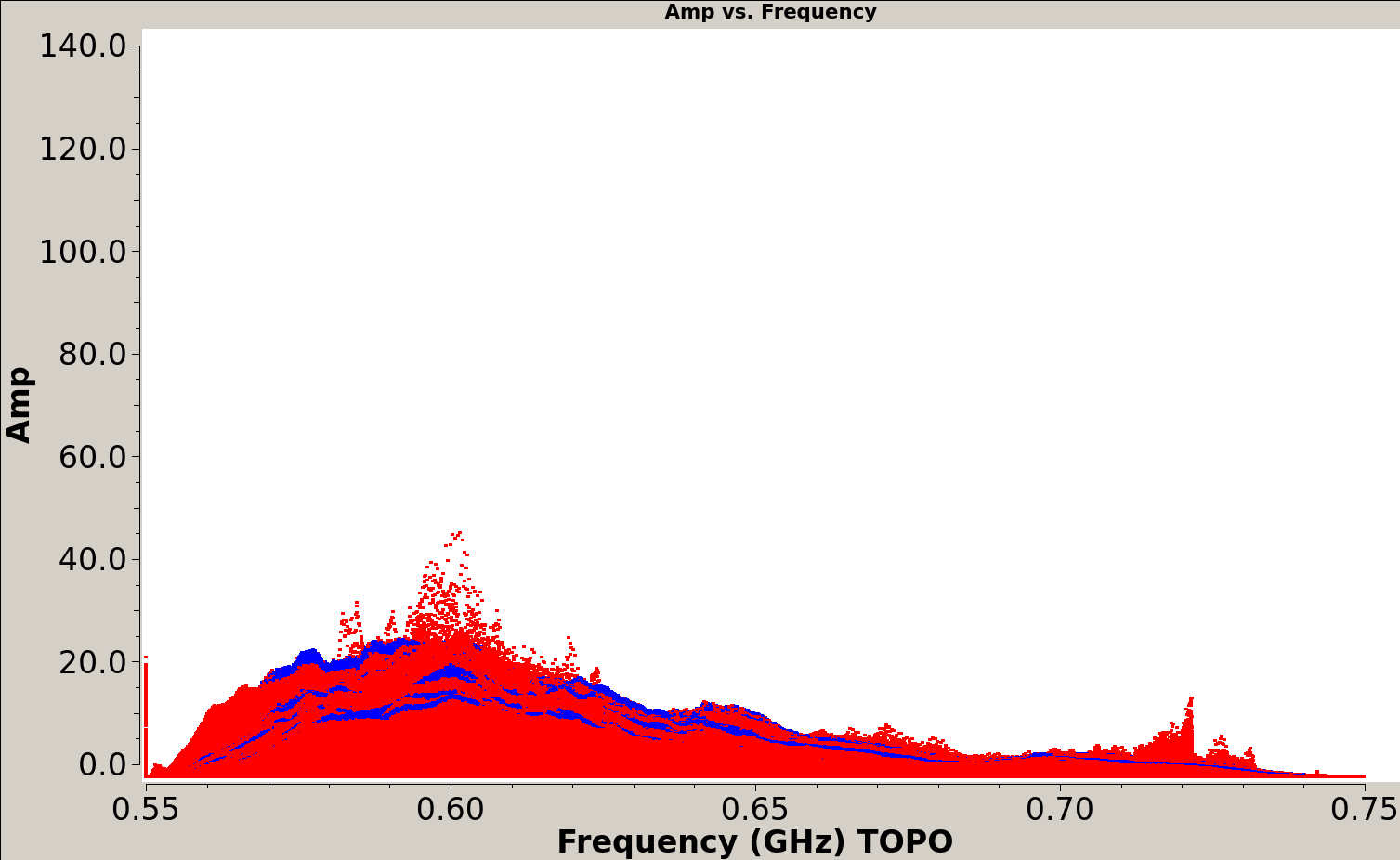}
\includegraphics[width=8cm]{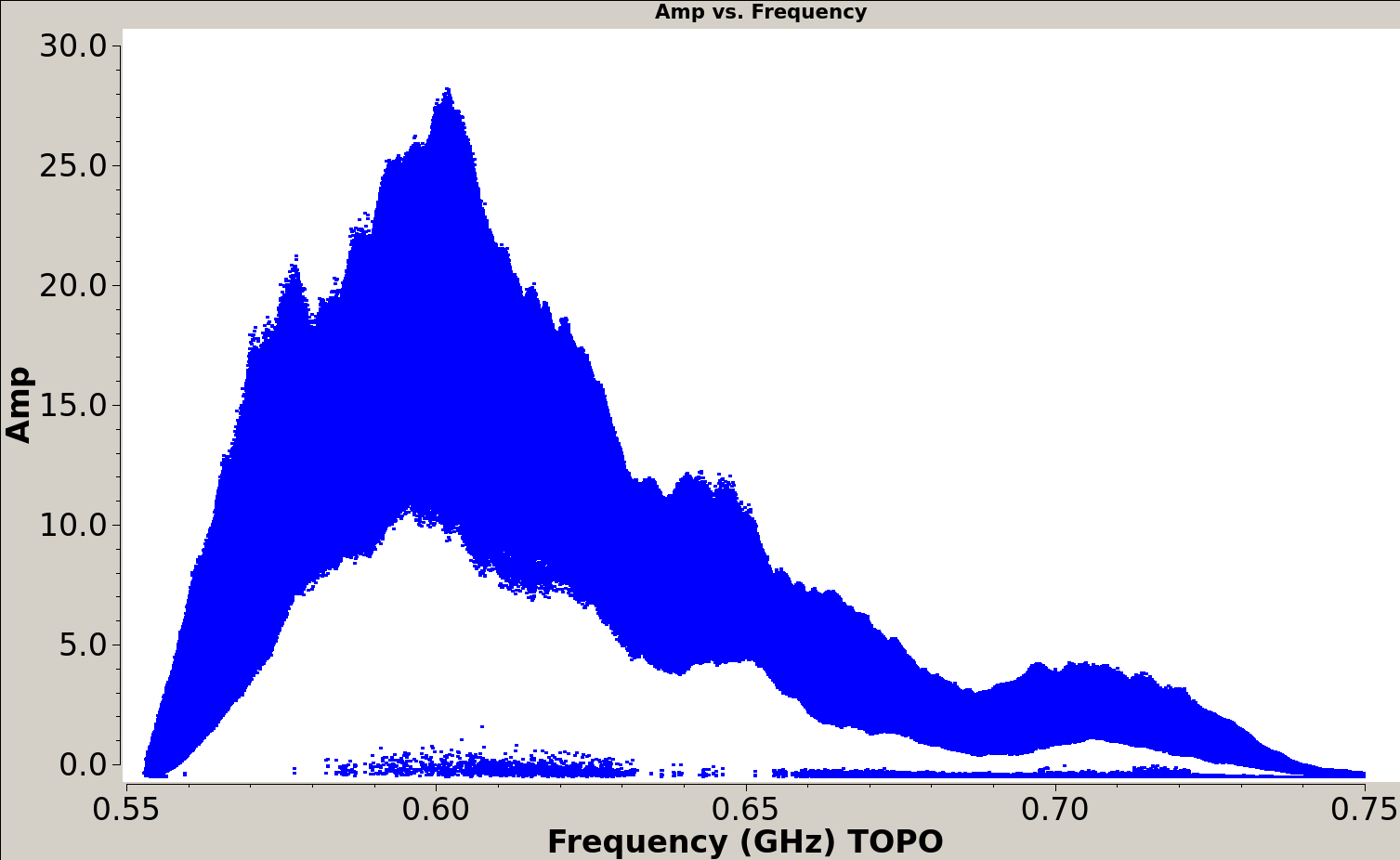}
\includegraphics[width=8cm]{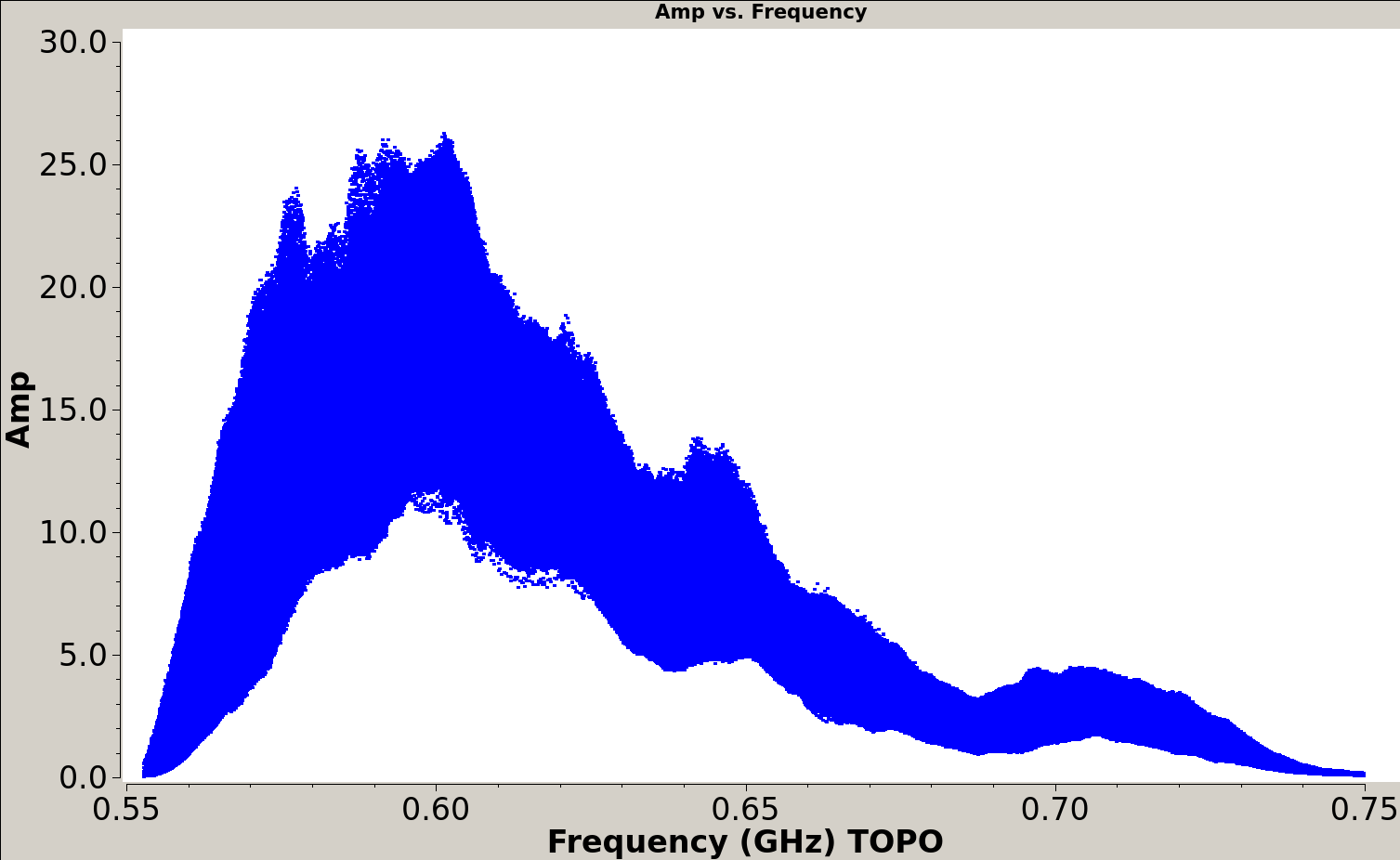}
\caption{\small Uncalibrated amplitude for the unpolarized calibrator 3C147 for a single scan (scan `5' on the left panel and scan `6' on the right) with respect to frequency ($0.55-0.75$~GHz) for the `RR' correlation with Walsh OFF (left) and ON (right) modes. The red points in the top panel mark the flagged data along with the blue points, which remain unflagged after running the flagging routines in our calibration pipeline. The bottom panel shows the uncalibrated data after removing the flagged data. The uncalibrated amplitude ranges shown here are from 0 to 140 in the top panel, 0 to 30 in the bottom panel, in arbitrary units.}
\label{fig:ampvsfreq3C147RR}
\end{figure}

\begin{figure}
\centering
\includegraphics[width=8cm]{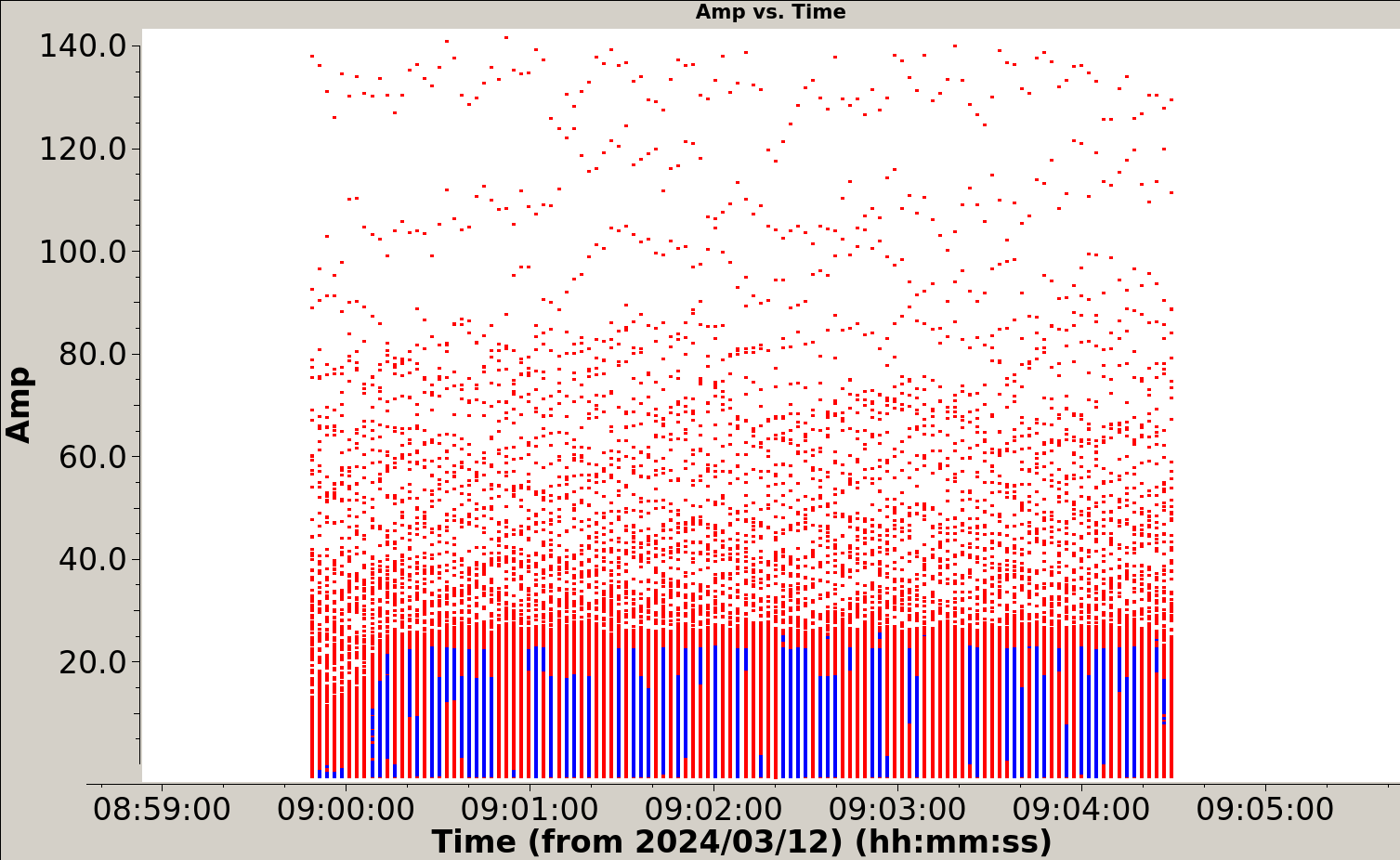}
\includegraphics[width=8cm]{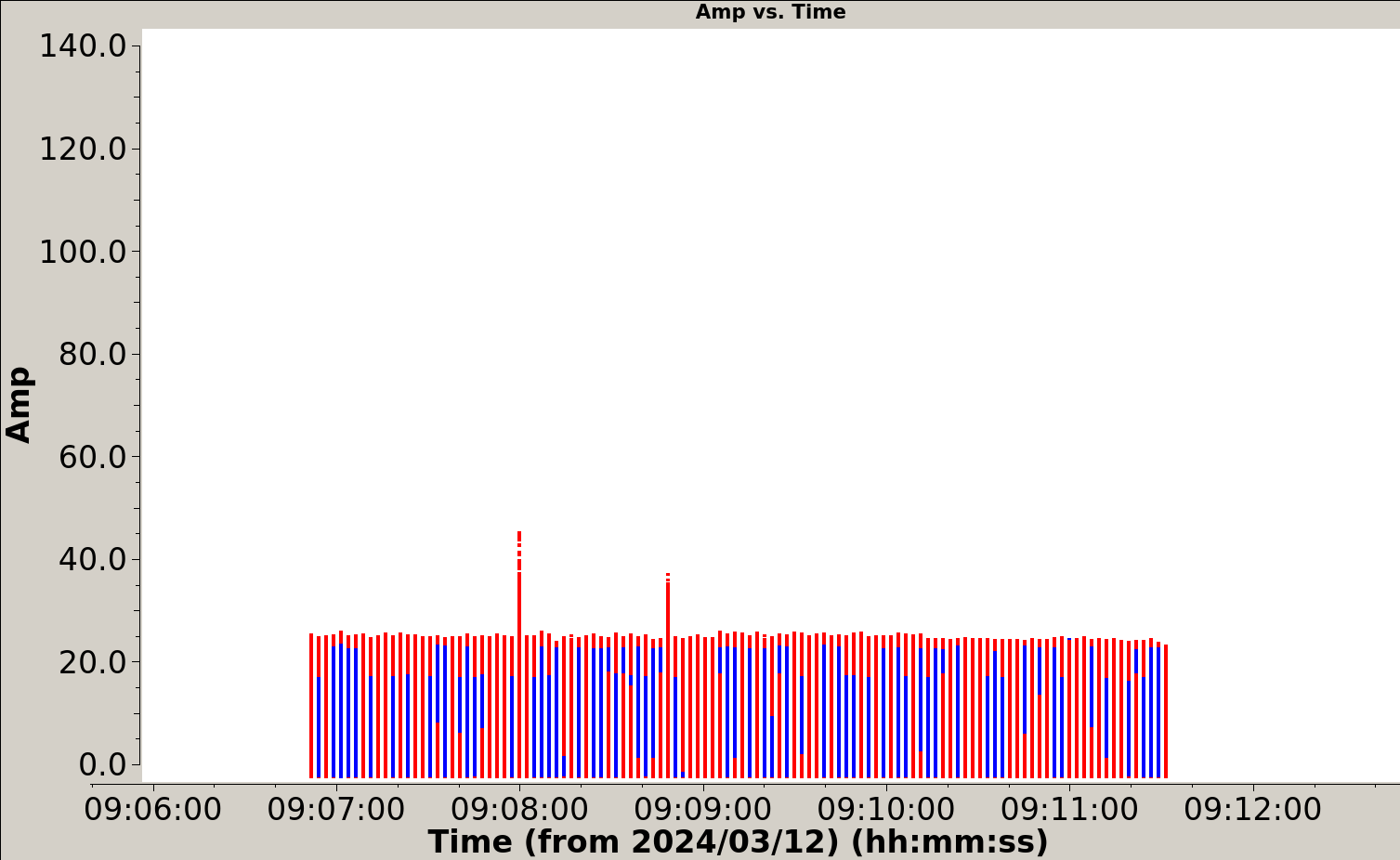}
\includegraphics[width=8cm]{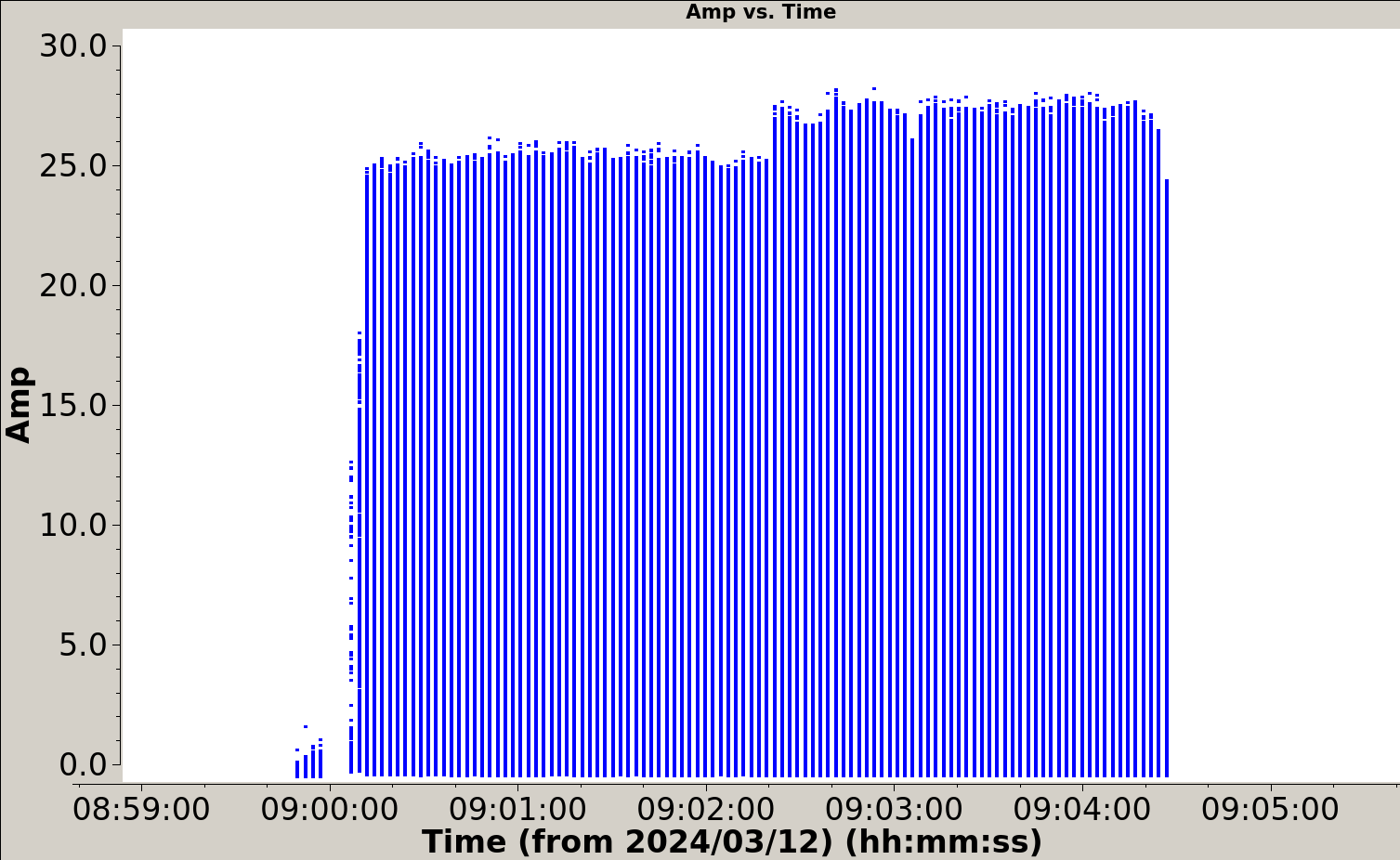}
\includegraphics[width=8cm]{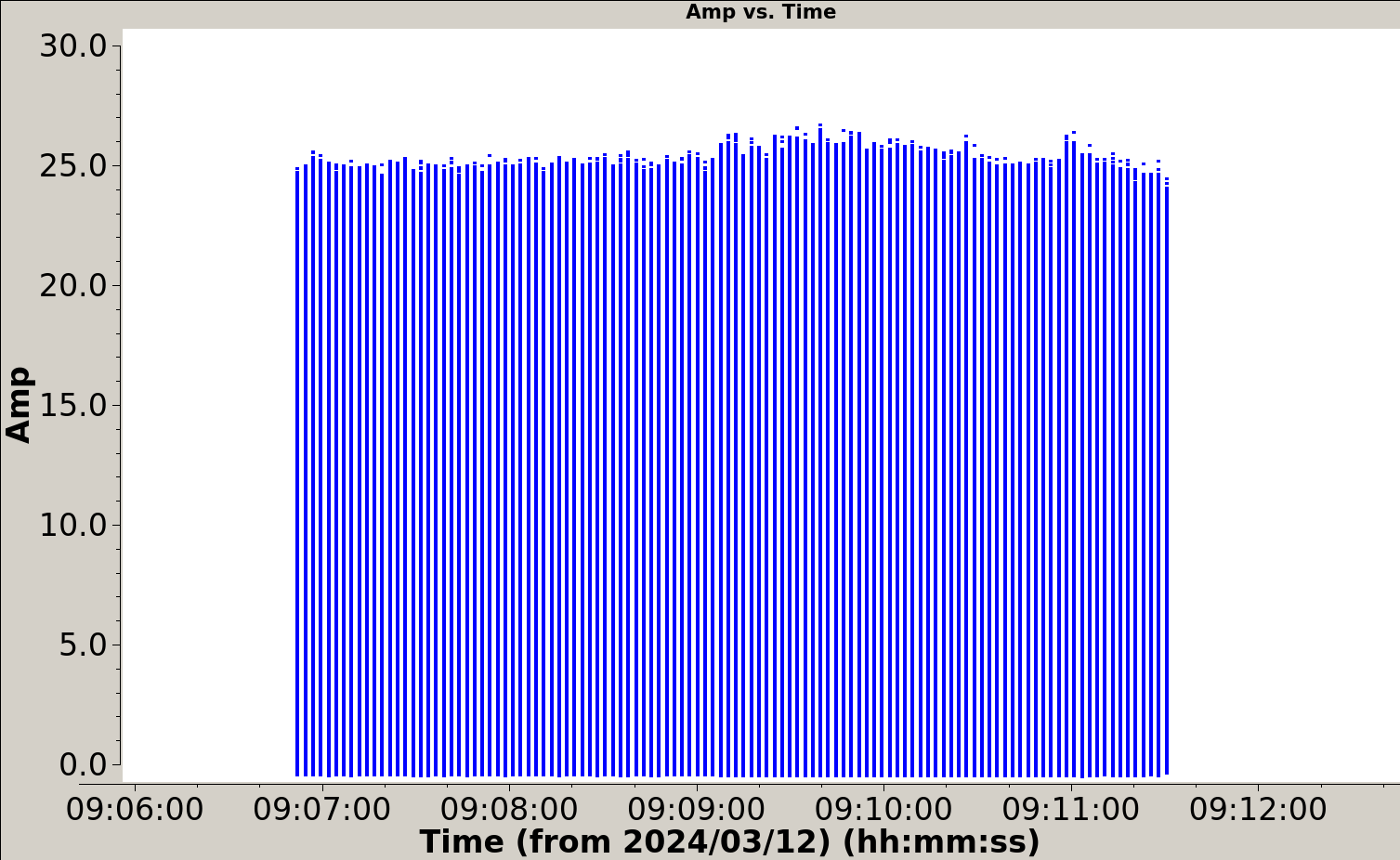}
\caption{\small Same as Figure \ref{fig:ampvsfreq3C147RR} but uncalibrated amplitudes plotted with respect to time. Data from a $\sim$4-min interval is shown.}
\label{fig:ampvstime3C147RR}
\end{figure}

The uncalibrated data show that the Walsh ON mode suppresses some of the high-amplitude features, particularly the narrow-band RFI patterns present in the visibilities. Figures \ref{fig:ampvsfreq3C147RR} and \ref{fig:ampvstime3C147RR} show the amplitude as a function of frequency and time, respectively, for the uncalibrated parallel-hand (RR) data, without (left) and with (right) the Walsh scheme. At this stage, we have passed the data through our initial flagging routines, including manual flagging (e.g., {\tt clip, quack}) and automated flagging (e.g., {\tt tfcrop, rflag}). Flagged\footnote{The recorded flag statistics summarise that in total $\sim62\%$ ($6.9\times10^9$ out of $1.1\times10^{10}$) and $\sim54\%$ ($6.1\times10^9$ out of $1.2\times10^{10}$) data were flagged for Walsh OFF and Walsh ON, respectively. In the Walsh OFF mode, 64\% and 58\% data were flagged for 3C138 and 3C147, respectively. In the Walsh ON mode, 56\% and 51\% data were flagged for 3C138 and 3C147, respectively.} and unflagged visibilities are shown in red and blue, respectively, demonstrating that our flagging procedures can effectively remove these high-amplitude outliers even in the absence of Walsh switching. However, Walsh phase switching provides an additional level of suppression by decorrelating unwanted correlated signals before correlation, thereby reducing their contribution to the measured visibilities. This can be particularly important for interference or cross-talk that is not readily distinguishable from the astronomical signal through post-correlation statistical flagging alone.

\subsection{Frequency dependent leakages}
\begin{figure}
    \centering
    \includegraphics[height=5cm]{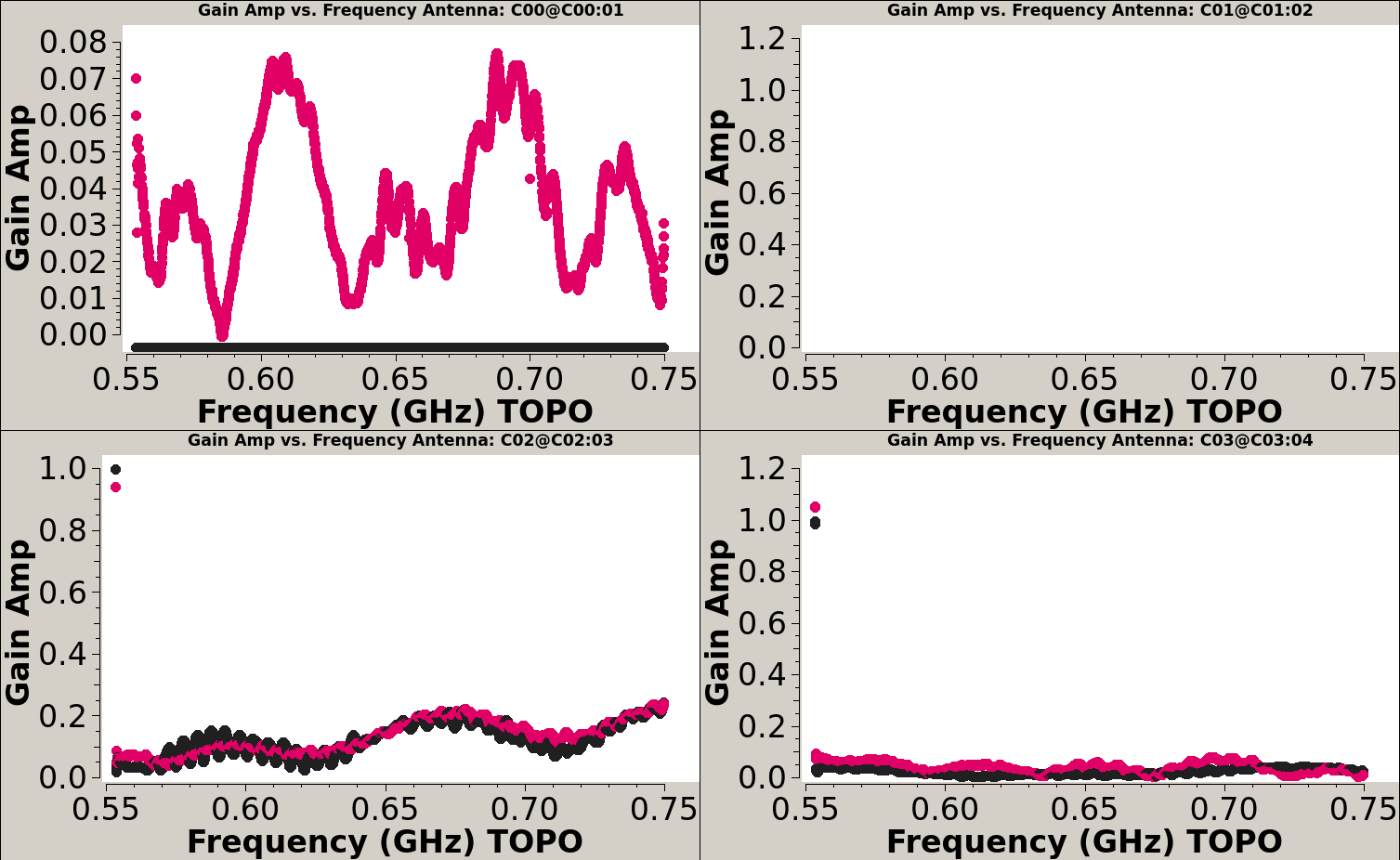}
    \hfill
    \includegraphics[height=5cm]{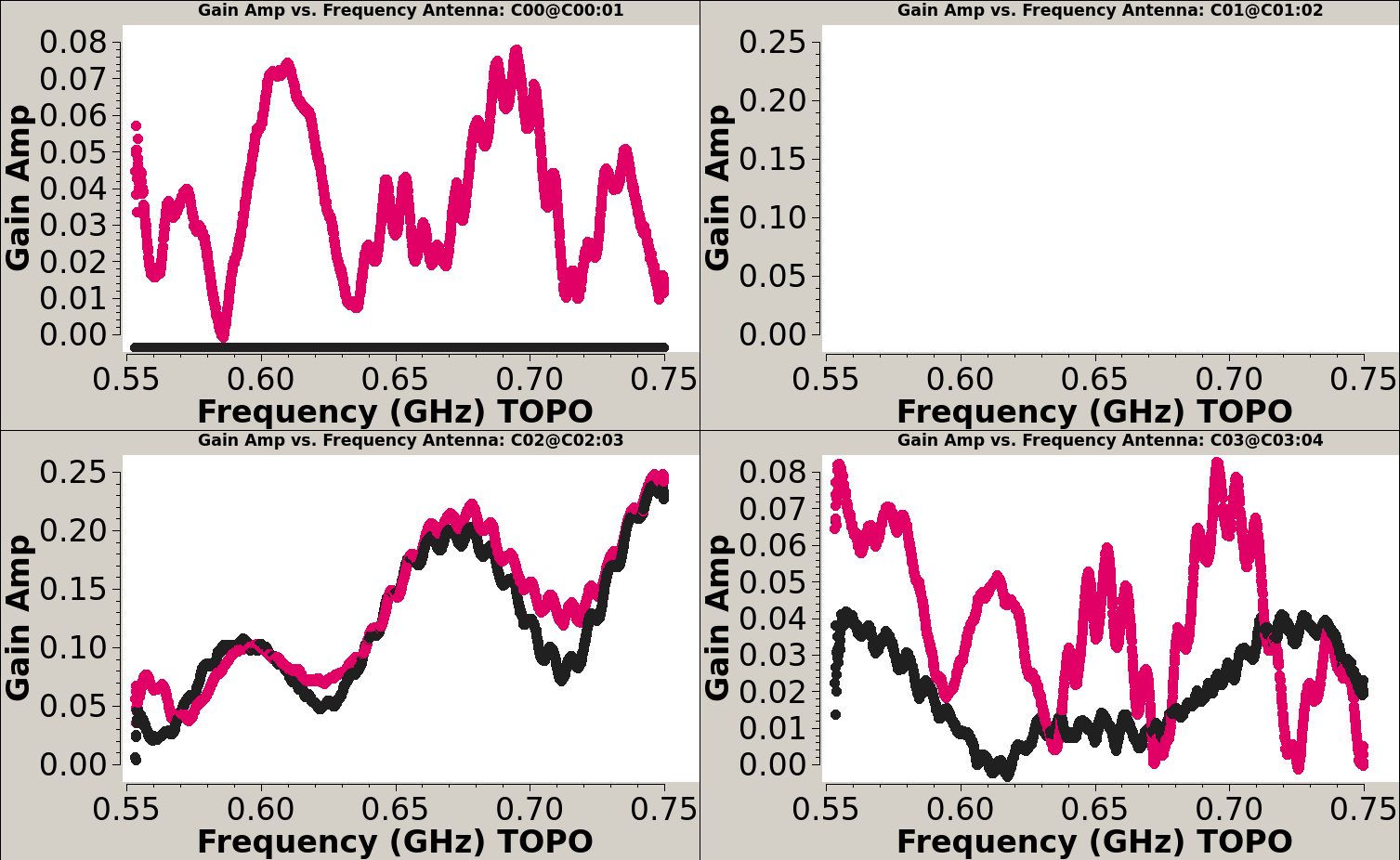}
    \includegraphics[height=5cm]{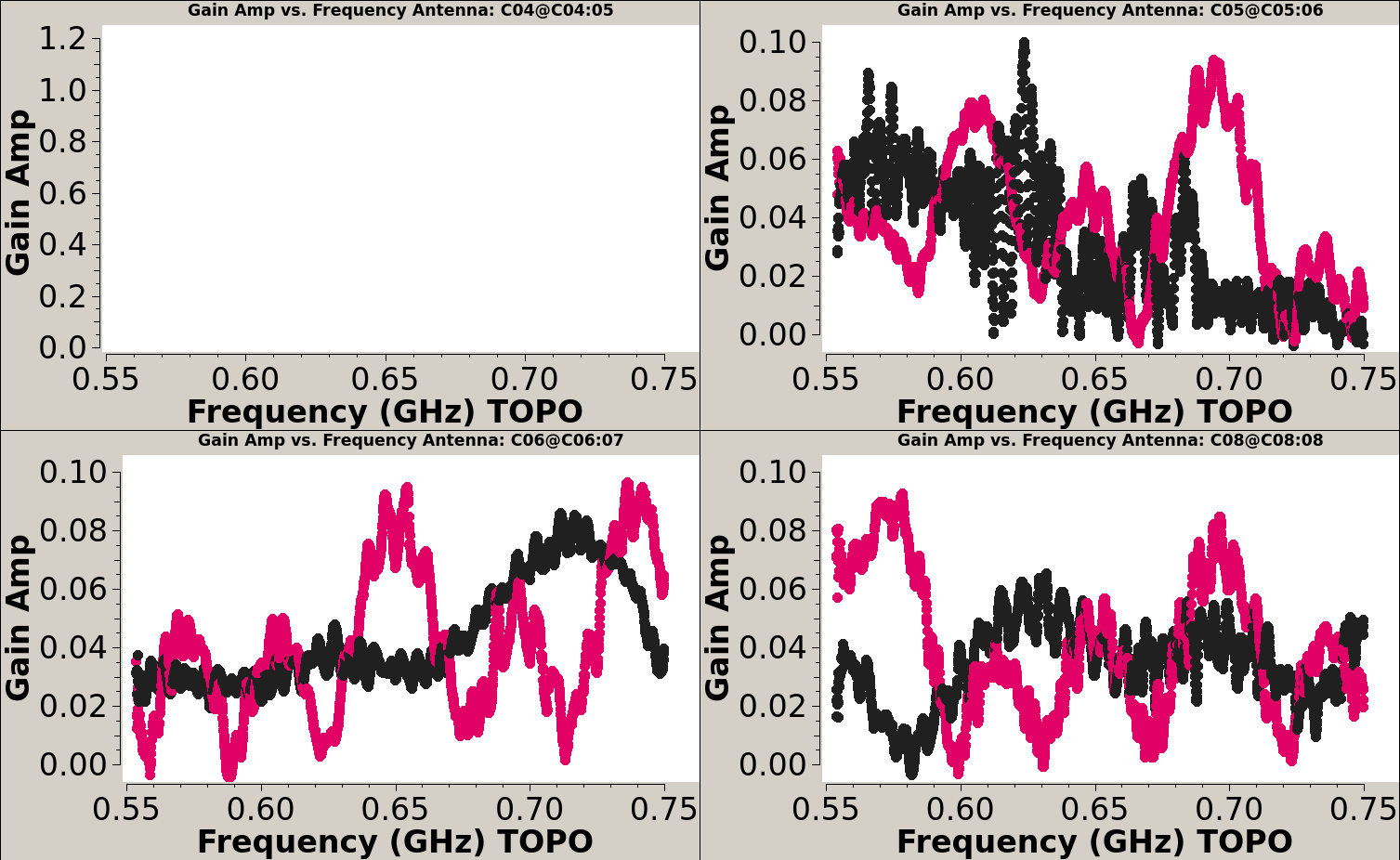}
     \hfill
    \includegraphics[height=5cm]{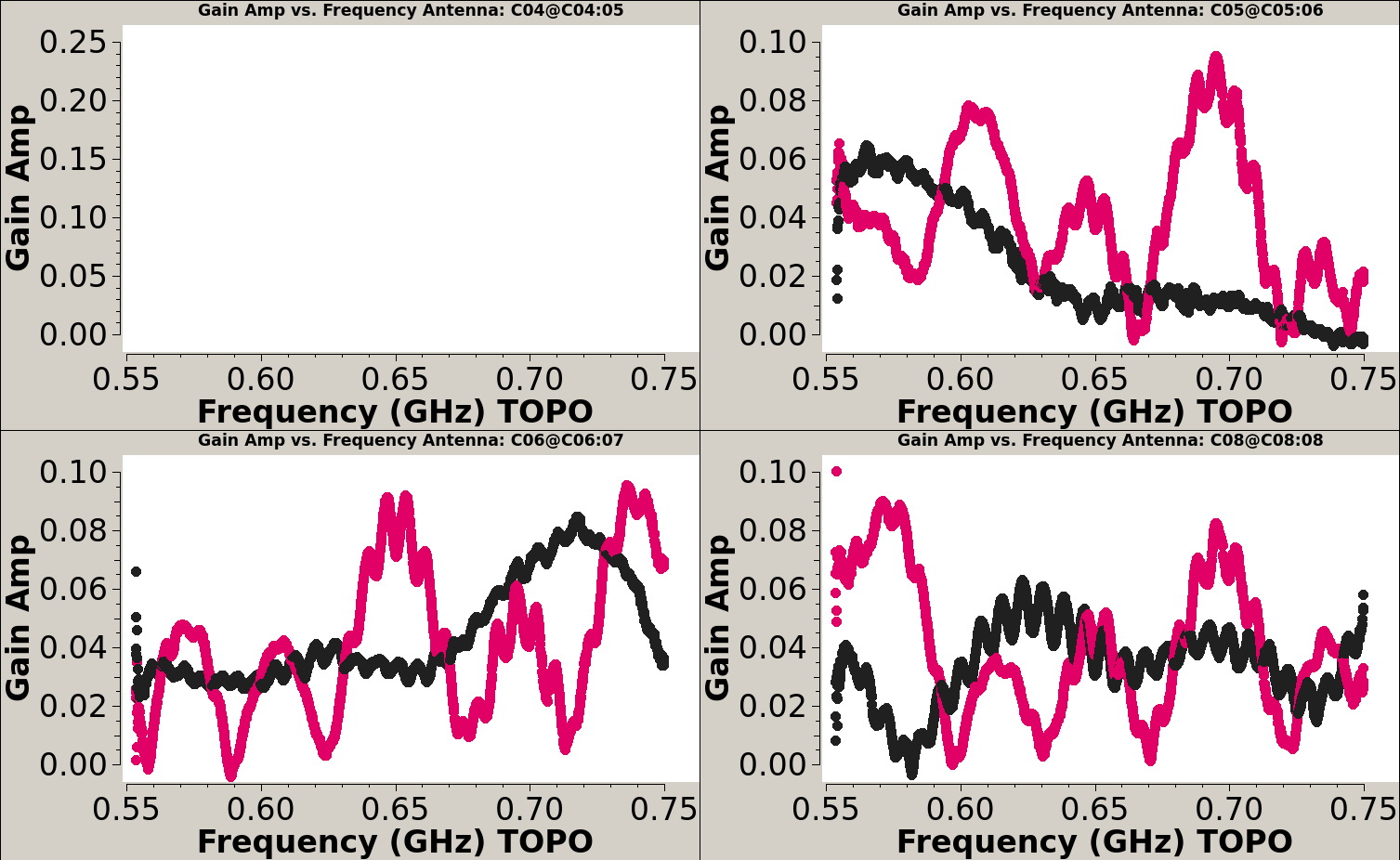}
    \includegraphics[height=5cm]{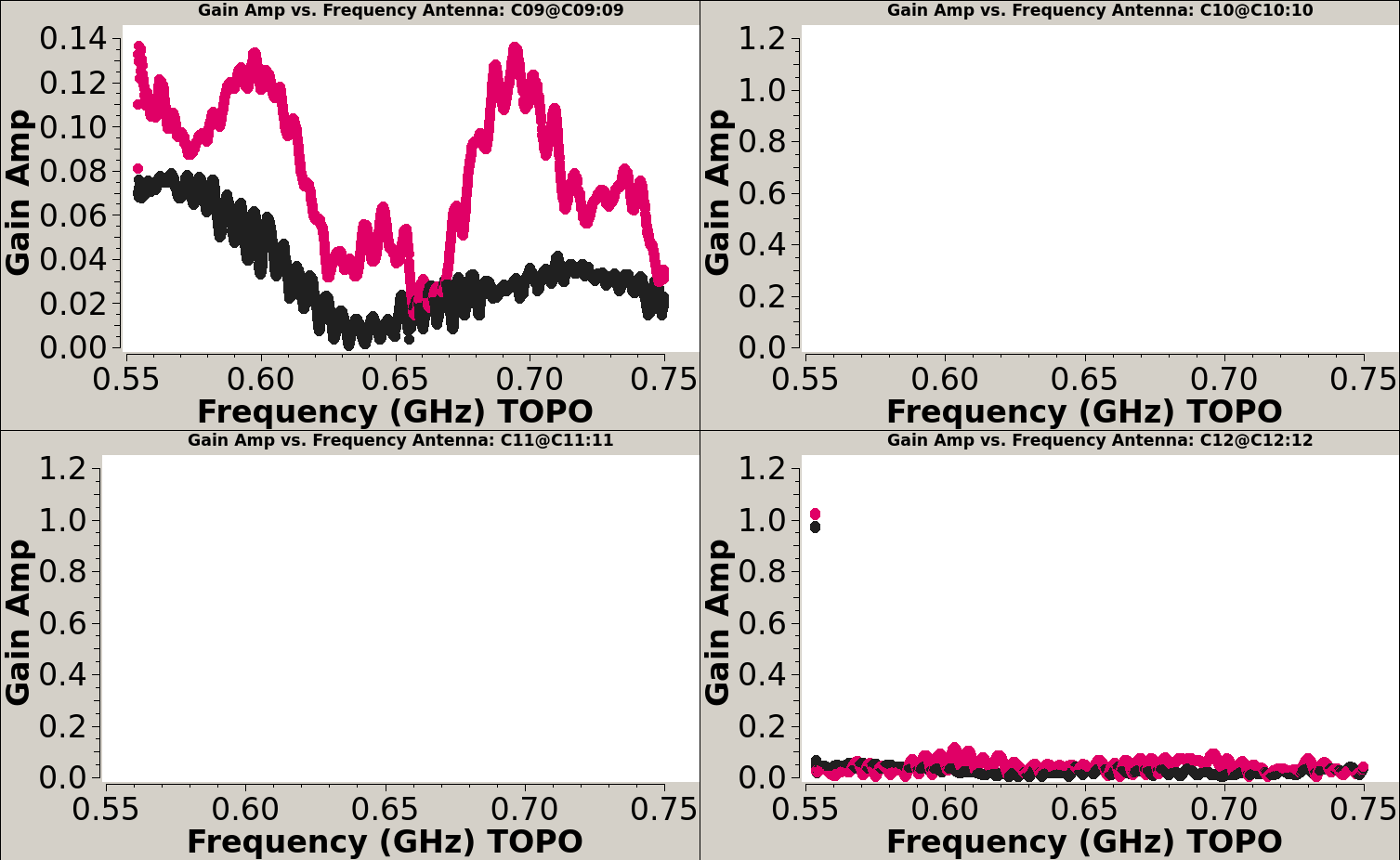}
     \hfill
    \includegraphics[height=5cm]{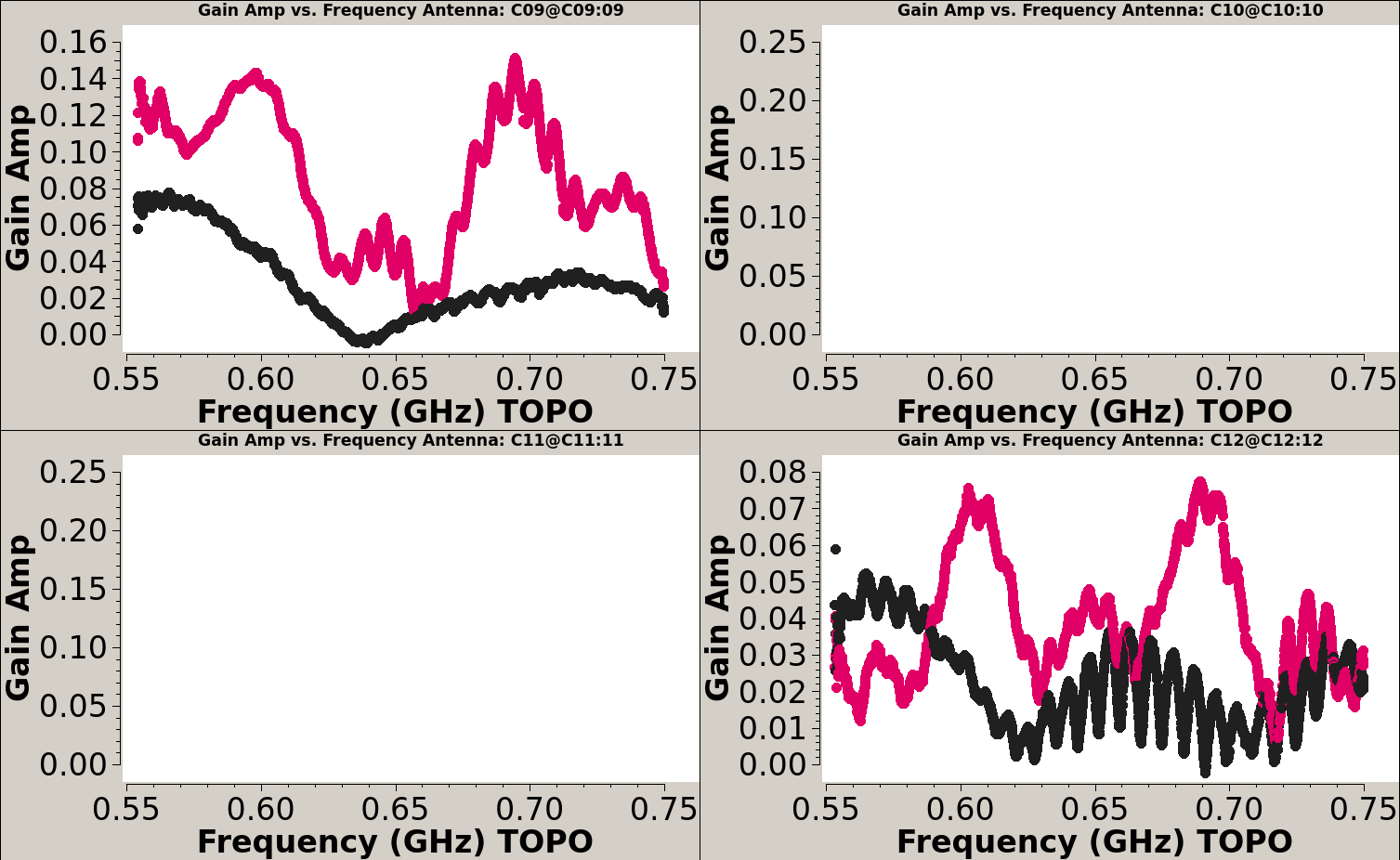}
    \includegraphics[height=5cm]{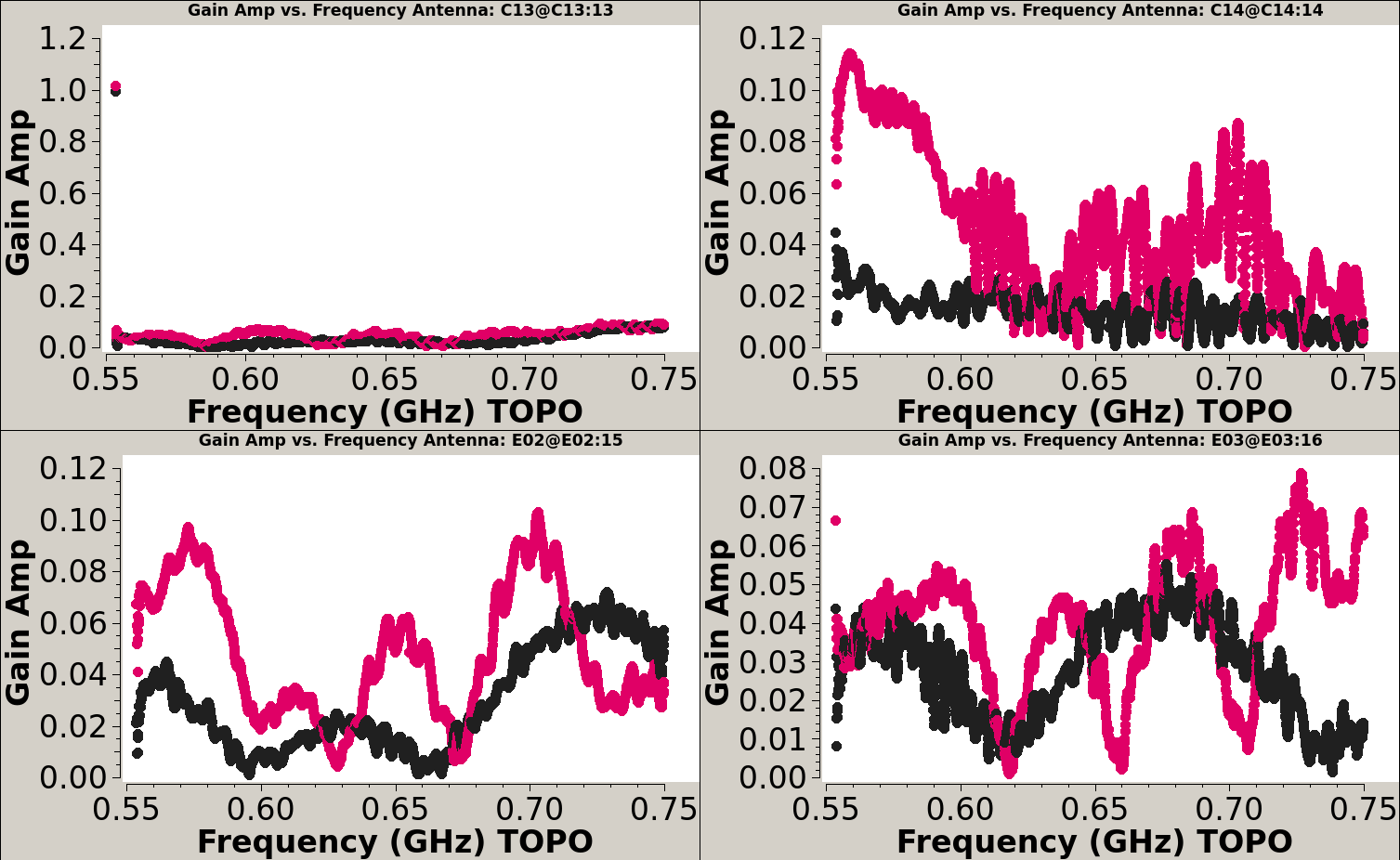}
     \hfill
    \includegraphics[height=5cm]{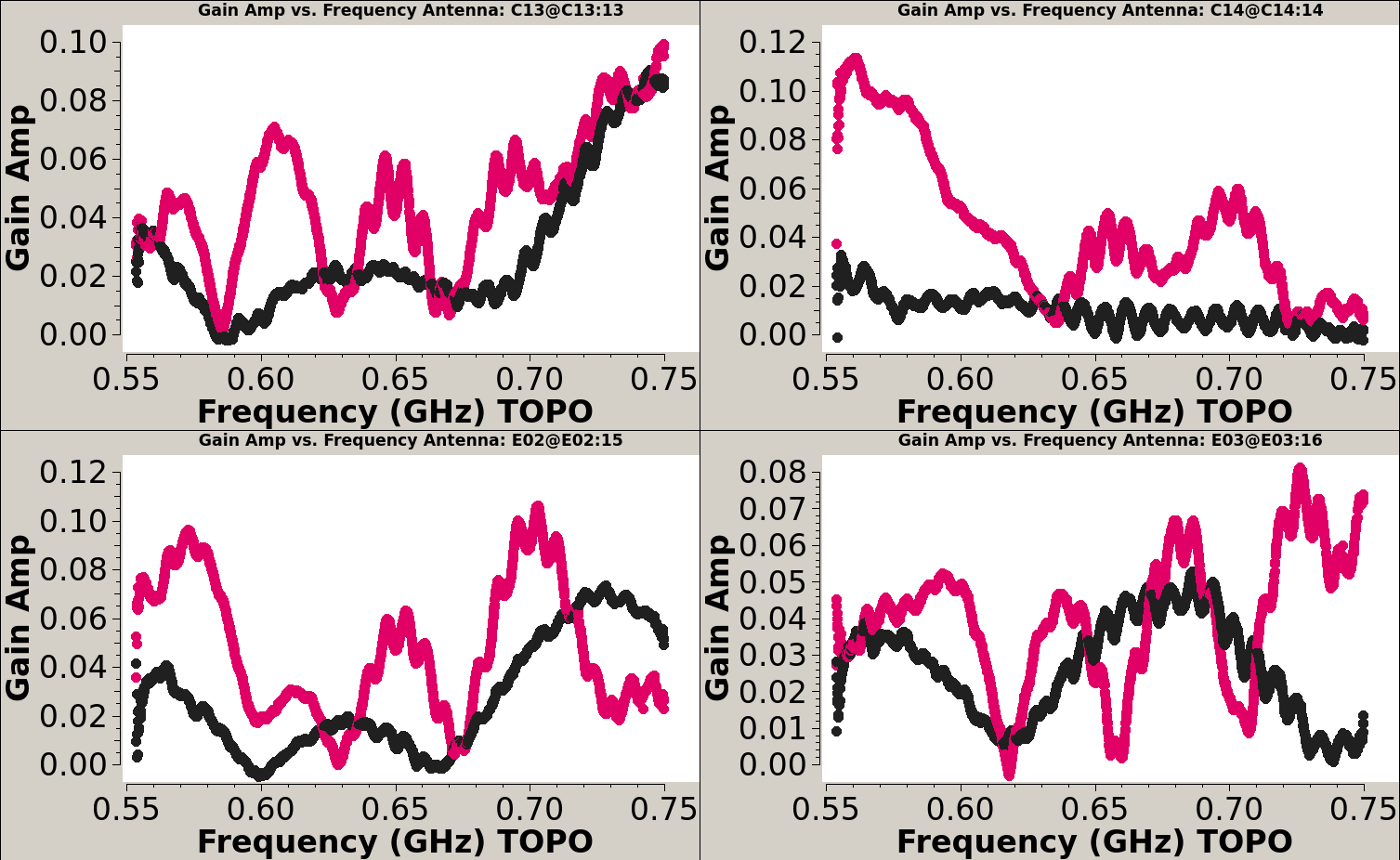}
\caption{\small Leakage amplitudes with respect to frequency (0.55$-$0.75~GHz) using the unpolarized calibrator 3C147 for antennas C00, C01, C02, C03, C04, C05, C06, C08, C09, C10, C11, C12, C13, C14, E02, E03 with Walsh OFF (left 2 columns) and Walsh ON (right 2 columns). The L and R correlations are coloured pink and black, respectively. C00 is the reference antenna for this experiment. The mean leakage amplitude for both Walsh OFF and Walsh ON is around 5-7\%, with the highest amplitudes going to 12$-$14\% at certain frequencies (exception: antenna C02). }
\label{fig:leakages}
\end{figure}

\begin{figure}    
    \includegraphics[height=5cm]{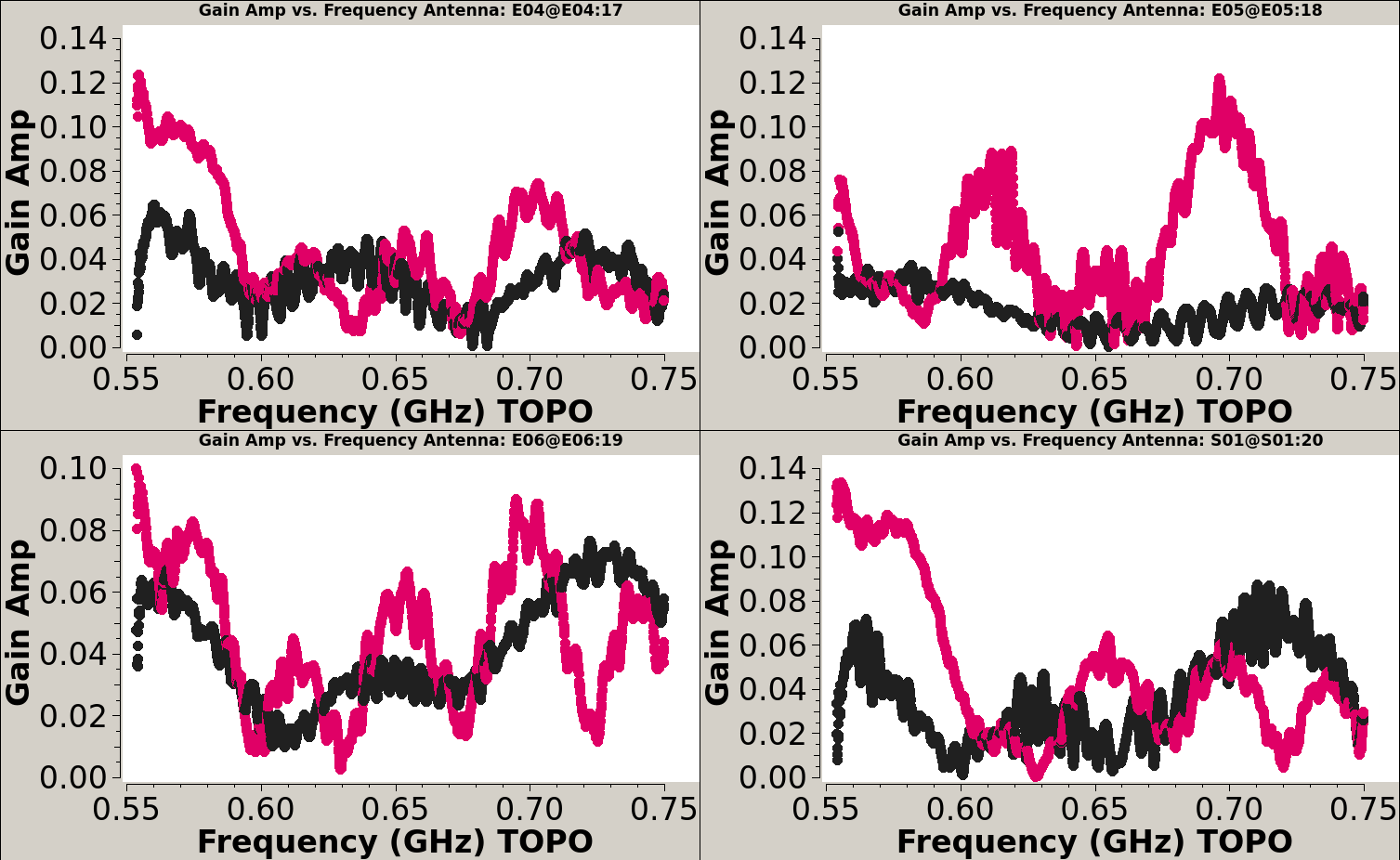}
     \hfill
    \includegraphics[height=5cm]{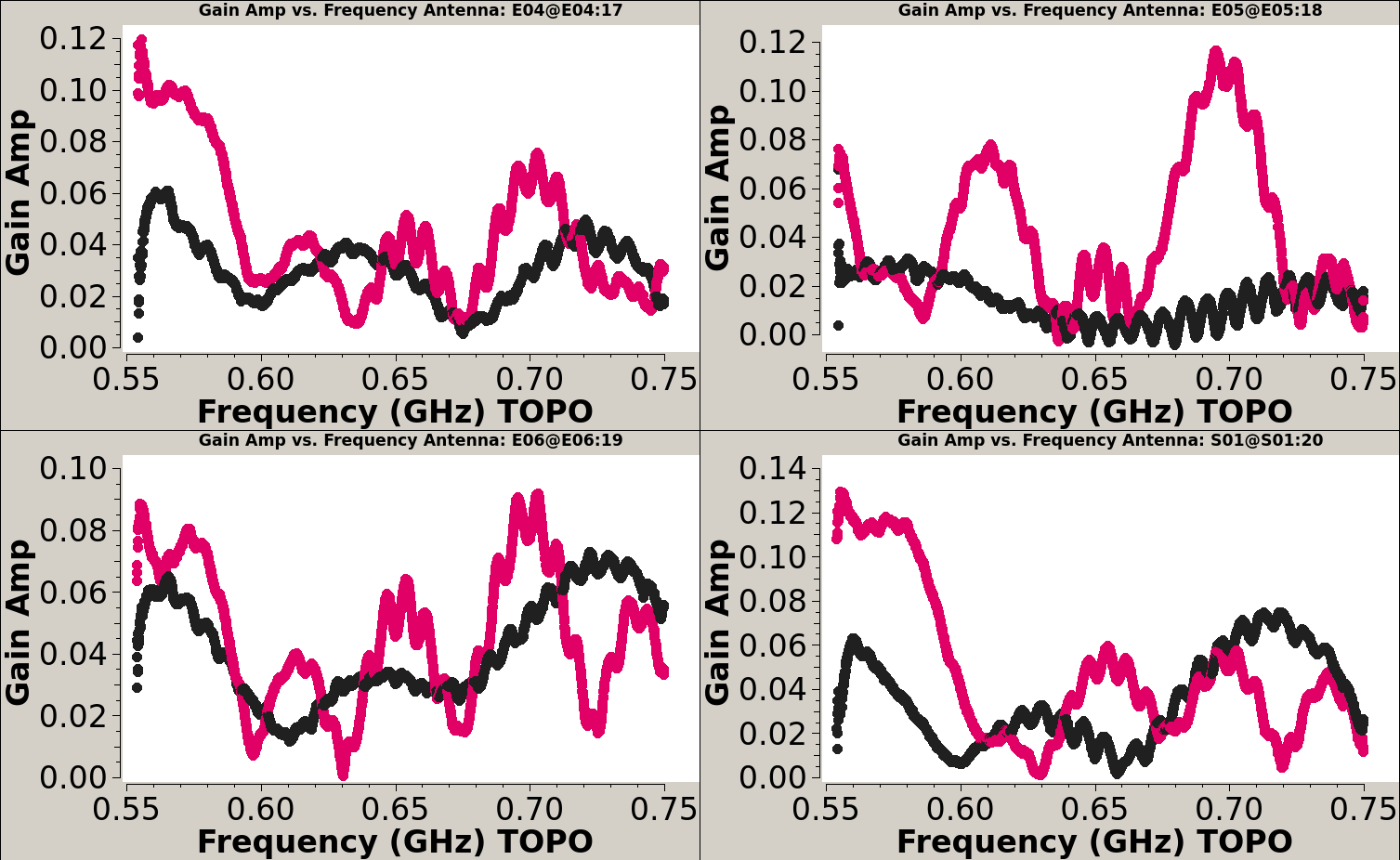}
    \includegraphics[height=5cm]{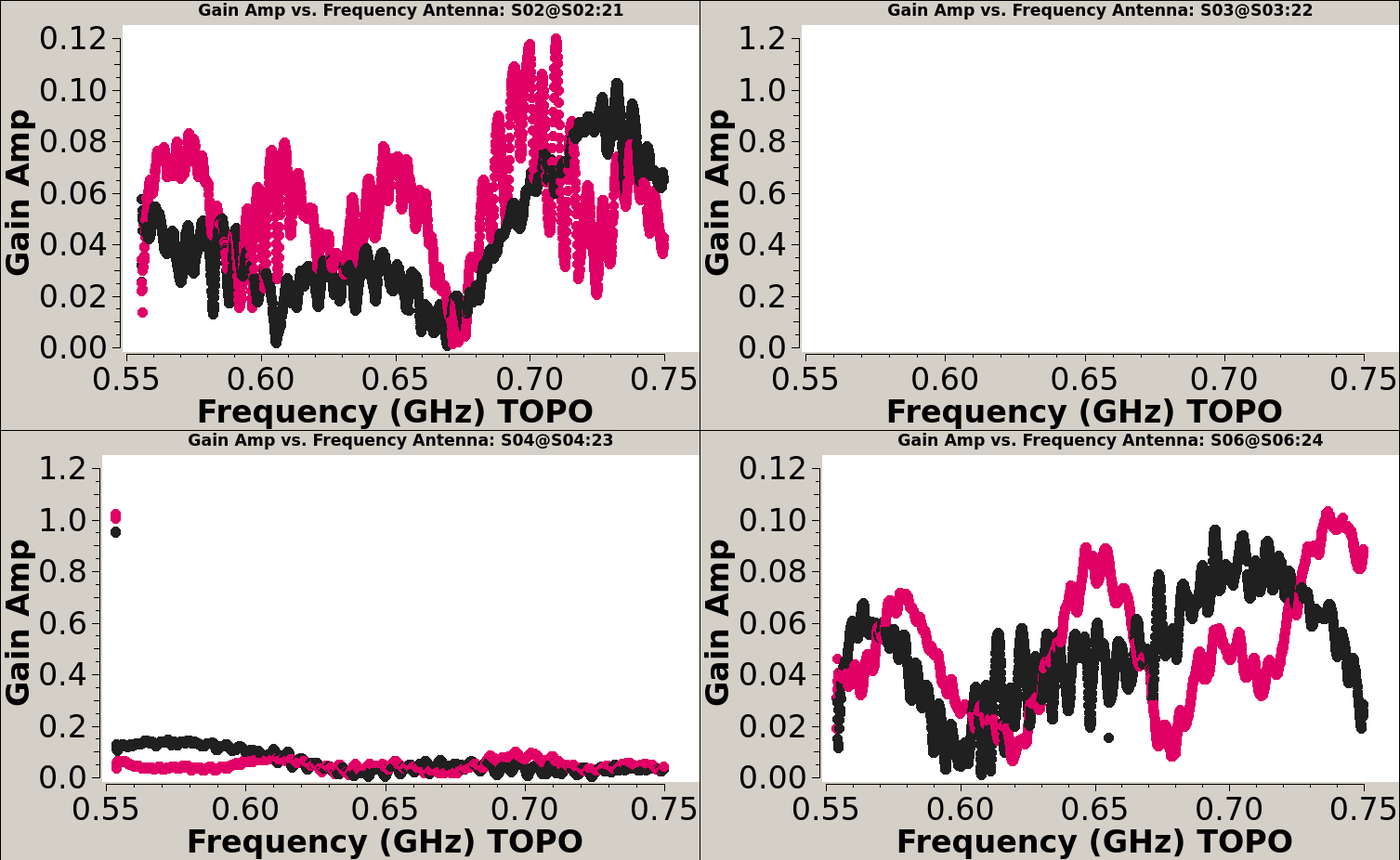}
     \hfill
    \includegraphics[height=5cm]{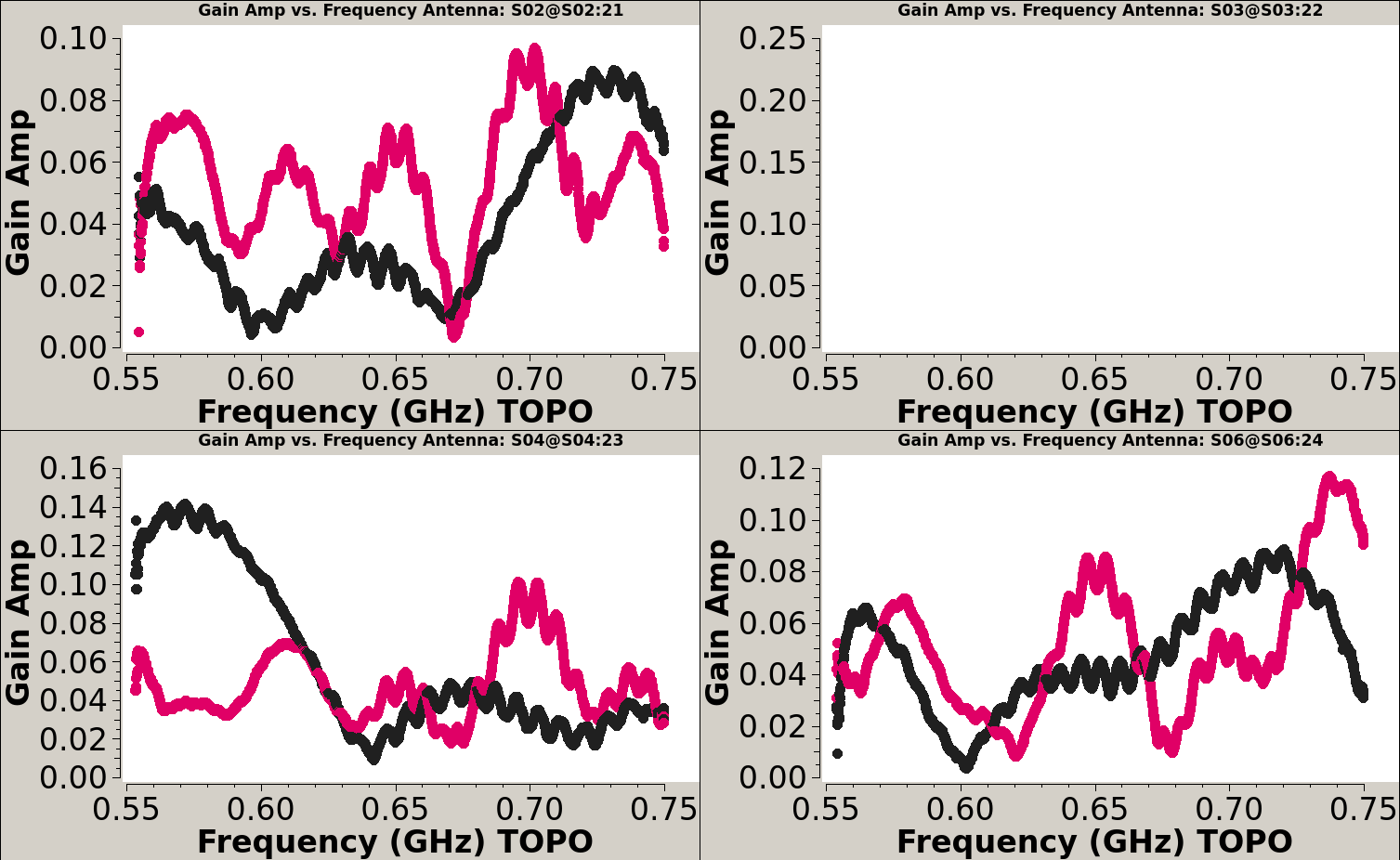}
    \includegraphics[height=5cm]{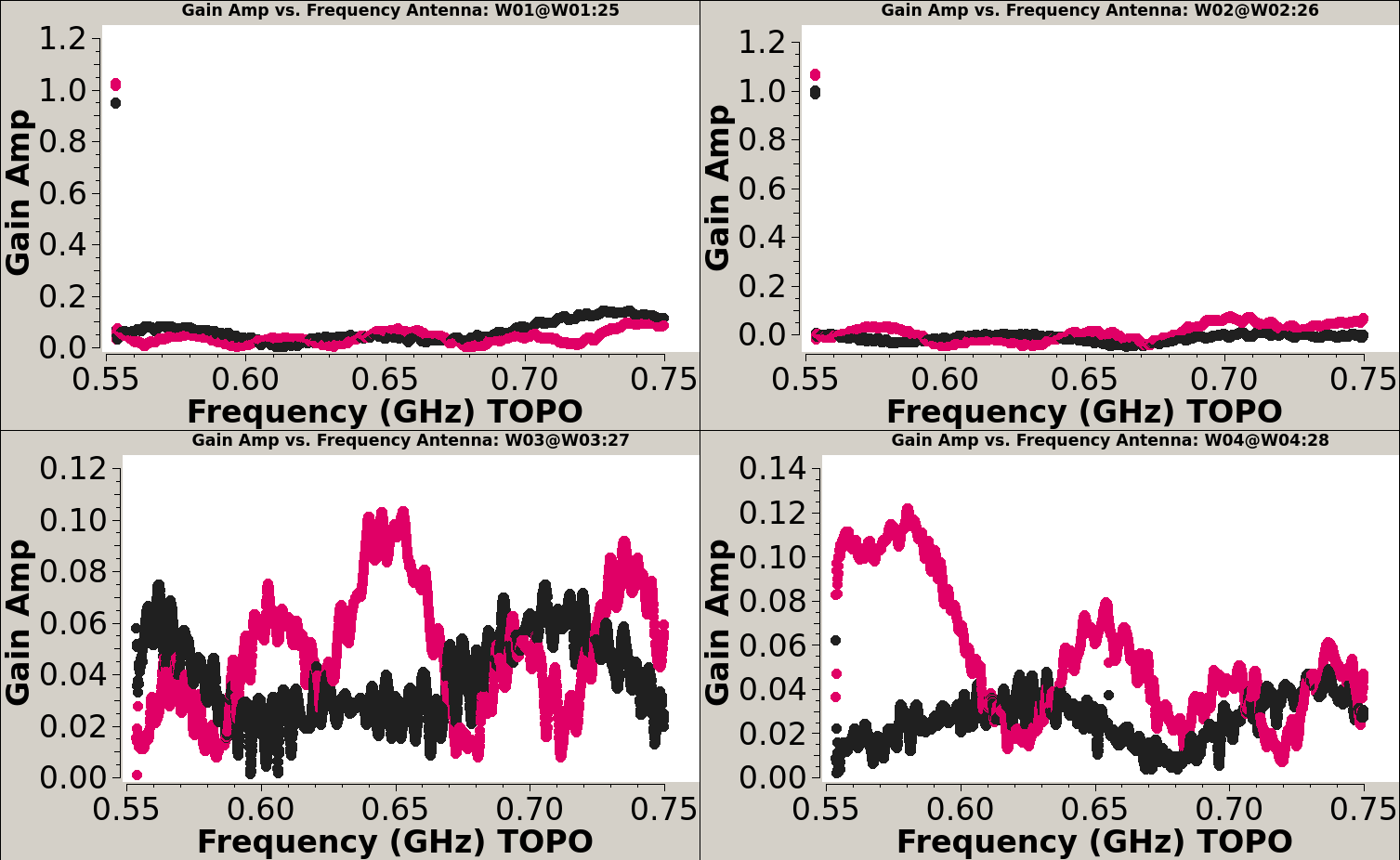}
     \hfill
    \includegraphics[height=5cm]{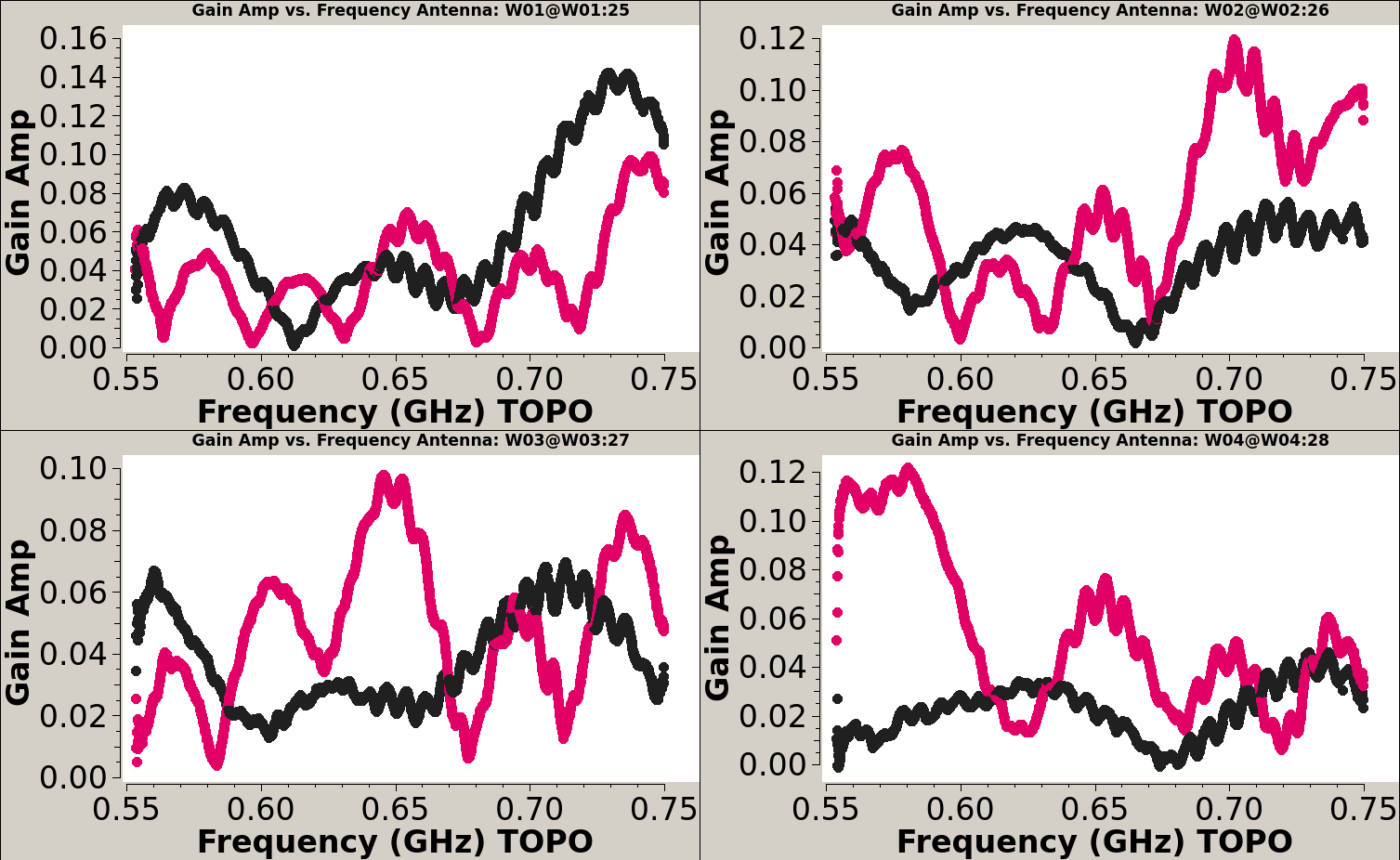}
    \includegraphics[height=2.5cm,trim=0 349 0 0,clip]{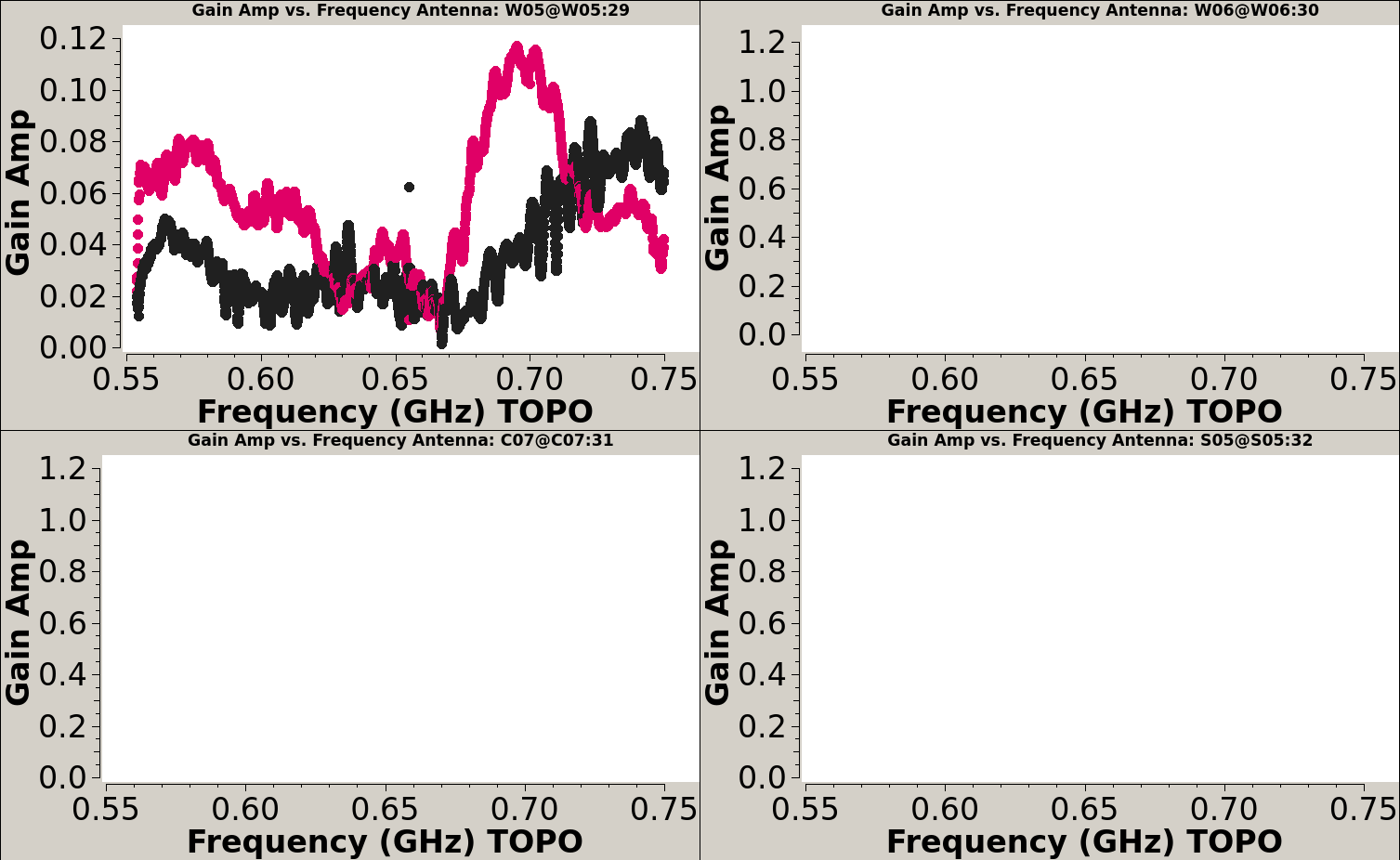}
     \hfill
    \includegraphics[height=2.5cm,trim=0 349 0 0,clip]{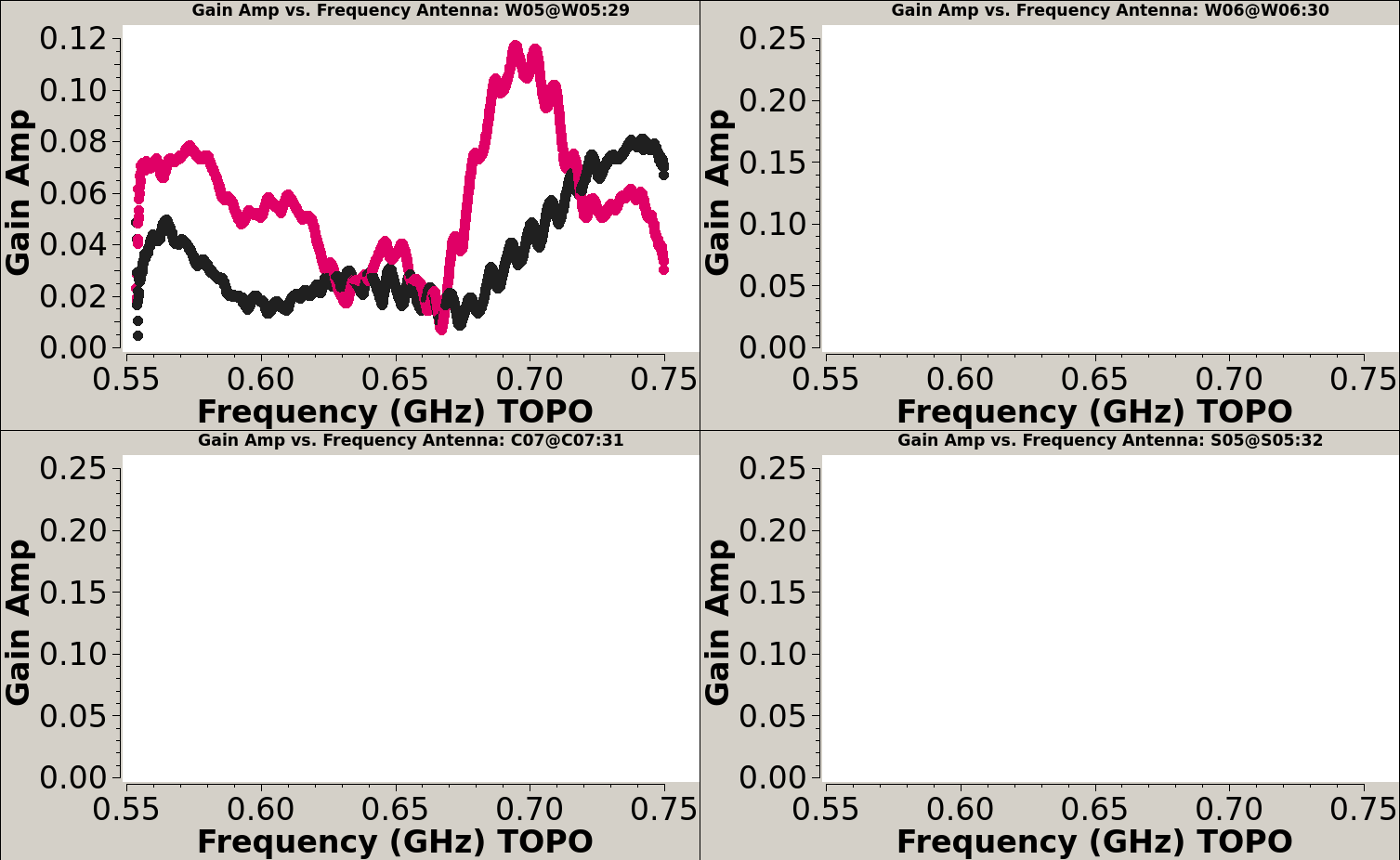}
\caption{\small Same as Figure \ref{fig:leakages} for antennas E04, E05, E06, S01, S02, S03, S04, S06, W01, W02, W03, W04, W05, W06. The mean leakage amplitude for both Walsh OFF and Walsh ON is around 5-7\%, with the highest amplitude going to 12$-$14\% at certain frequencies.}
    \label{fig:leakages_2}
\end{figure}

\begin{figure}
    \centering
    \setlength{\fboxsep}{0pt}
    \setlength{\fboxrule}{0.5pt}
    \fbox{\includegraphics[height= 9.8cm]{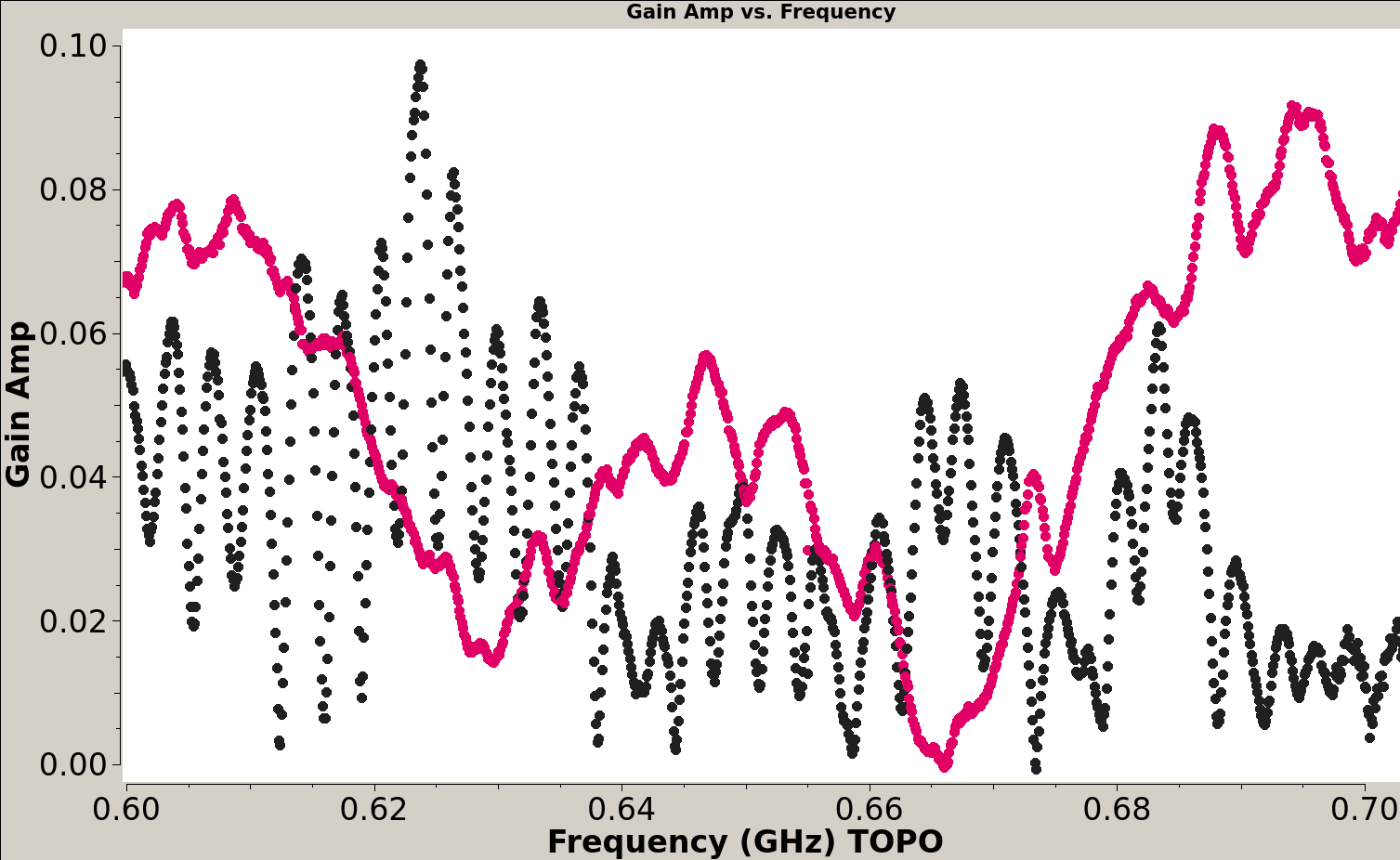}}
    \fbox{\includegraphics[height= 9.8cm]{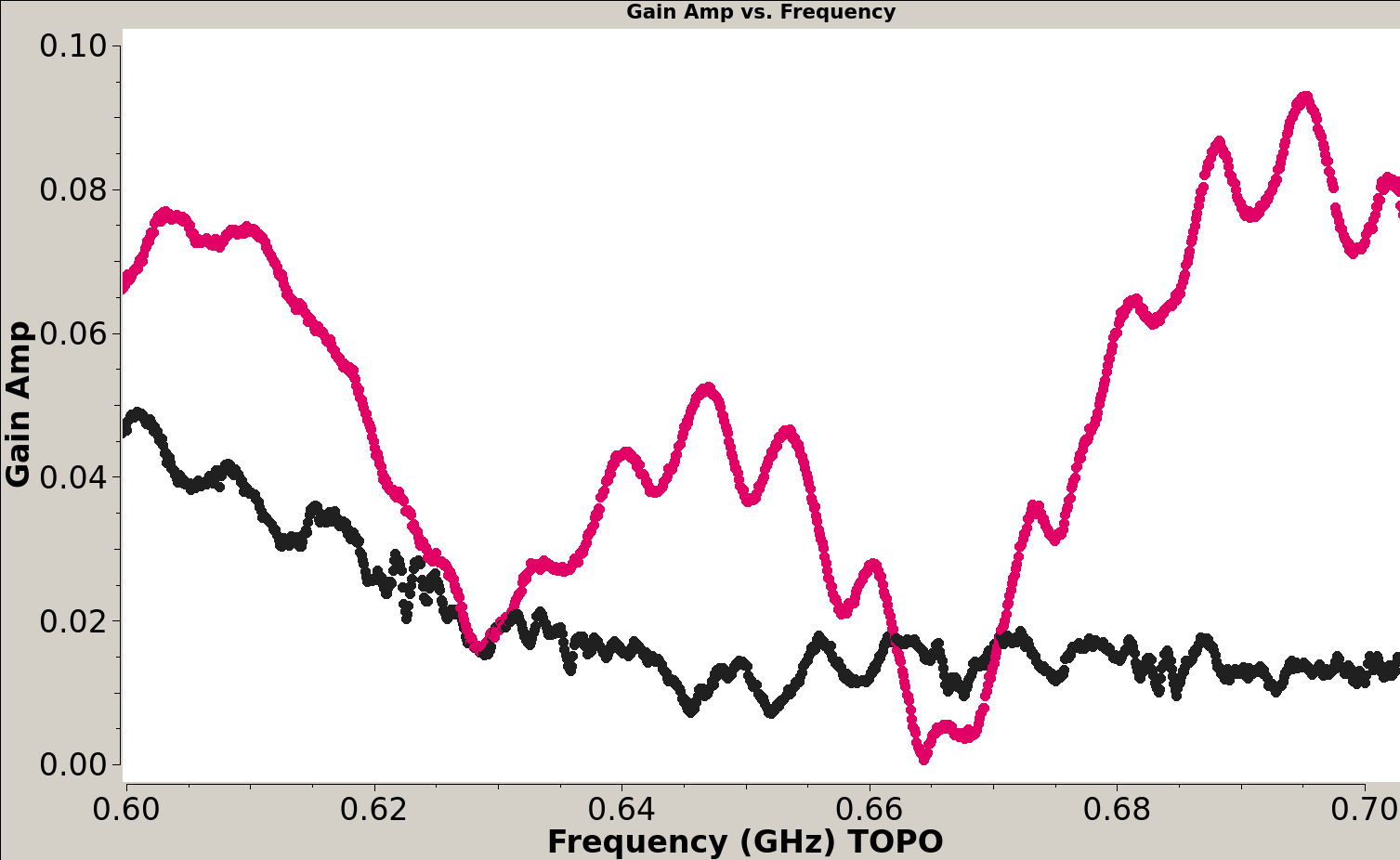}}
\caption{\small A zoomed-in view of the leakage amplitudes using the unpolarized calibrator 3C147 for antenna C05 in Walsh OFF (top) and Walsh ON (bottom) mode, with leakage amplitudes ranging from 0 to 10\% over the frequency range 0.6$-$0.7~GHz.}
    \label{fig:leakge_C05}
\end{figure}

\begin{figure}
    \centering
    \setlength{\fboxsep}{0pt}
    \setlength{\fboxrule}{0.5pt}
    \fbox{\includegraphics[height= 9.8cm]{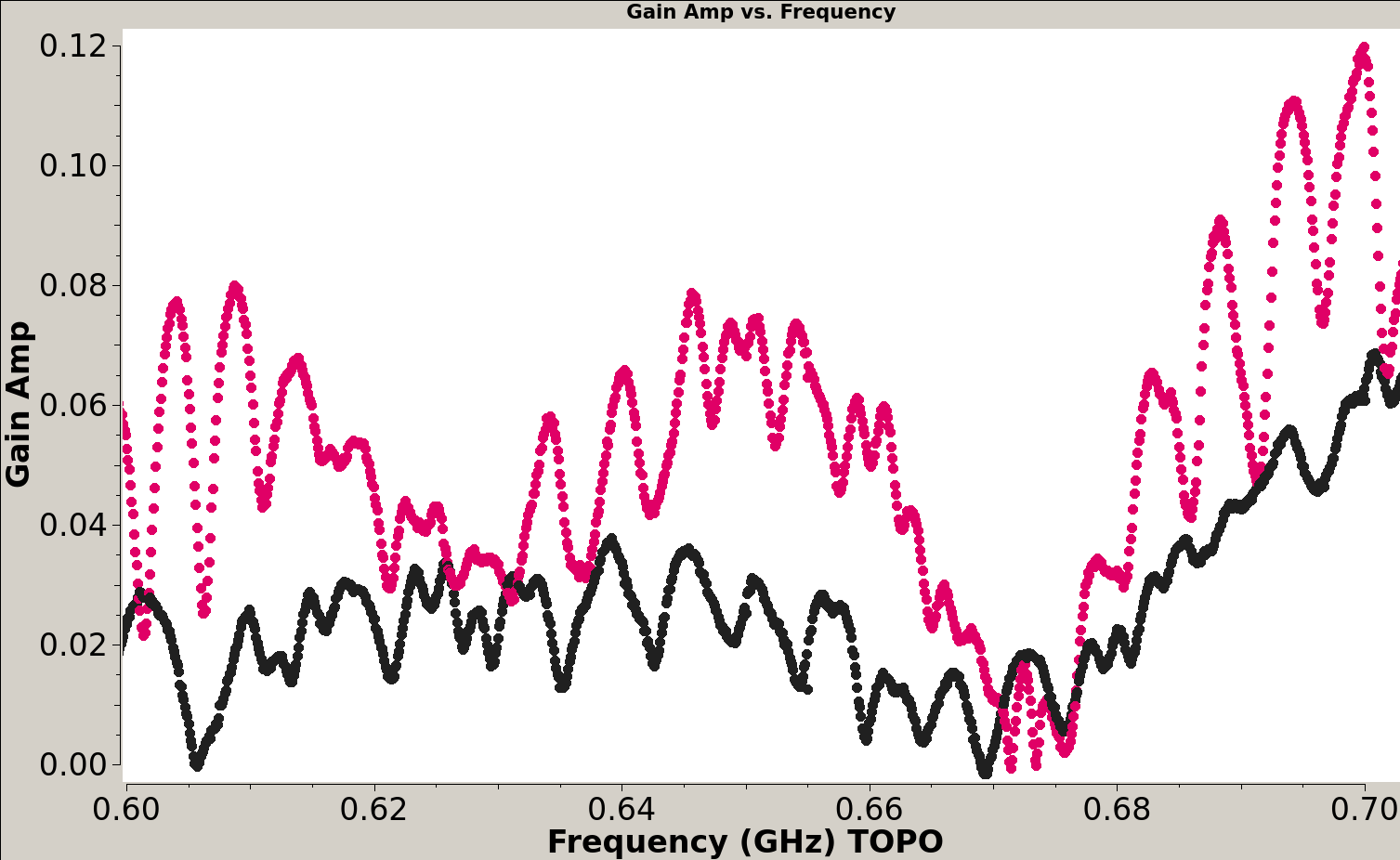}}
    \fbox{\includegraphics[height= 9.8cm]{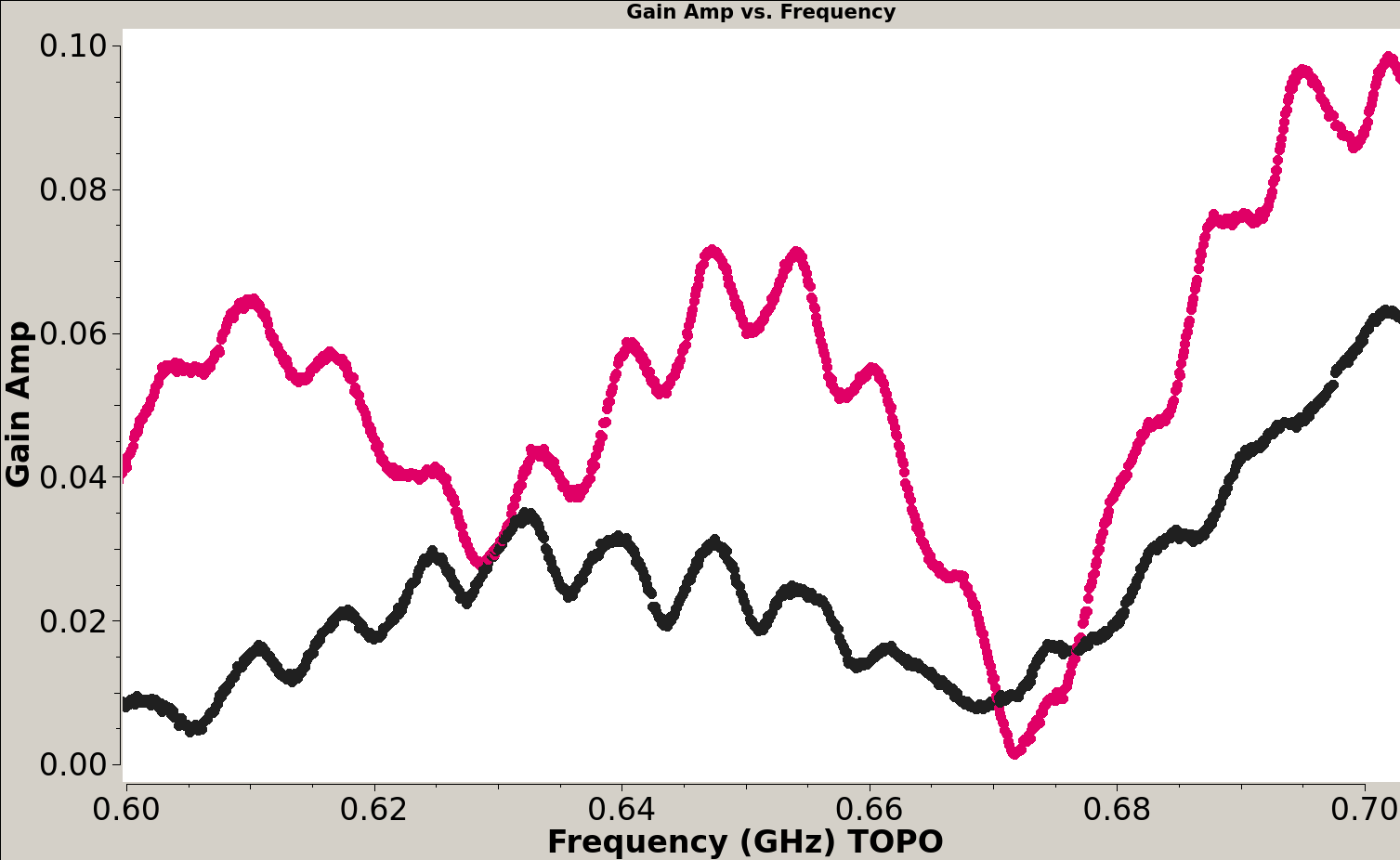}}
\caption{\small A zoomed-in view of the leakage amplitudes using the unpolarized calibrator 3C147 for antenna S02 in Walsh OFF (top) and Walsh ON (bottom) mode, with leakage amplitudes ranging from 0 to 10\% over the frequency range 0.6$-$0.7~GHz.}
    \label{fig:leakge_S02}
\end{figure}

A major difference was observed in the leakage amplitudes produced using the {\tt polcal} task in {\tt CASA} using the unpolarized calibrator 3C147. We observe leakages with large oscillatory patterns (also observed in the VLA with smaller amplitudes; see \href{https://library.nrao.edu/public/memos/evla/EVLAM_134.pdf}{EVLA memos 134} \& \href{https://library.nrao.edu/public/memos/evla/EVLAM_209.pdf}{209}) with a mean amplitude of 5$-$7\% and a maximum amplitude of 20\%, as are typically seen in band-4 observations \citep[see typical uGMRT D-term plots in][]{Kharb2023}. Without the application of the Walsh scheme, we find small `wiggles' on top of the large oscillatory patterns, and in some cases, very high leakage amplitude data points as well. The introduction of the Walsh scheme in the circuit path reduces these outliers and suppresses the tiny fluctuations in hundreds of kHz to tens of MHz frequency ranges. Figures~\ref{fig:leakages} to \ref{fig:leakge_S02} present the leakage amplitudes for the antennas in this experiment, both with Walsh OFF (left panels in figures \ref{fig:leakages}, \ref{fig:leakages_2}) and Walsh ON (right panels in figures \ref{fig:leakages}, \ref{fig:leakages_2}).

\subsection{Calibrated visibilities}
\begin{figure}
    \centering
    \setlength{\fboxsep}{0pt}
    \setlength{\fboxrule}{0.5pt}
    \fbox{\includegraphics[height=5cm]{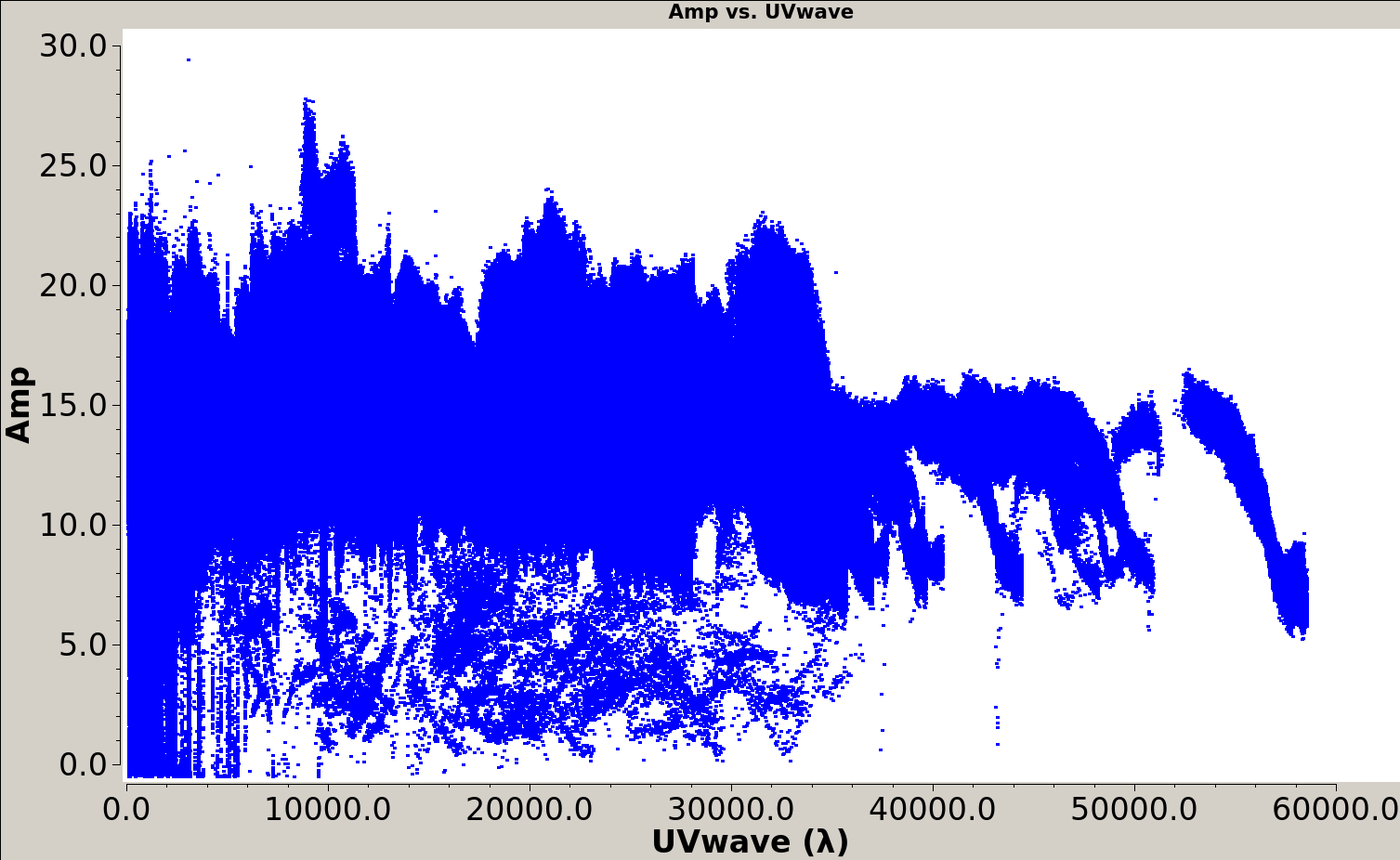}}
    \fbox{\includegraphics[height=5cm]{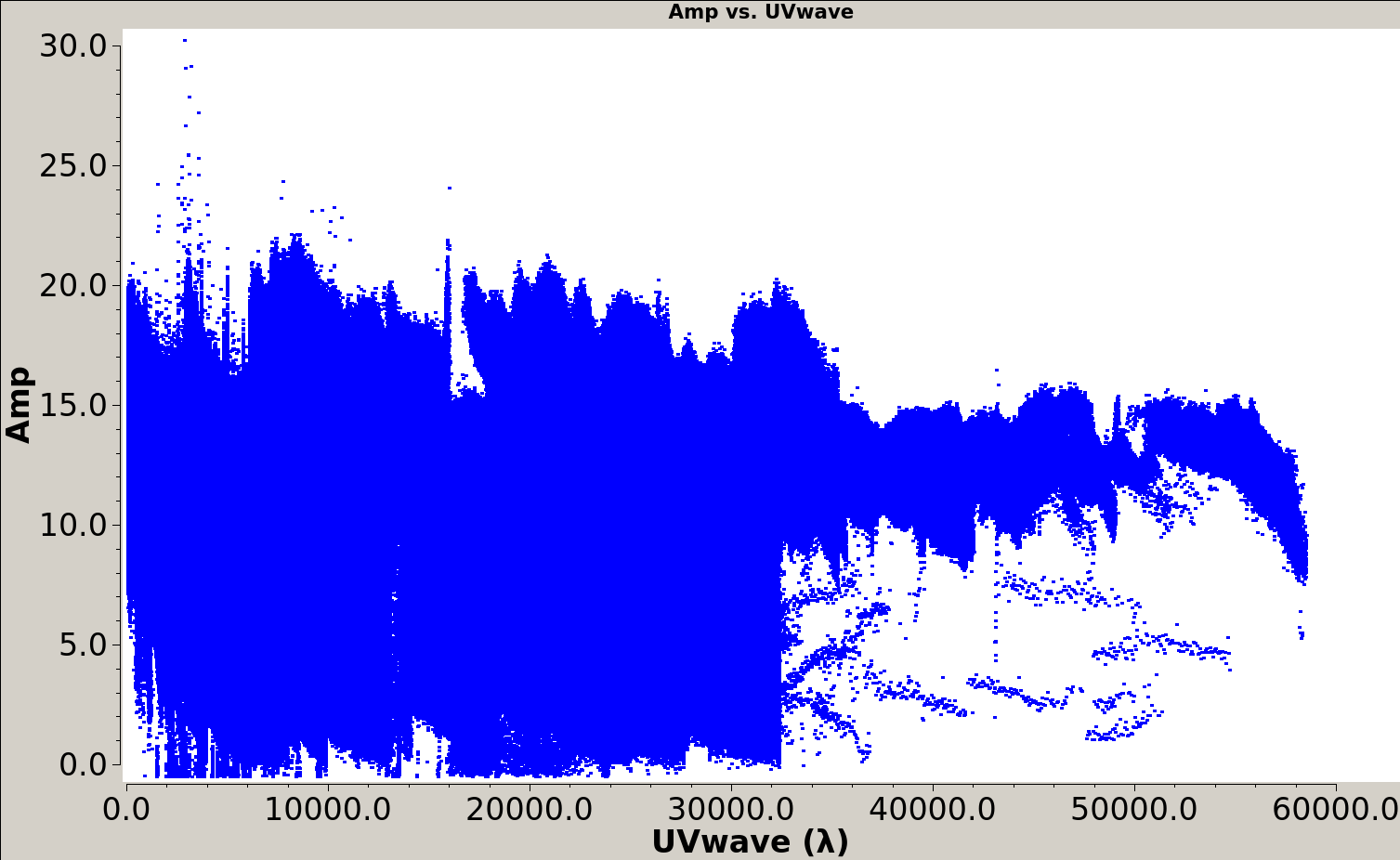}}
    \fbox{\includegraphics[height=5cm]{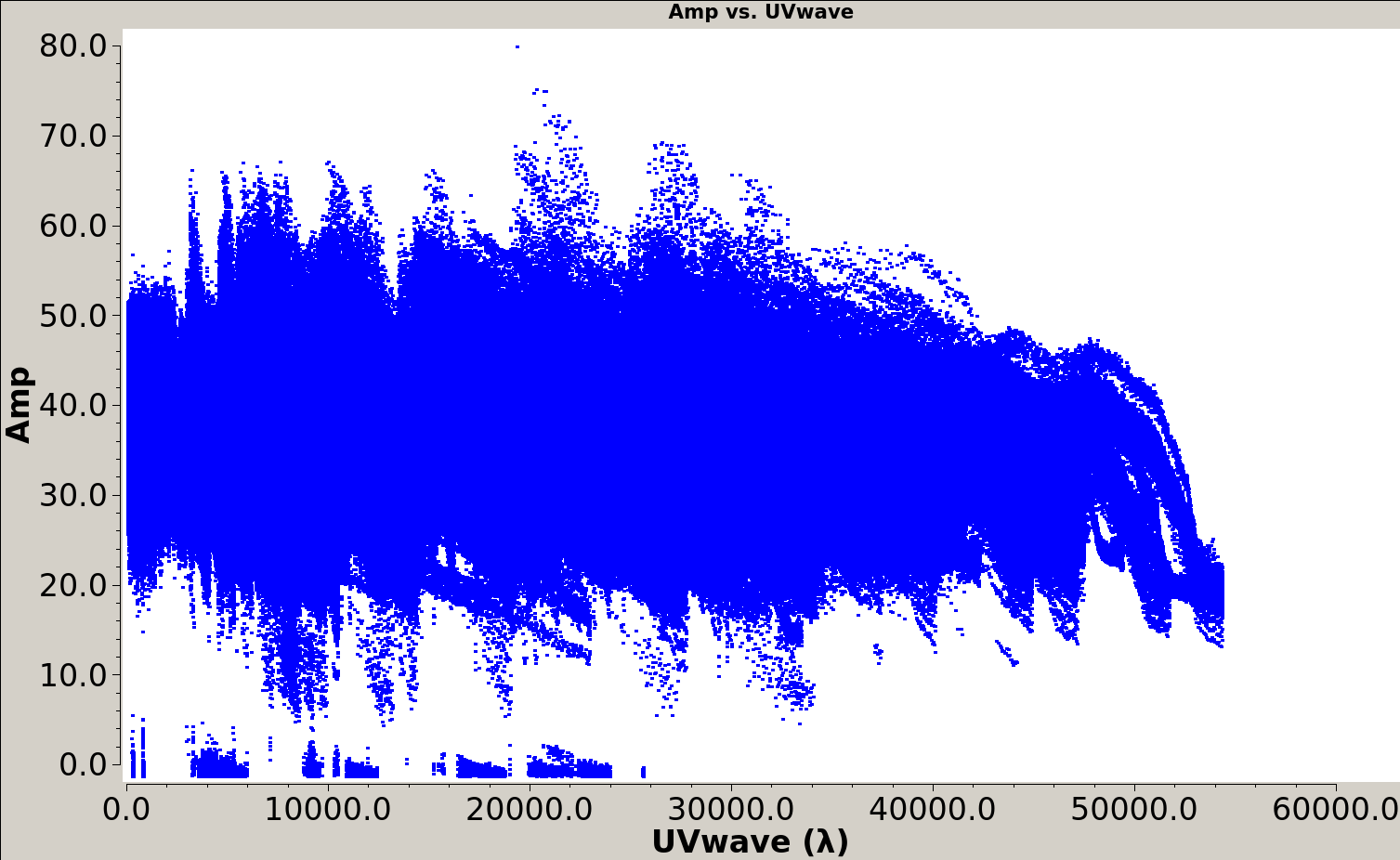}}
    \fbox{\includegraphics[height=5cm]{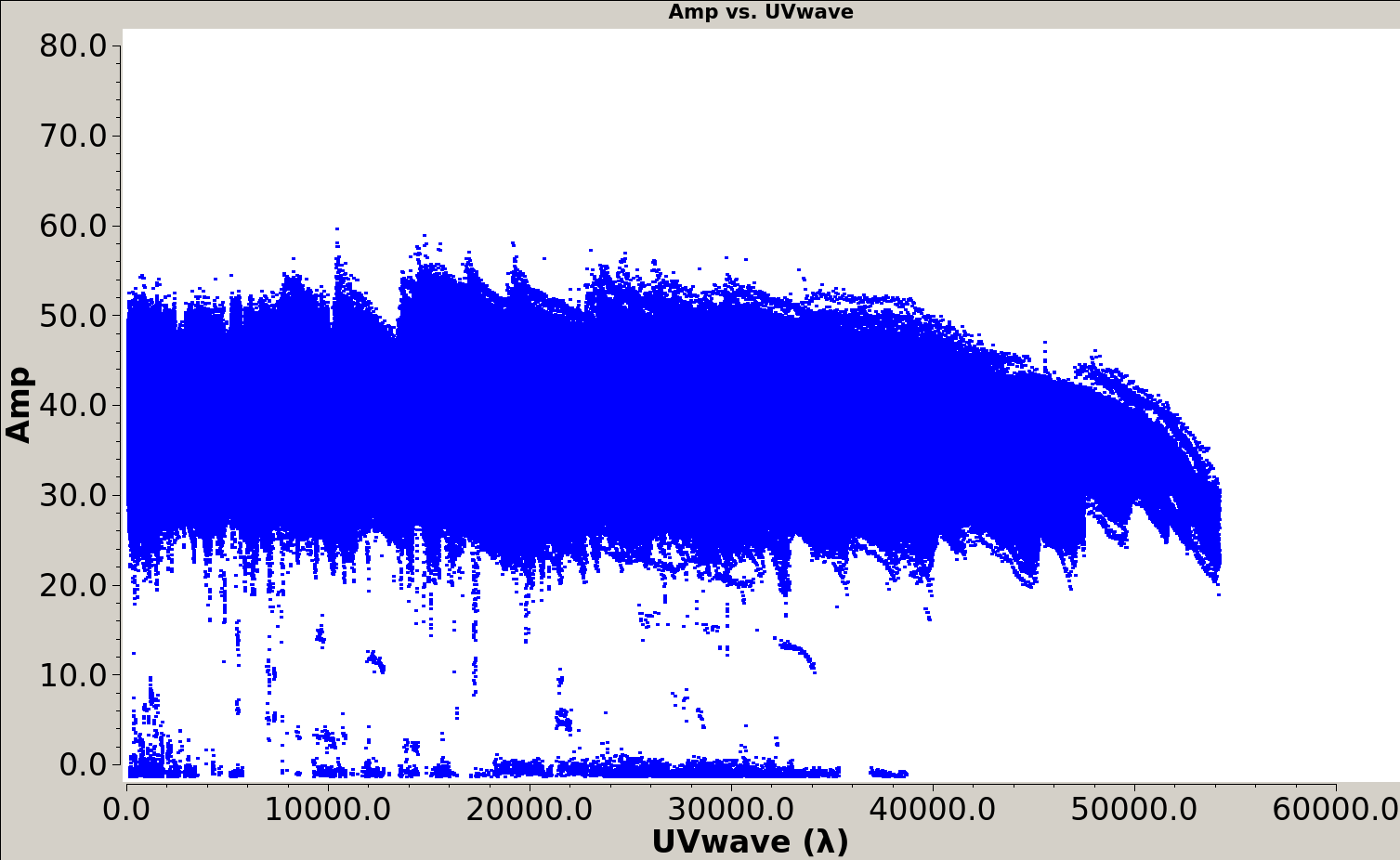}}
    \caption{\small Top: calibrated amplitude versus UV-distance (in terms of wavelength) for RR and LL correlations with Walsh OFF (left) and Walsh ON (right) for 3C138. The UVwave ranges from 0 to 60 k$\lambda$. The calibrated RR and LL amplitudes are plotted from 0$-$30 Jy. Bottom: Calibrated amplitude (RR, LL) vs. UV-distance for 3C147 with Walsh OFF (left) and Walsh ON (right). The amplitudes are plotted from 0$-$80 Jy.}
    \label{fig:ampvsuvwave3C147138RRLL}
\end{figure}

\begin{figure}
    \centering
    \setlength{\fboxsep}{0pt}
    \setlength{\fboxrule}{0.5pt}
    \fbox{\includegraphics[height=5cm]{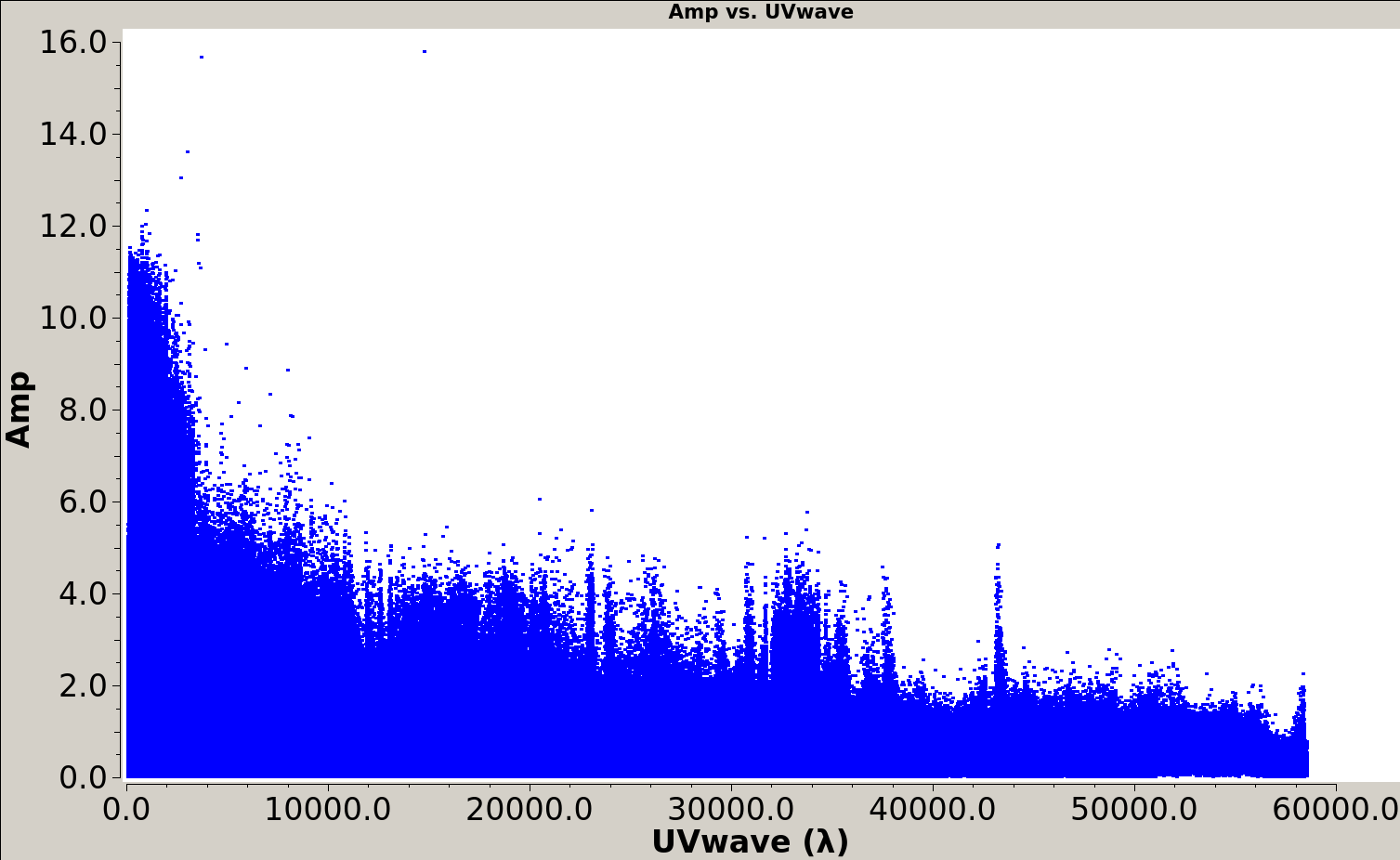}}
    \fbox{\includegraphics[height=5cm]{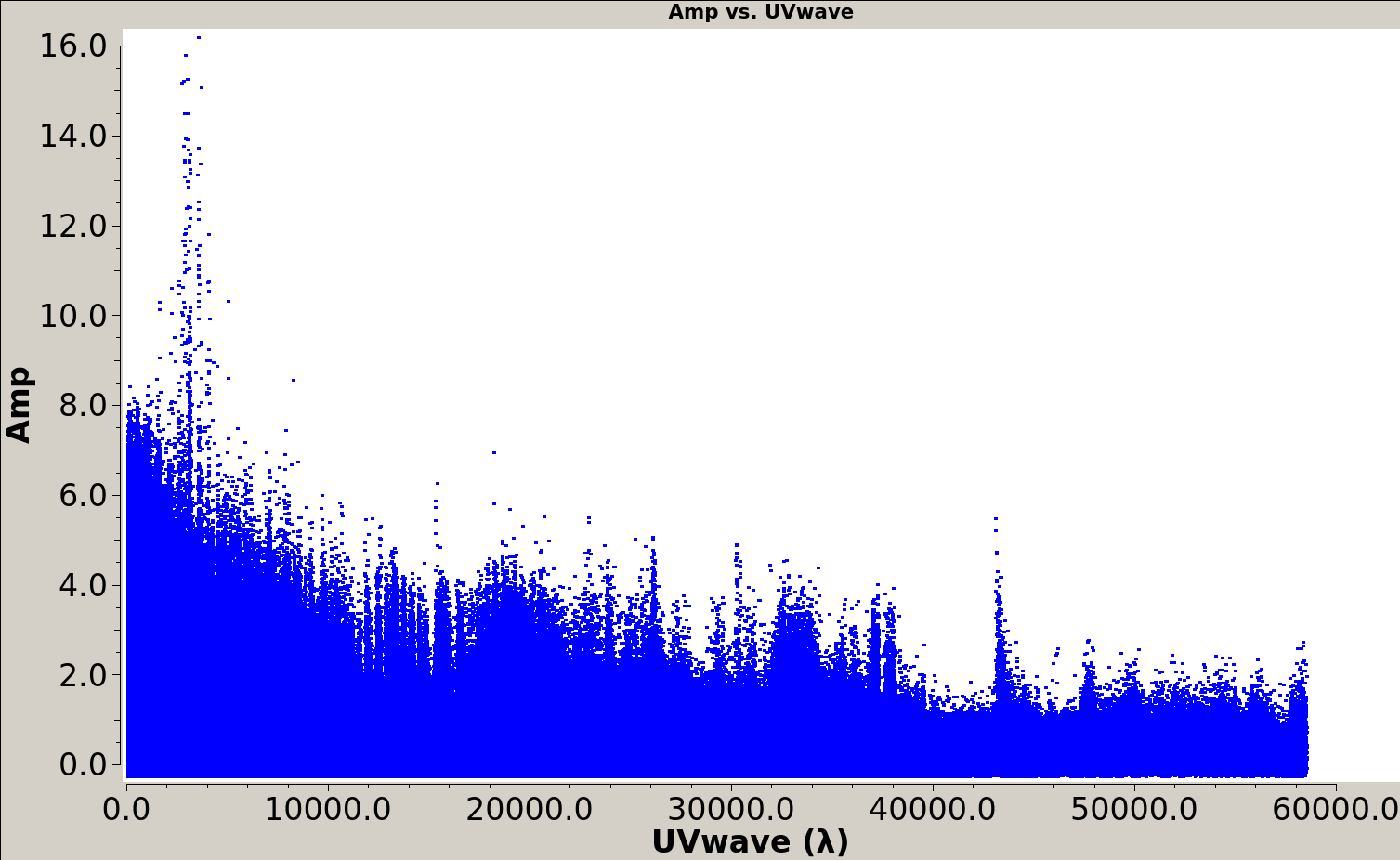}}
    \fbox{\includegraphics[height=5cm]{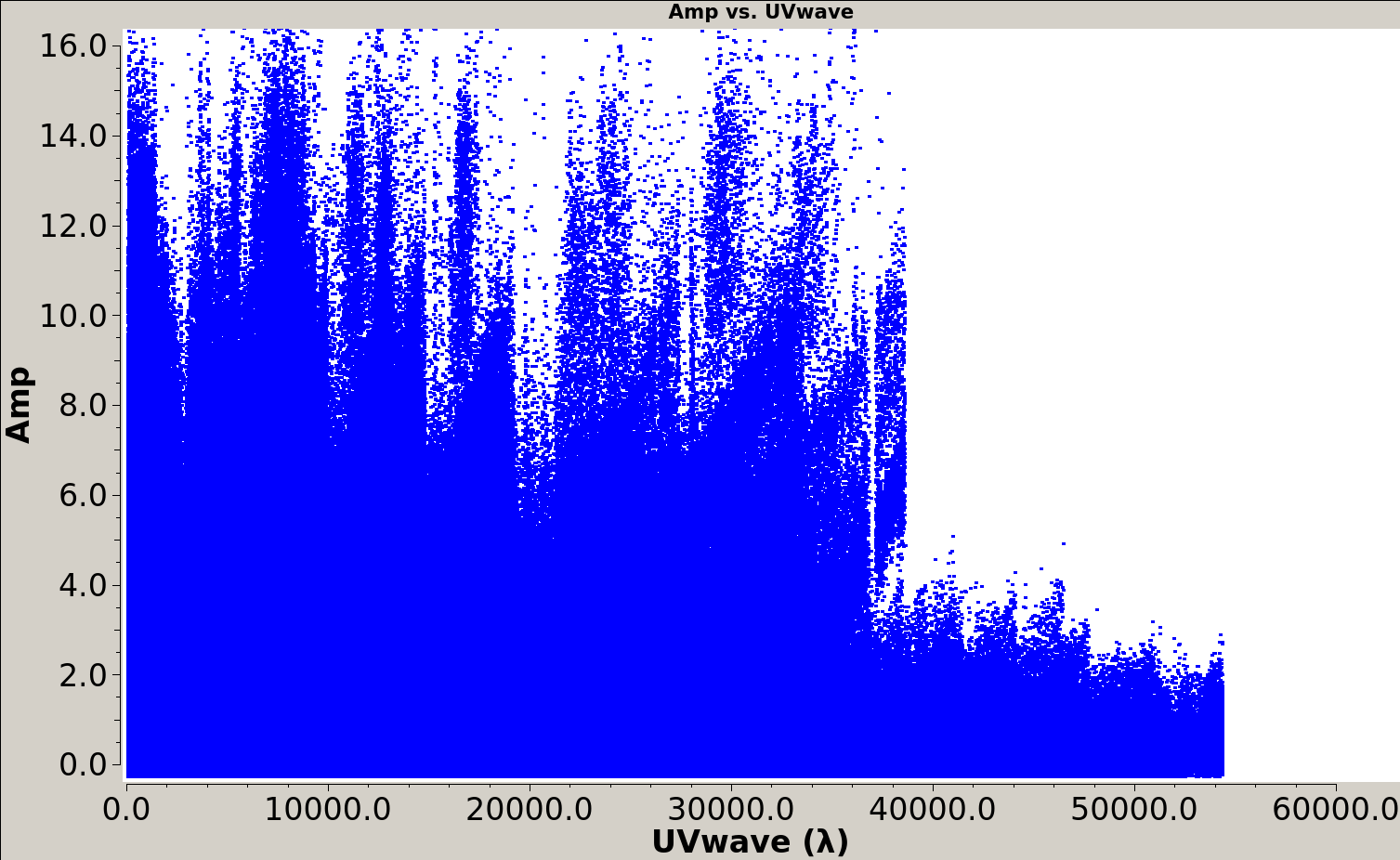}}
    \fbox{\includegraphics[height=5cm]{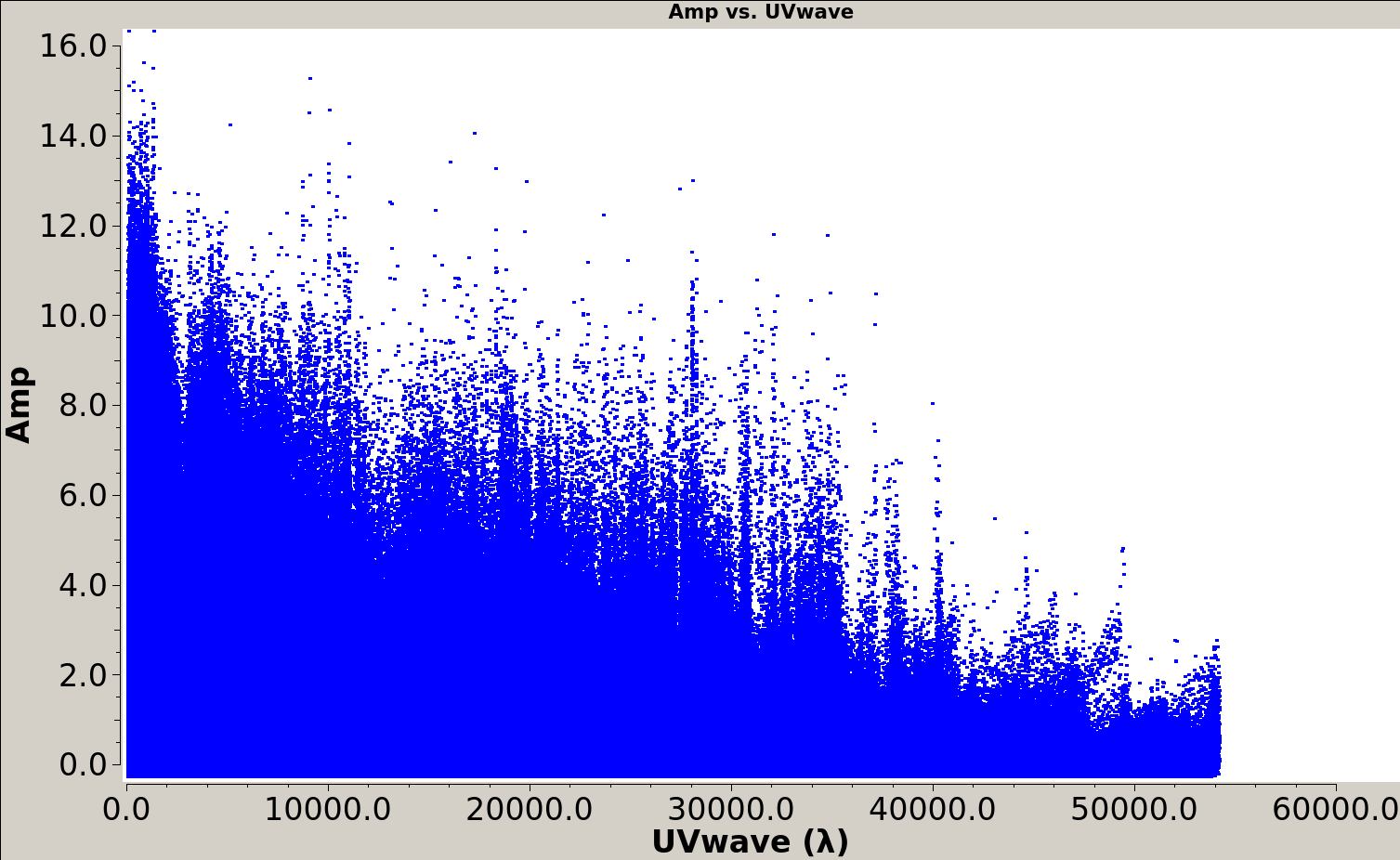}}
    \caption{\small UV-distance (in terms of wavelength) versus calibrated amplitude for RL and LR correlations, with Walsh OFF (left) and Walsh ON (right) for 3C138 (top) and 3C147 (bottom). The UVwave ranges from 0 to 60 k$\lambda$. The calibrated RL and LR amplitudes are plotted from 0$-$16 Jy in all panels.}
    \label{fig:ampvsuvwave3C147138RLLR}
\end{figure}

Figures \ref{fig:ampvsuvwave3C147138RRLL} and \ref{fig:ampvsuvwave3C147138RLLR} depict that the calibrated gain amplitudes for both the polarized source 3C138 (top panel) and the unpolarized source 3C147 (bottom panel) show a suppression of outlier data points with the Walsh ON mode (right) compared to the Walsh OFF mode (left) for both parallel and cross-hand correlations, respectively. However, the thermal noise seems to remain similar even for the Walsh ON mode, marking no drastic improvement or difference in the calibrated visibilities with the Walsh scheme enabled. 

\subsection{Polarization data}
\label{sec:poln}
\begin{figure}
    \centering
    \includegraphics[width=0.48\textwidth]{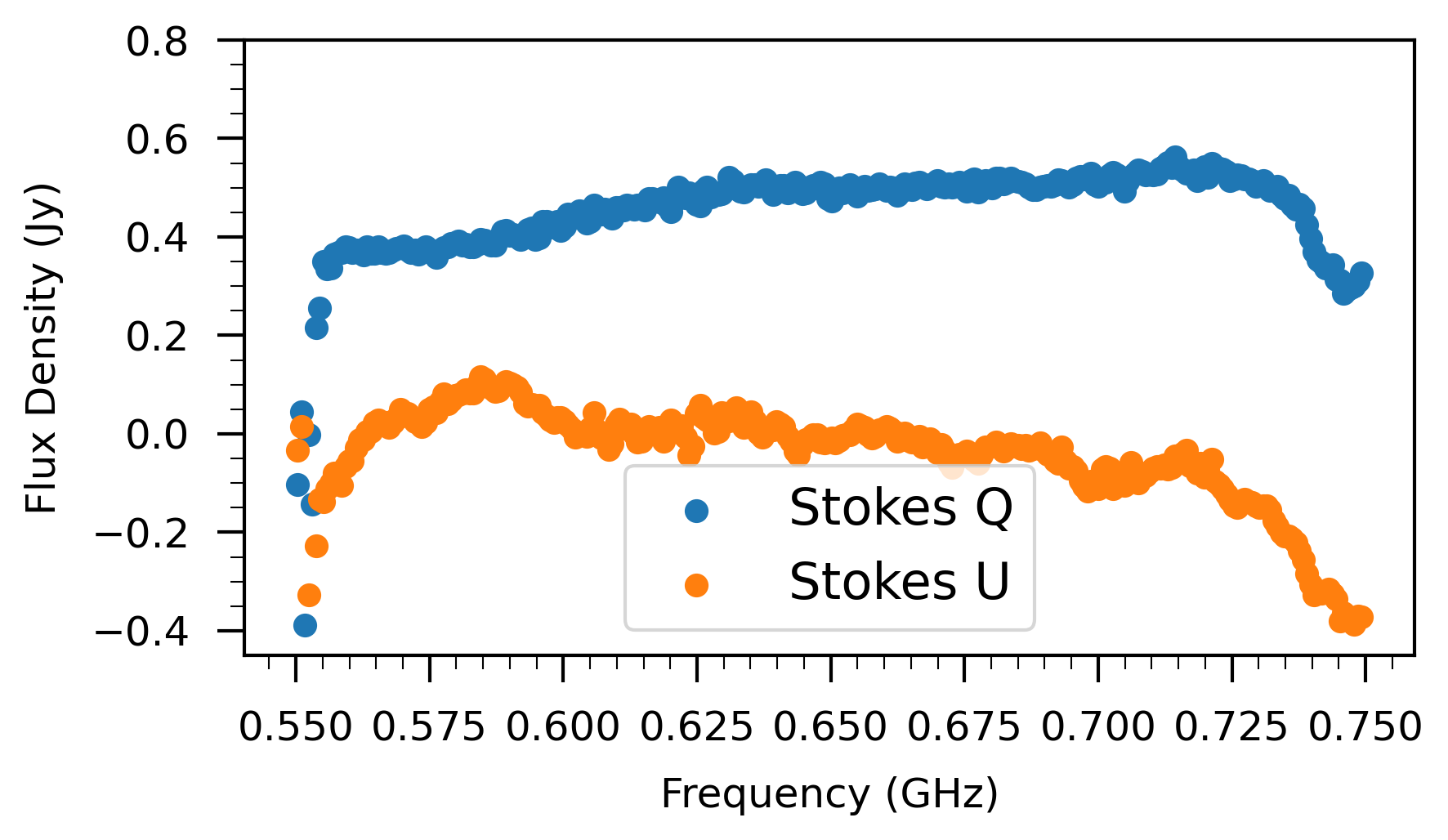}
    \includegraphics[width=0.48\textwidth]{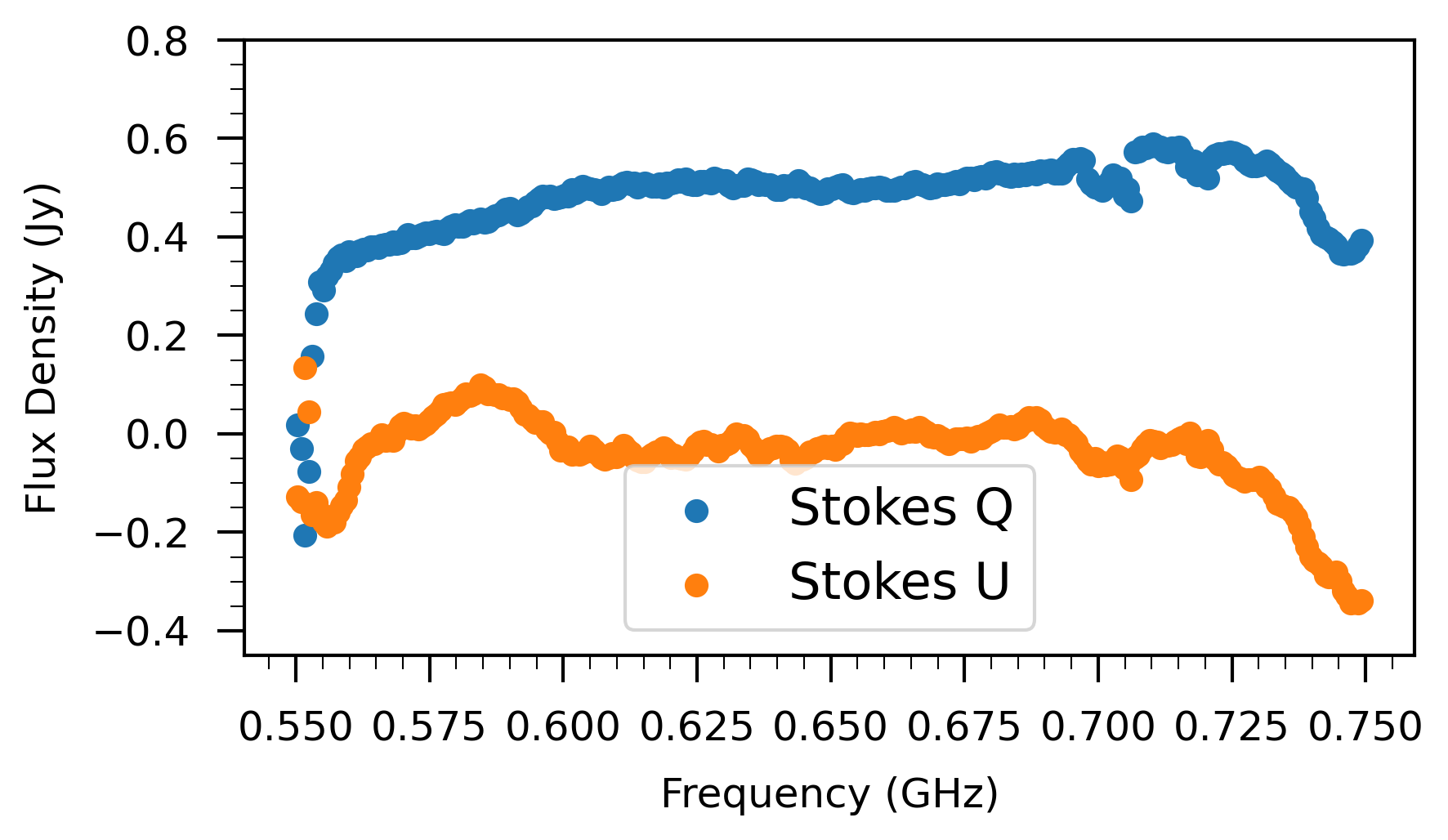}
    \includegraphics[width=0.48\textwidth]{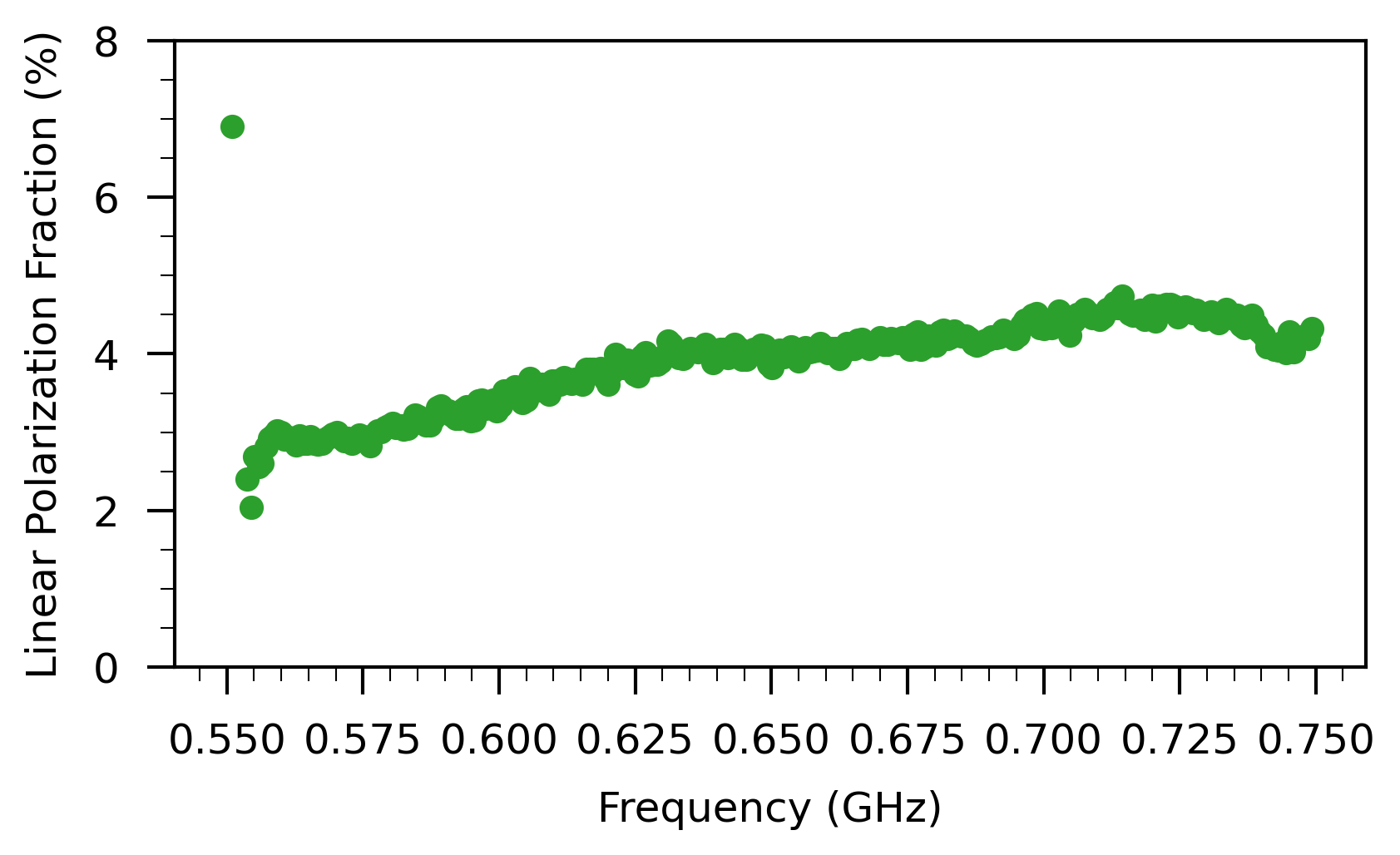}
    \includegraphics[width=0.48\textwidth]{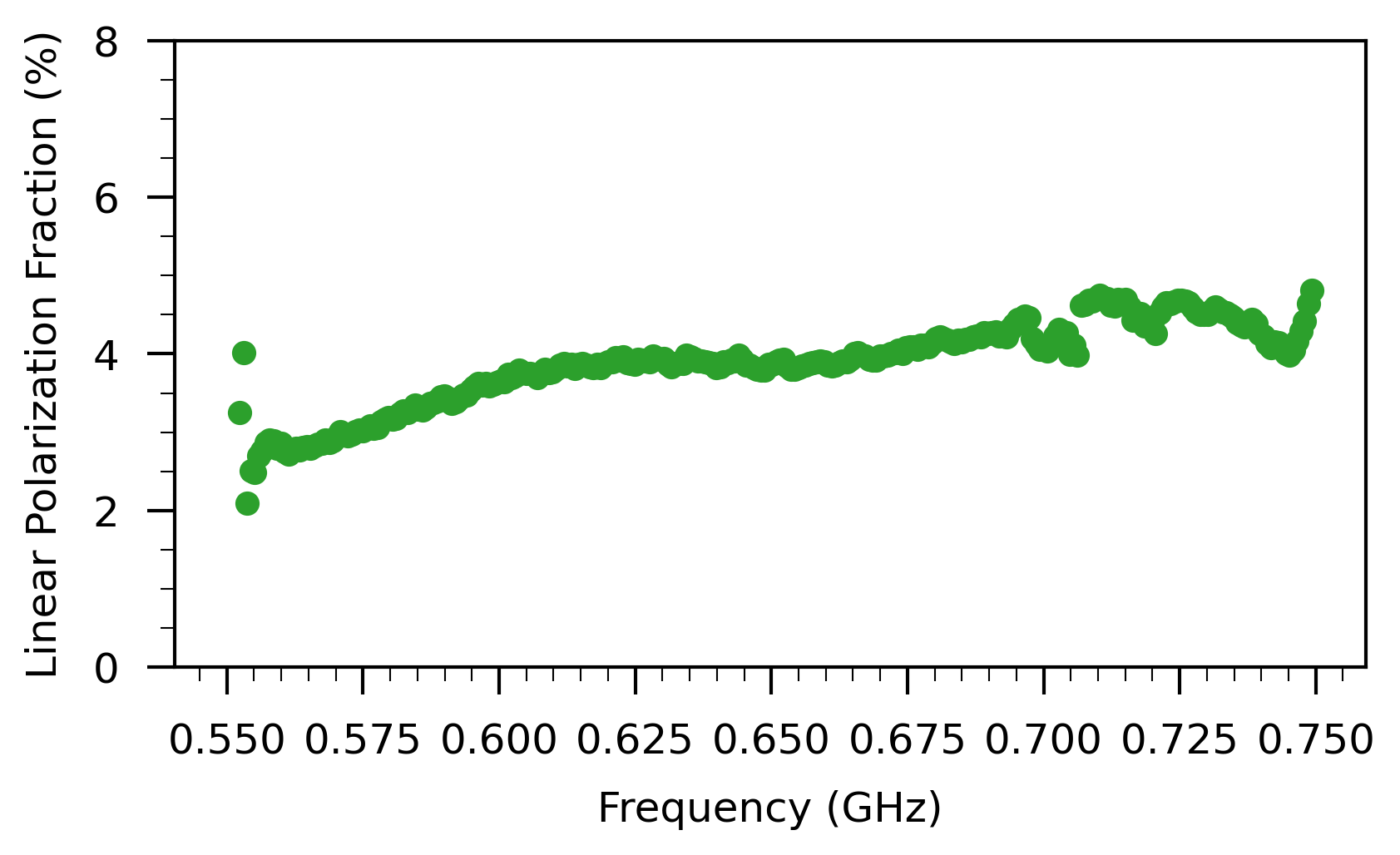}
    \caption{\small Top: Stokes Q and U for 3C138 obtained from the calibrated RL and LR amplitudes with Walsh OFF (left) and Walsh ON (right) plotted with frequency. Bottom: Linear polarization fraction in percentage versus frequency with Walsh OFF (left) and Walsh ON (right) for 3C138.}
    \label{fig:polarization_3C138}
\end{figure}

\begin{figure}
    \centering
    \includegraphics[width=0.48\textwidth]{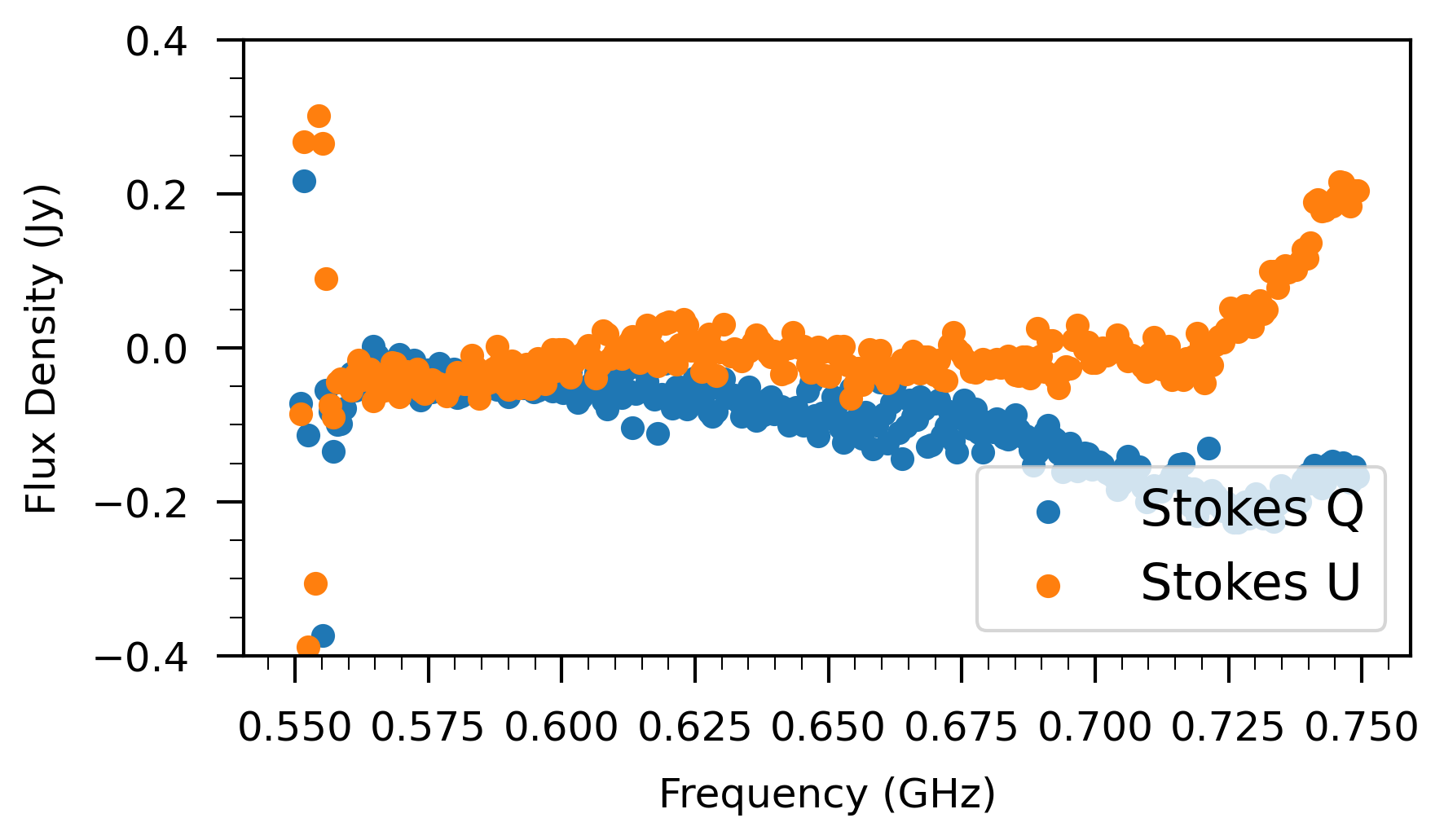}
    \includegraphics[width=0.48\textwidth]{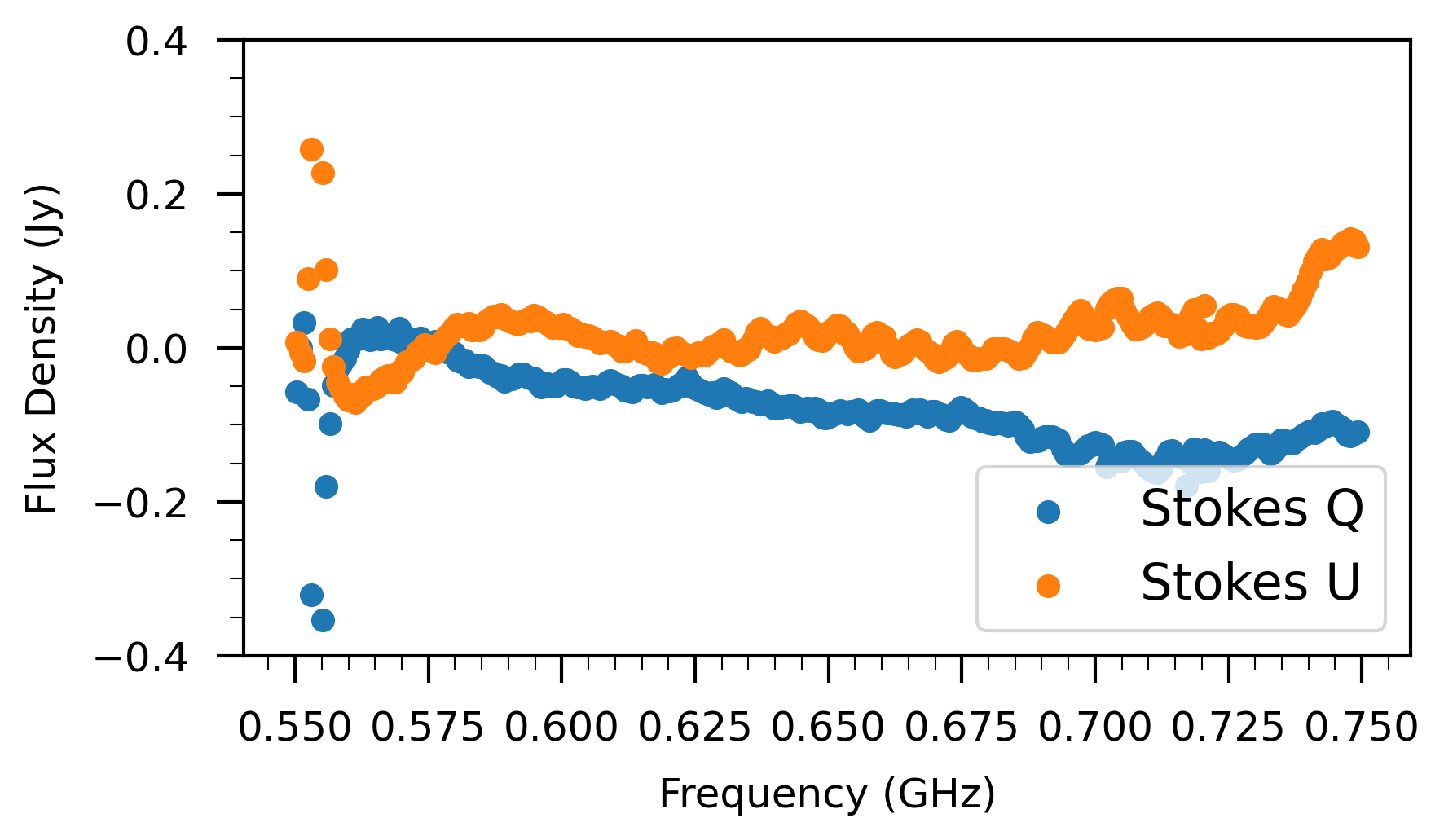}
    \includegraphics[width=0.48\textwidth]{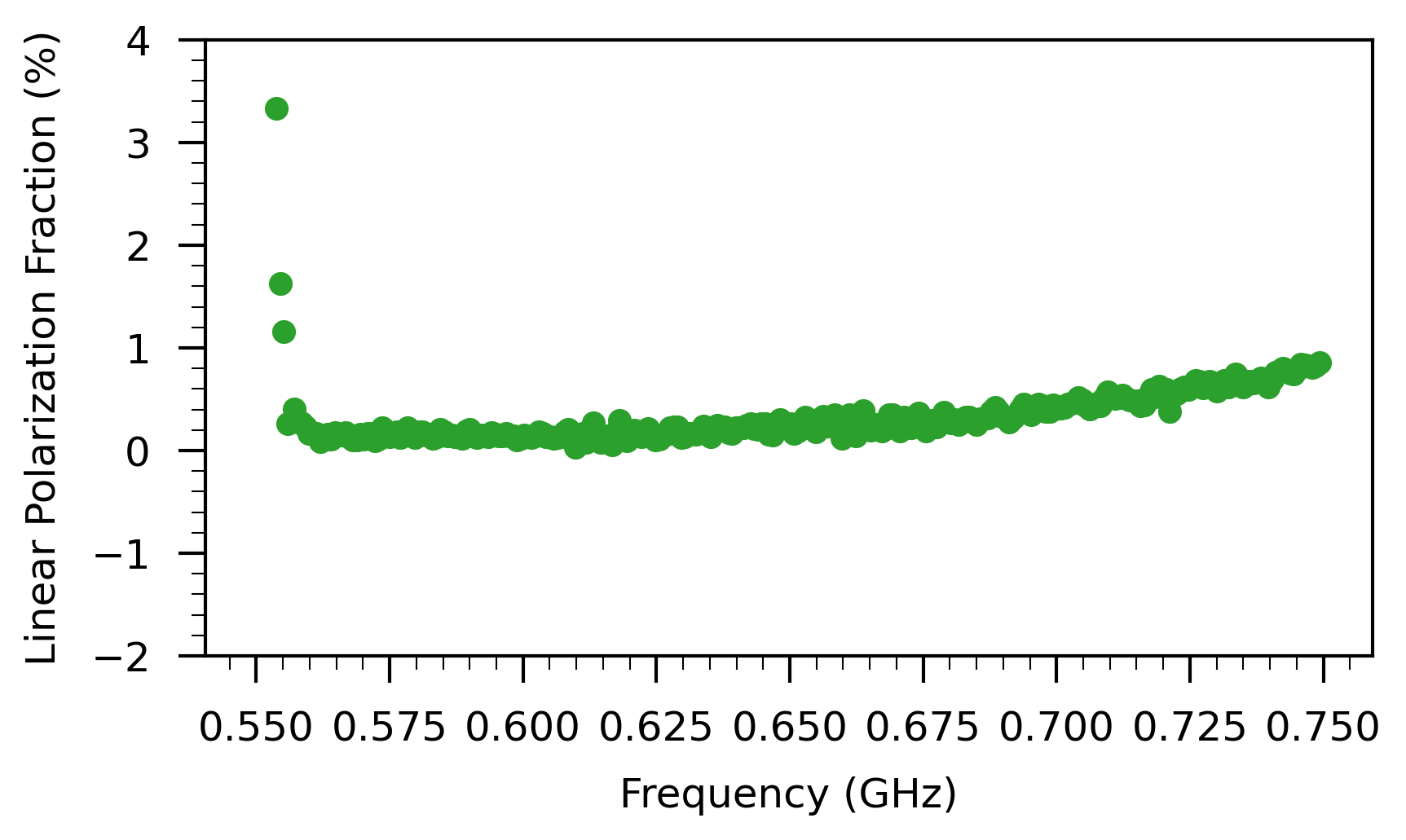}
    \includegraphics[width=0.48\textwidth]{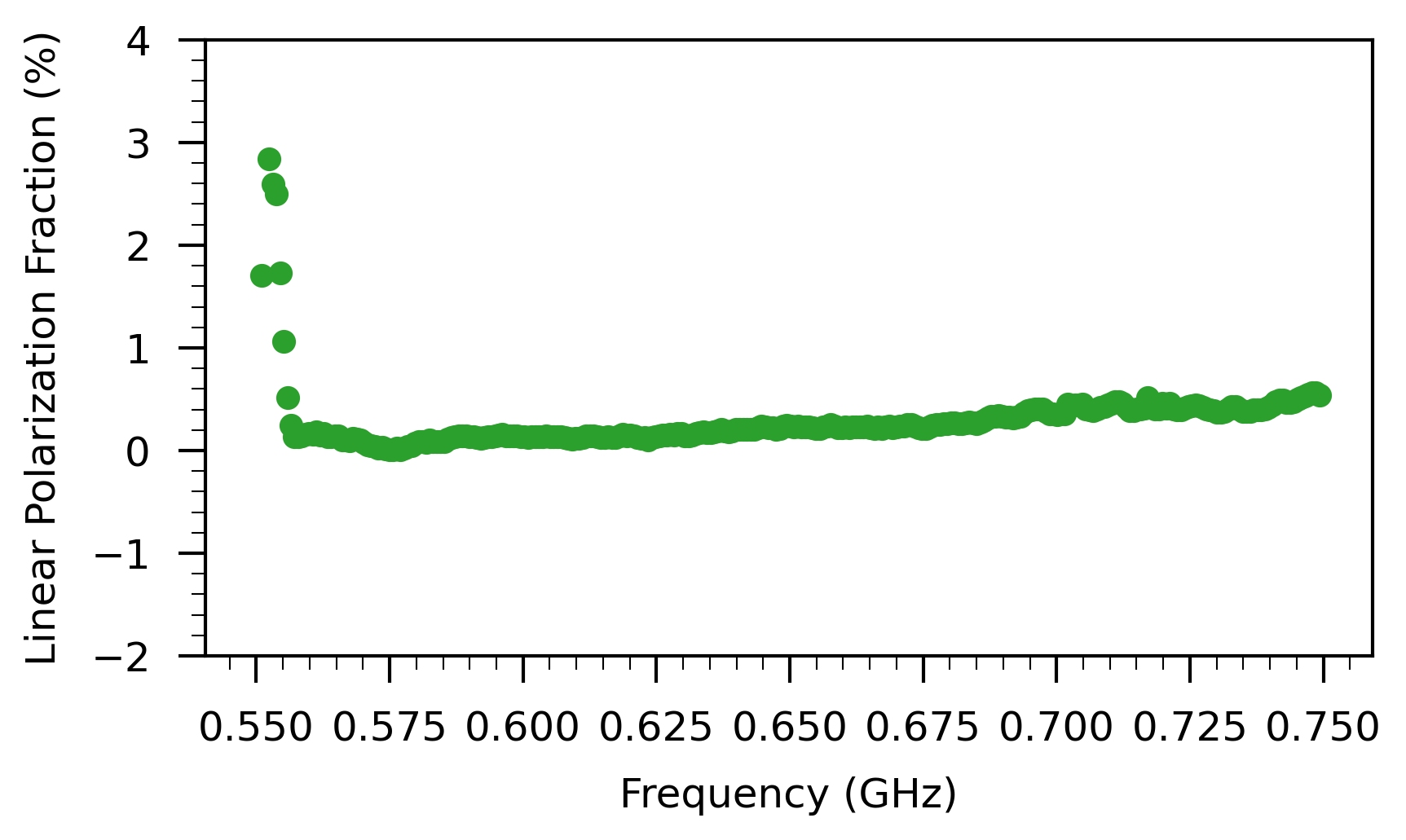}
    \caption{\small Same as Figure \ref{fig:polarization_3C138} but for 3C147.}
    \label{fig:polarization_3C147}
\end{figure}

To further assess the impact of Walsh switching on the frequency-dependent stability of full-Stokes measurements, we compared the calibrated Stokes Q, U, and linear polarization fraction (FPOL) data of both sources  (3C138 and 3C147) obtained with Walsh switching ON  and OFF (see Figures \ref{fig:polarization_3C138} \& \ref{fig:polarization_3C147}). For each frequency channel, the calibrated complex visibilities in the circular polarization basis were used to derive the Stokes parameters according to 
$I=(RR+LL)/2$, $Q=(RL+LR)/2$, and $U=(RL-LR)/(2i)$. 
The complex Stokes visibilities were then vector-averaged over the selected baselines and time samples, and the real part of the averages is reported as the measured Stokes' values. 
When all the baselines were considered, the averaged visibilities of 3C138 underestimate the Stokes' values by a factor of $4-5$. When restricting to shorter baselines ($<$1 km), the averaging yielded values consistent with the reported values in the literature. This is consistent with the source being partially resolved in uGMRT images. 

\begin{figure}
    \centering
    \includegraphics[width=0.24\textwidth]{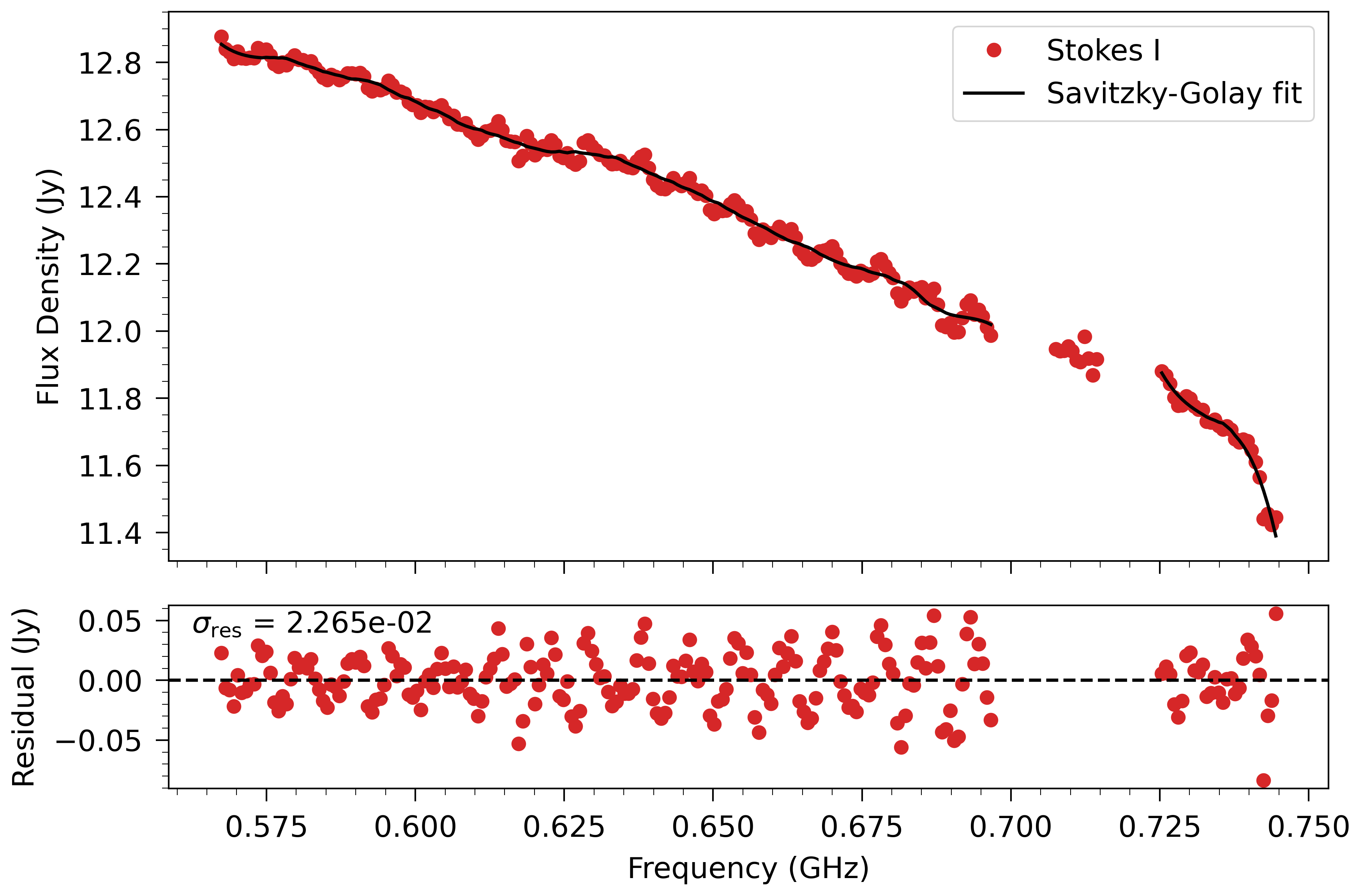}
    \includegraphics[width=0.24\textwidth]{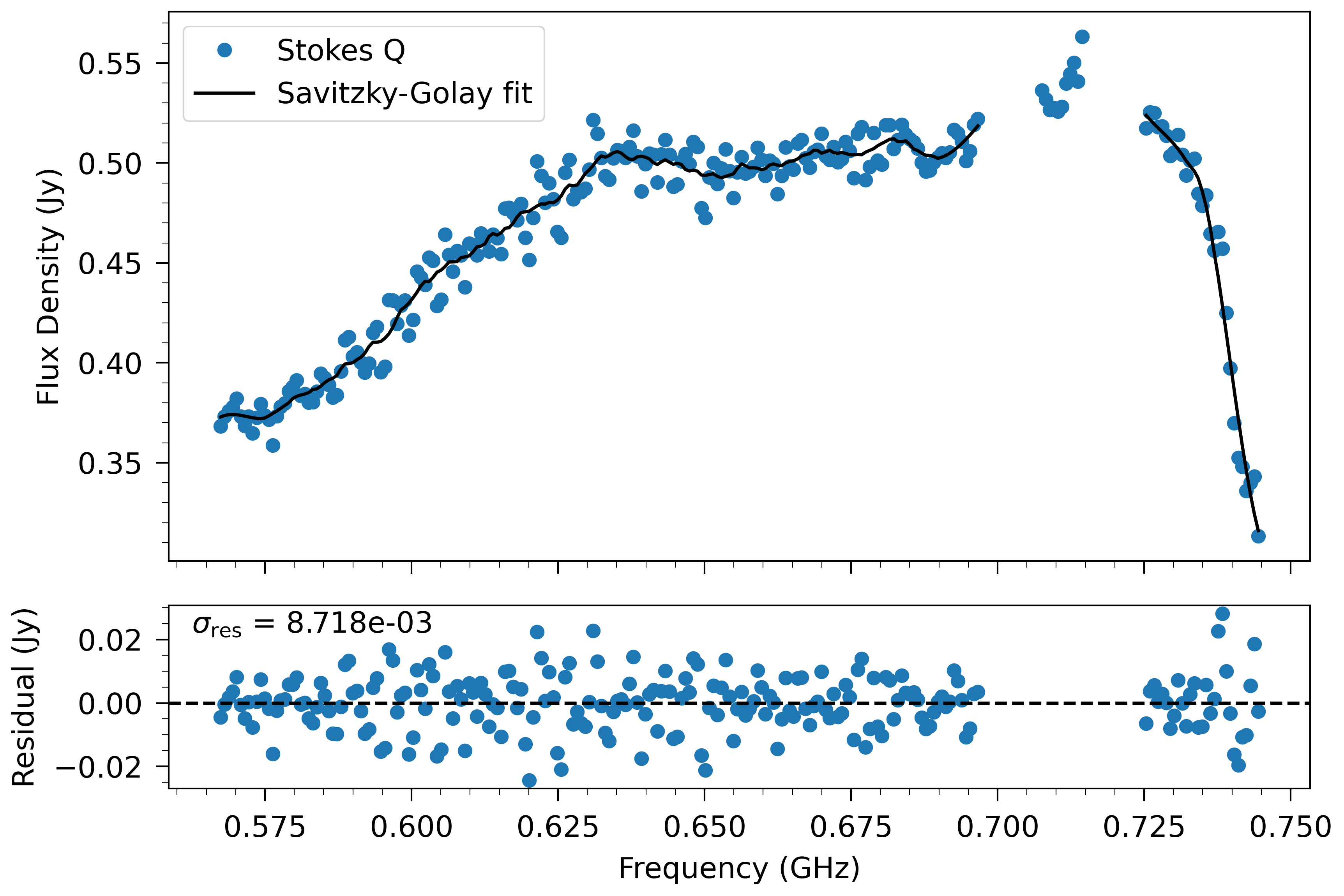}
    \includegraphics[width=0.24\textwidth]{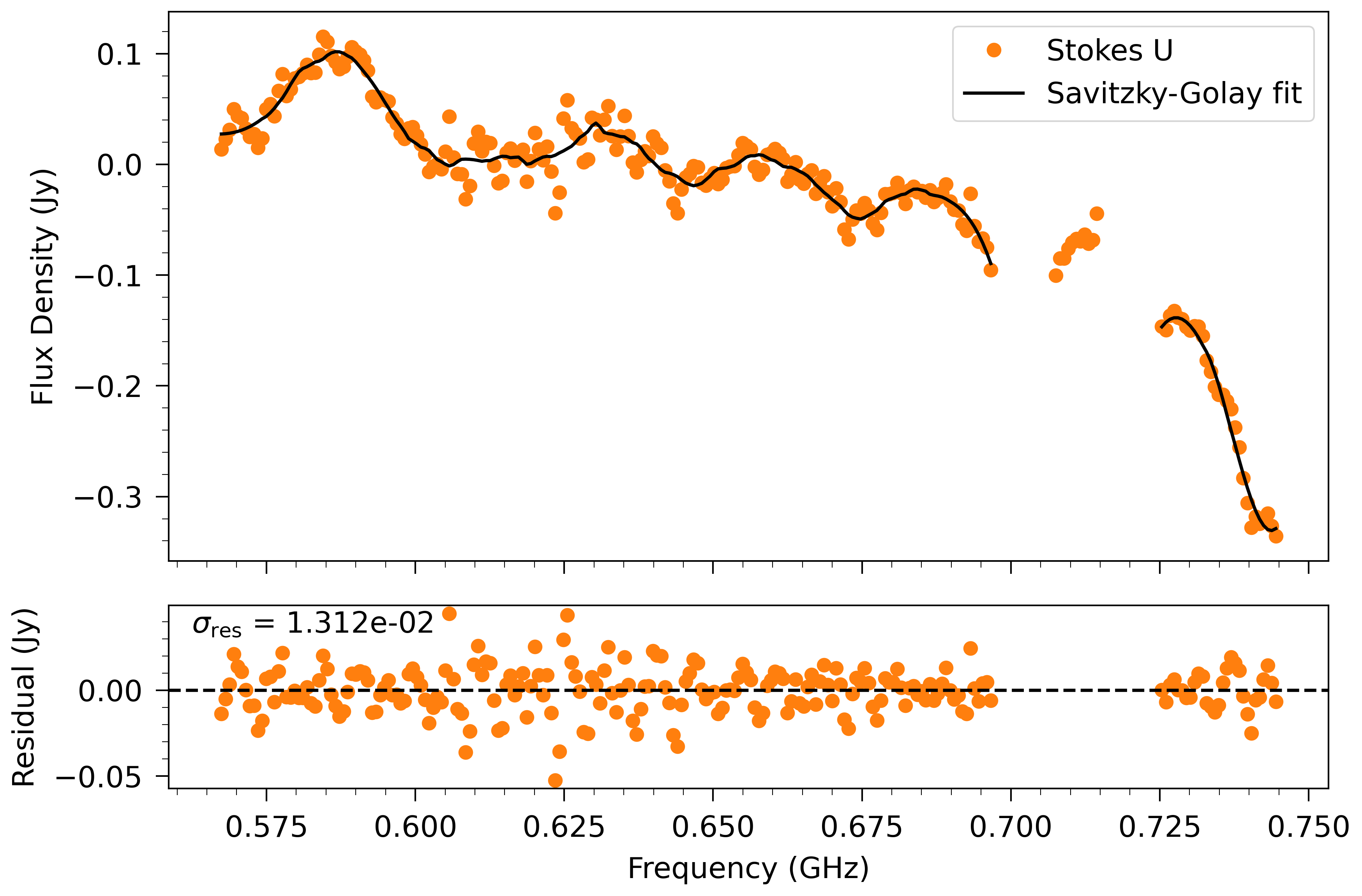}
    \includegraphics[width=0.24\textwidth]{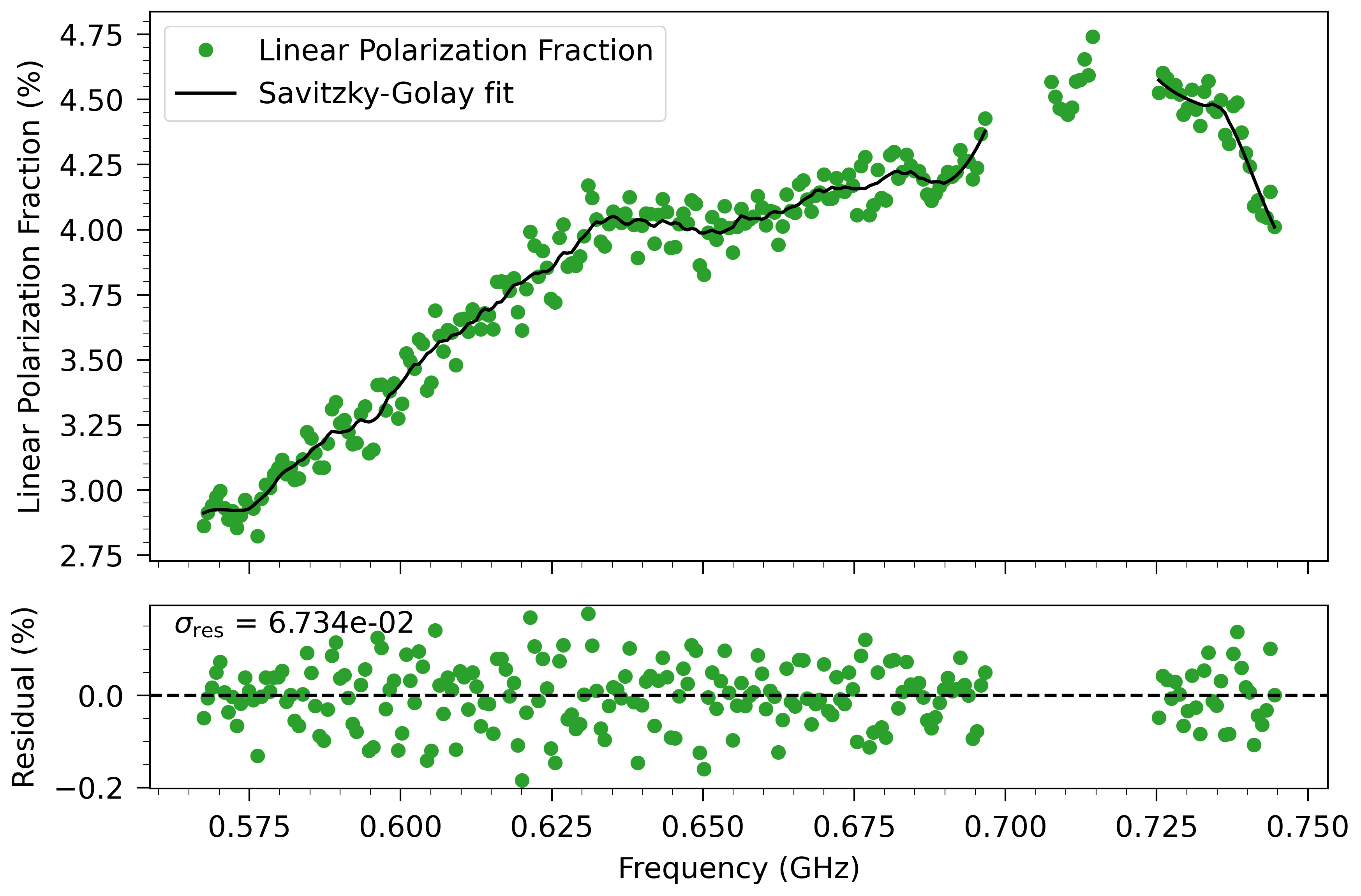}
    \includegraphics[width=0.24\textwidth]{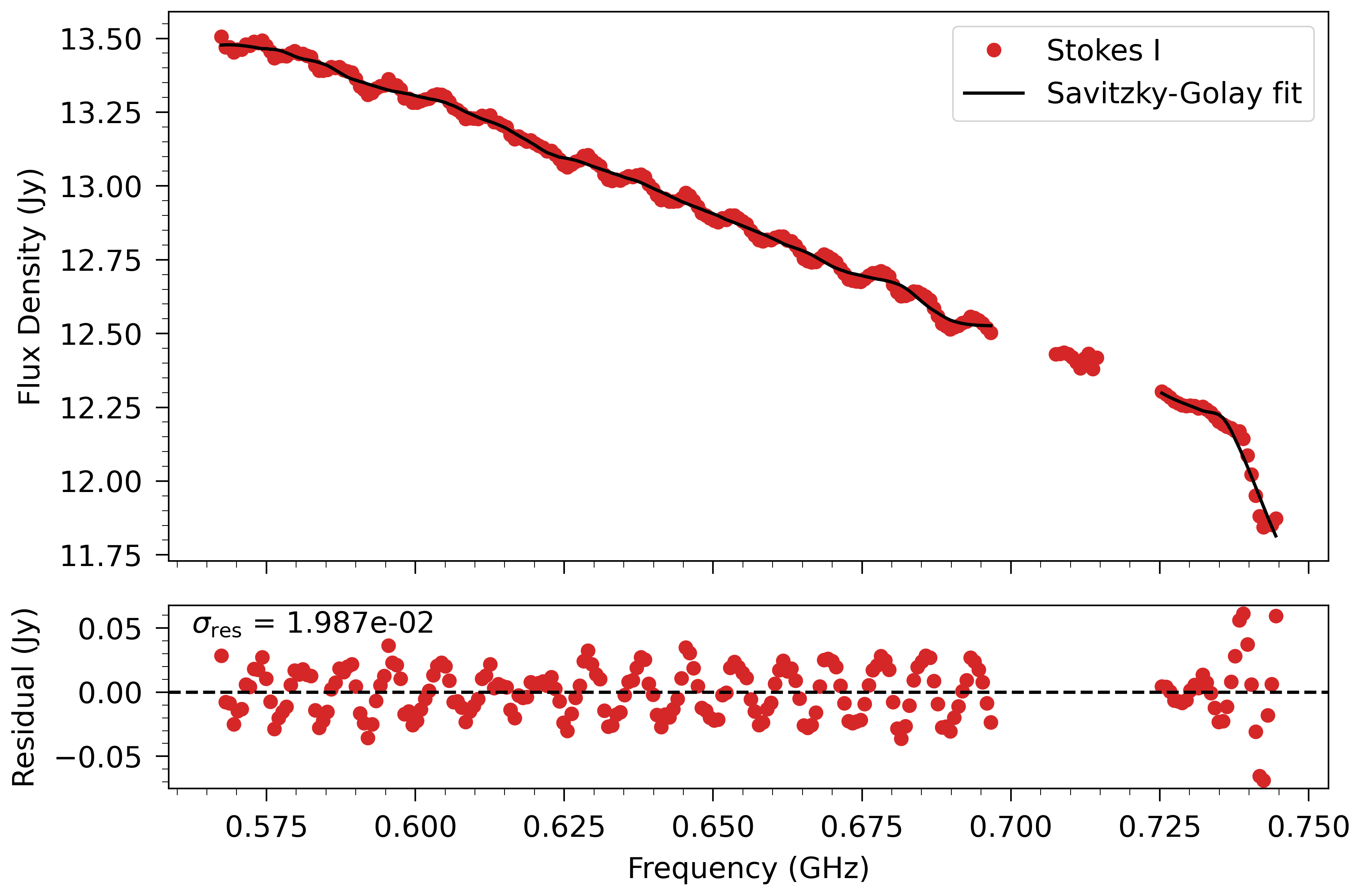}
    \includegraphics[width=0.24\textwidth]{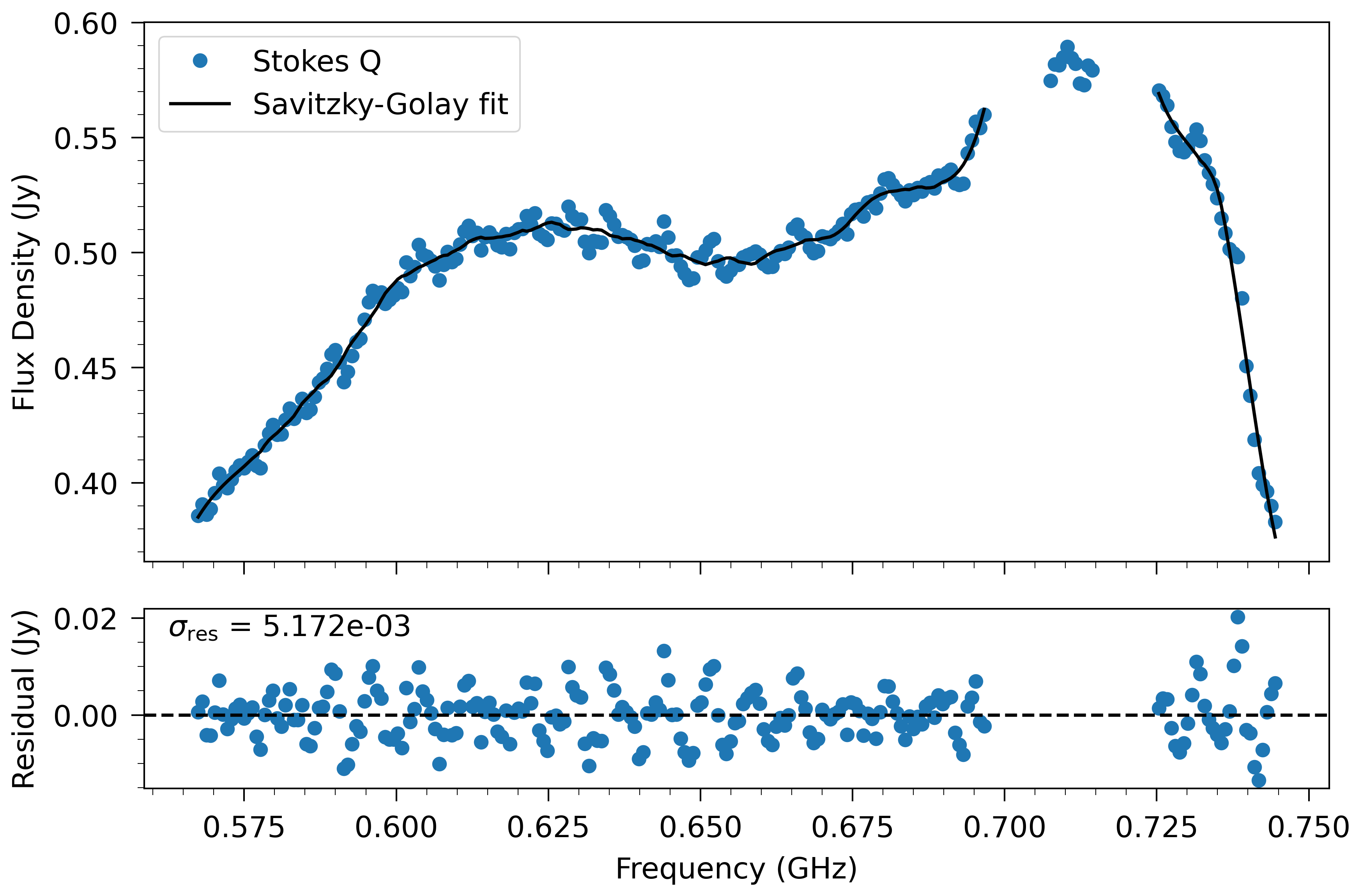}
    \includegraphics[width=0.24\textwidth]{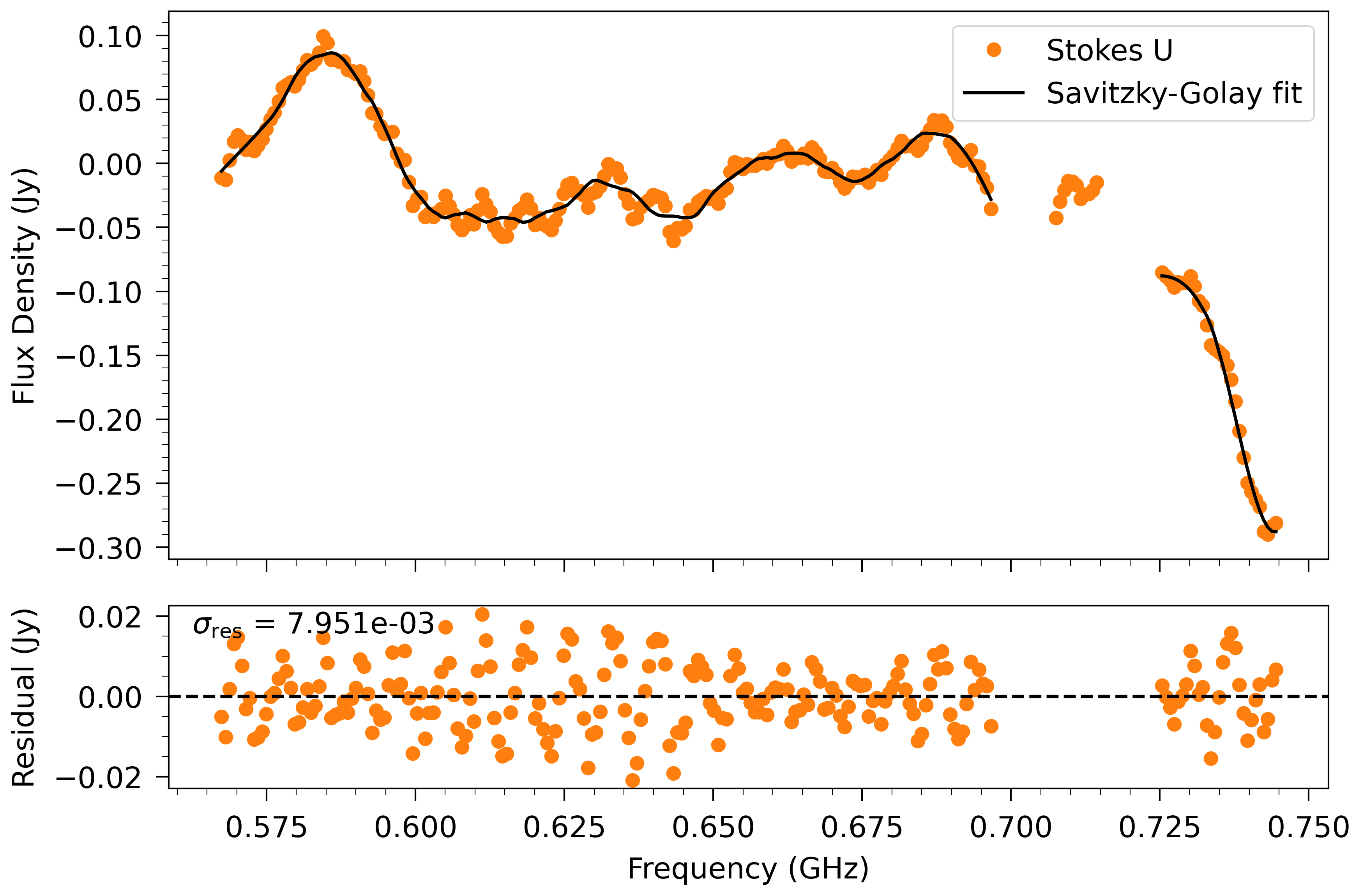}
    \includegraphics[width=0.24\textwidth]{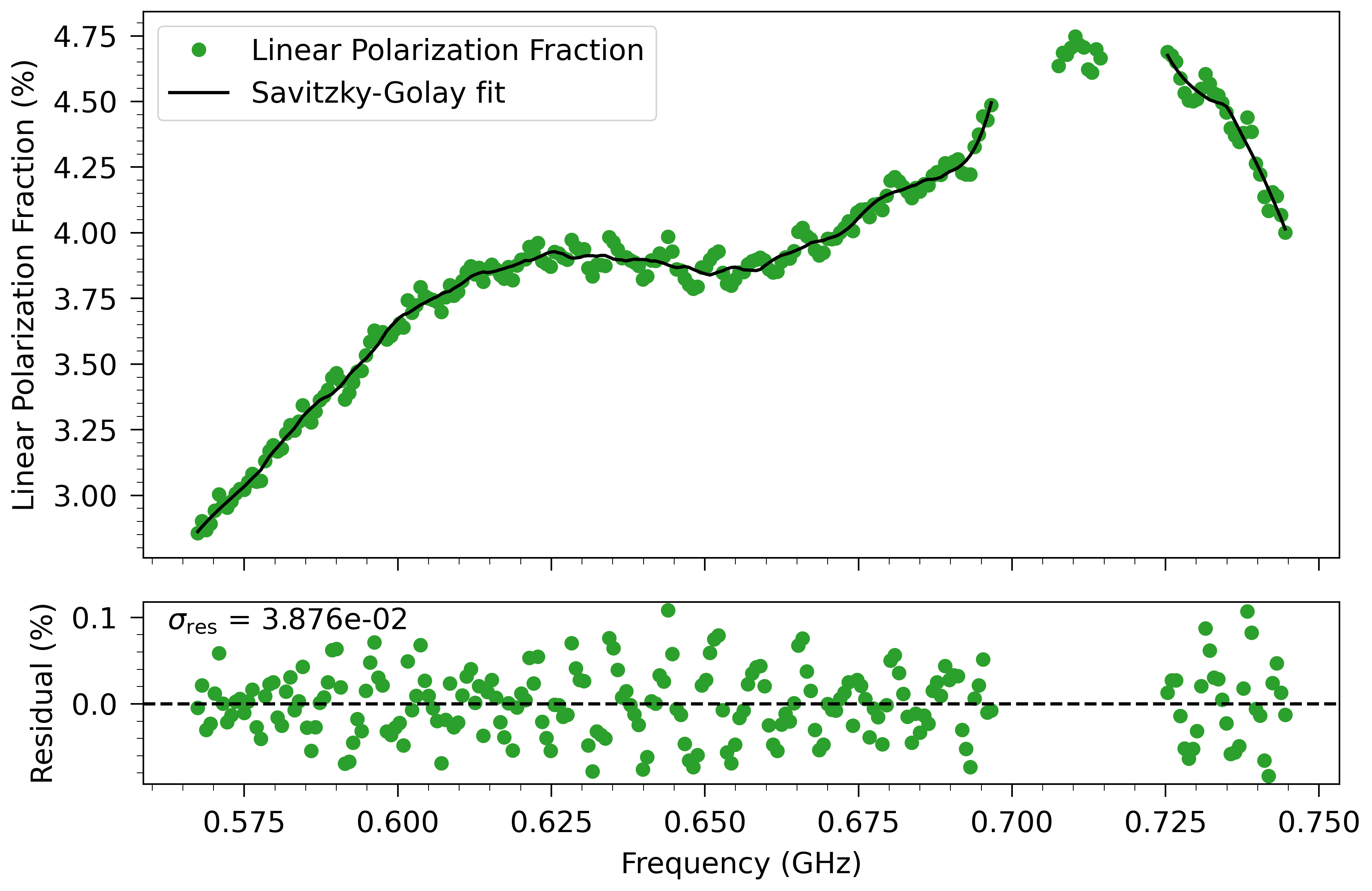}    
\caption{\small Stokes I, Q, U, and FPOL data of 3C138 with Walsh OFF (top) and Walsh ON (bottom) after removing the initial 25 edge channels affected by spectral response. The colored points represent the measured values (red: Stokes I, blue: Stokes Q, orange: Stokes U, green: FPOL, while the black curves show the Savitzky–Golay fits applied. The bottom panels display the residuals of the fits, highlighting the remaining frequency-dependent variations and their $\sigma_{\mathrm{res}}$ values. The residual spectra provide an estimate of the amplitude of frequency-dependent instrumental fluctuations after calibration.}
    \label{fig:3C138_IQUf_residuals}
\end{figure}

\begin{figure}
    \centering
    \includegraphics[width=0.24\textwidth]{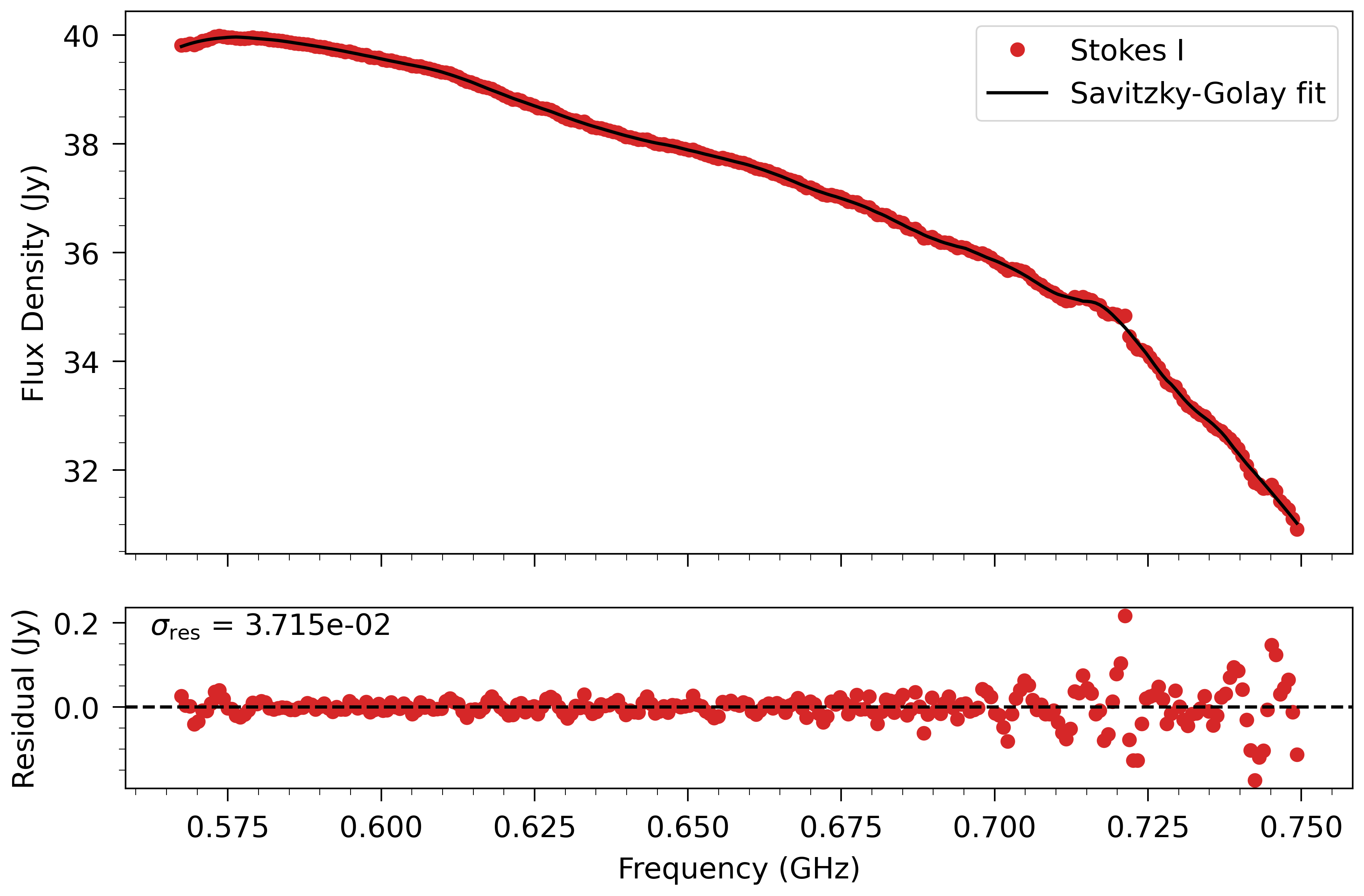}
    \includegraphics[width=0.24\textwidth]{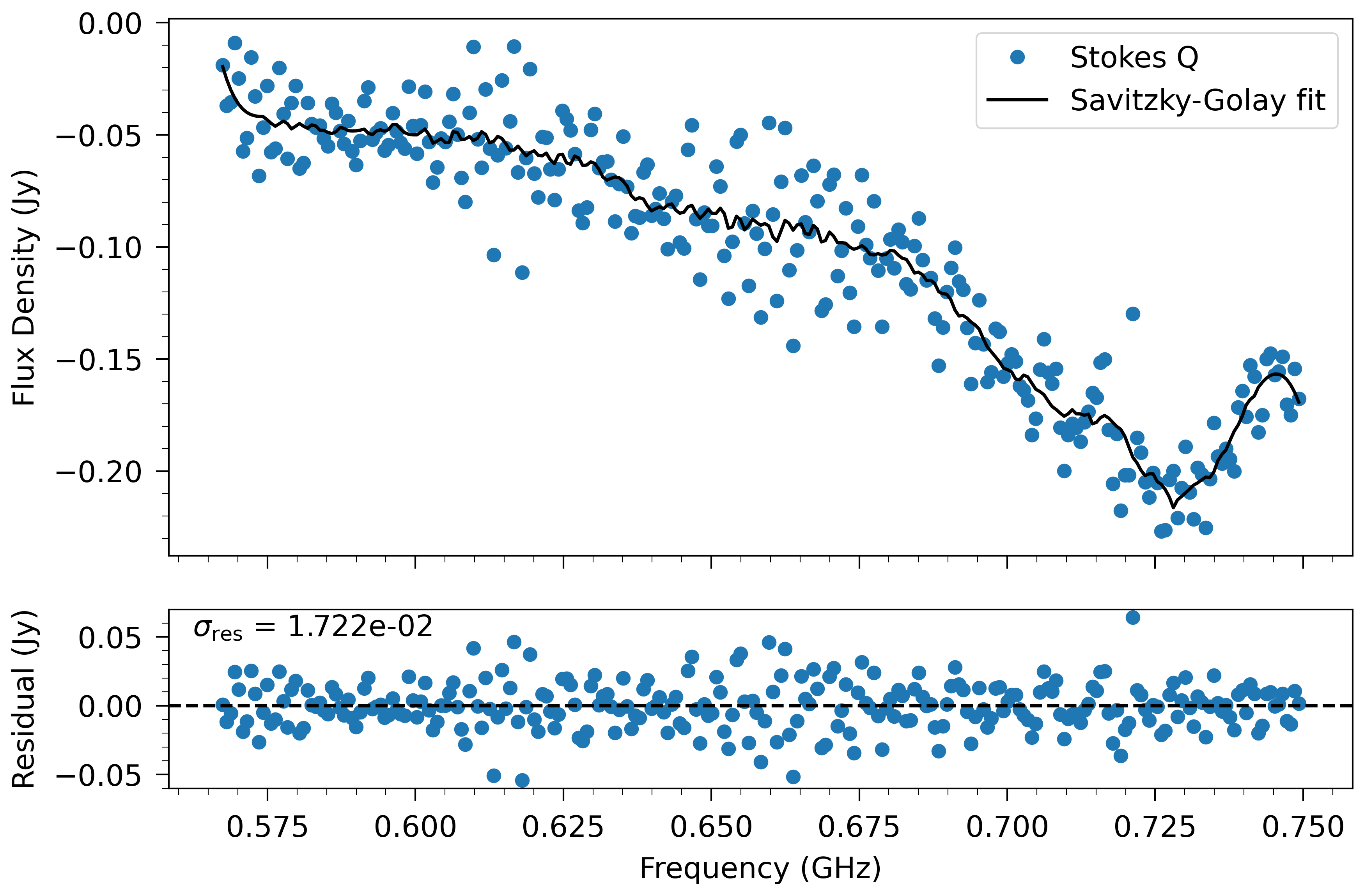}
    \includegraphics[width=0.24\textwidth]{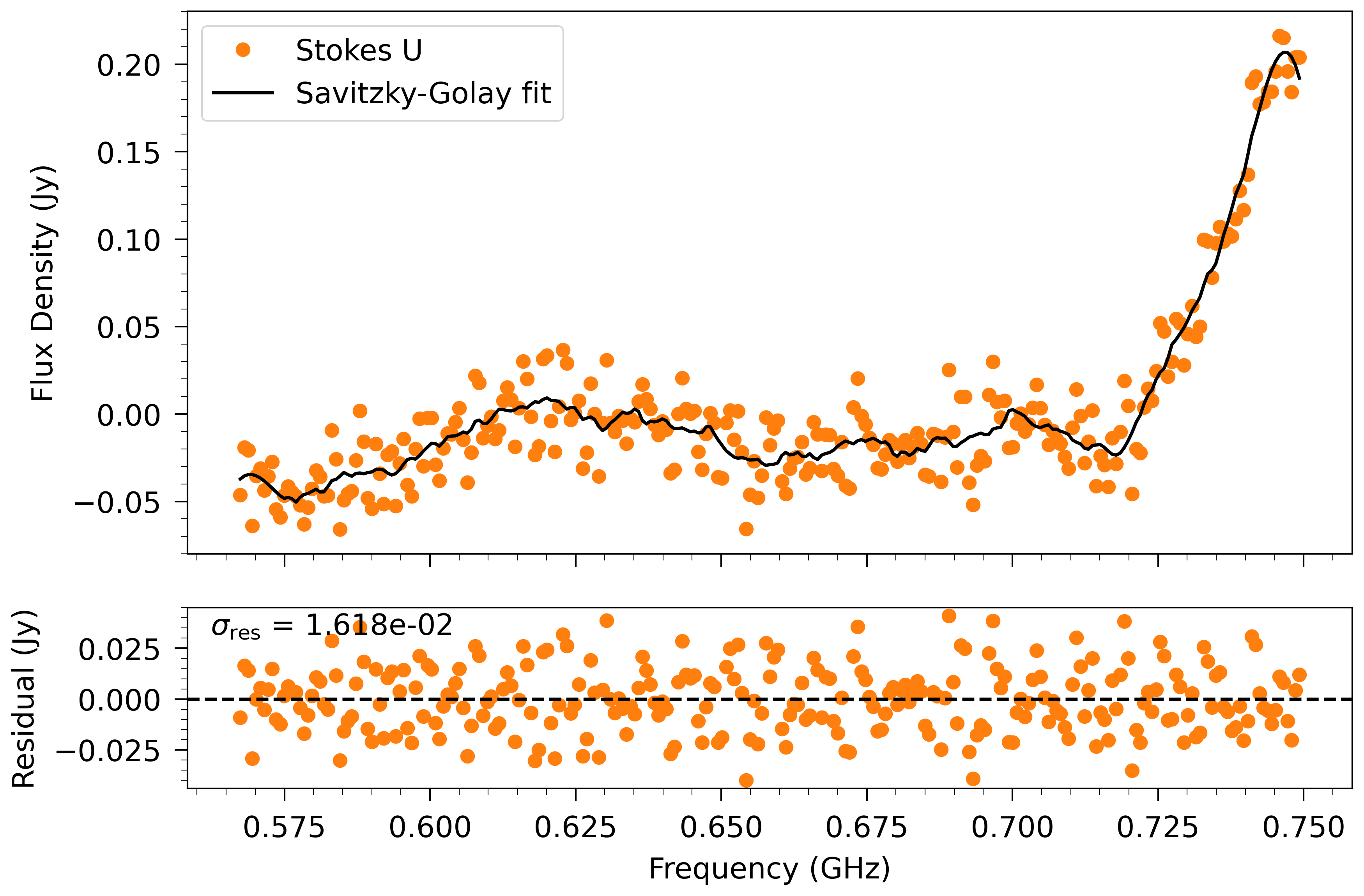}
    \includegraphics[width=0.24\textwidth]{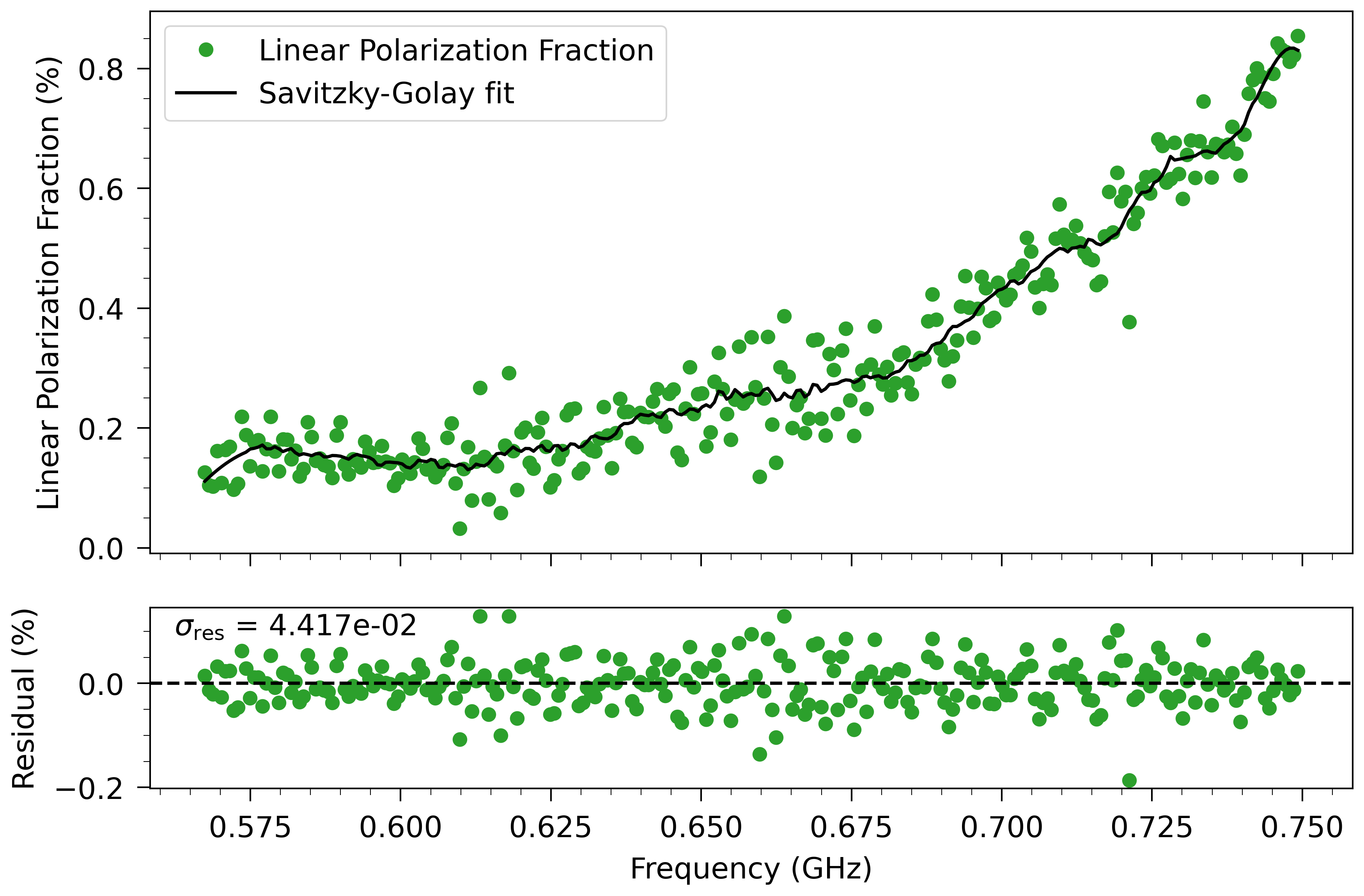}
    \includegraphics[width=0.24\textwidth]{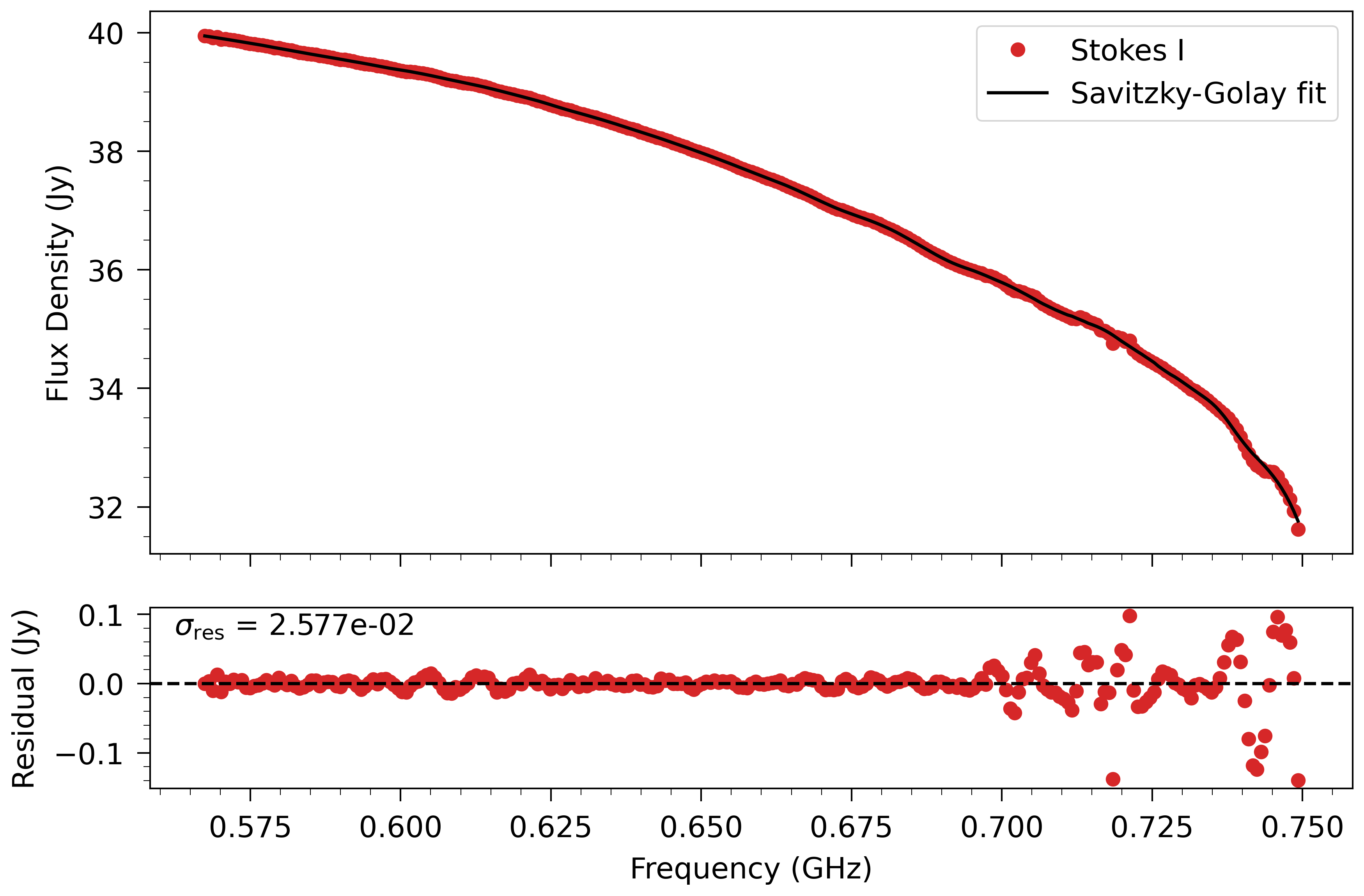}
    \includegraphics[width=0.24\textwidth]{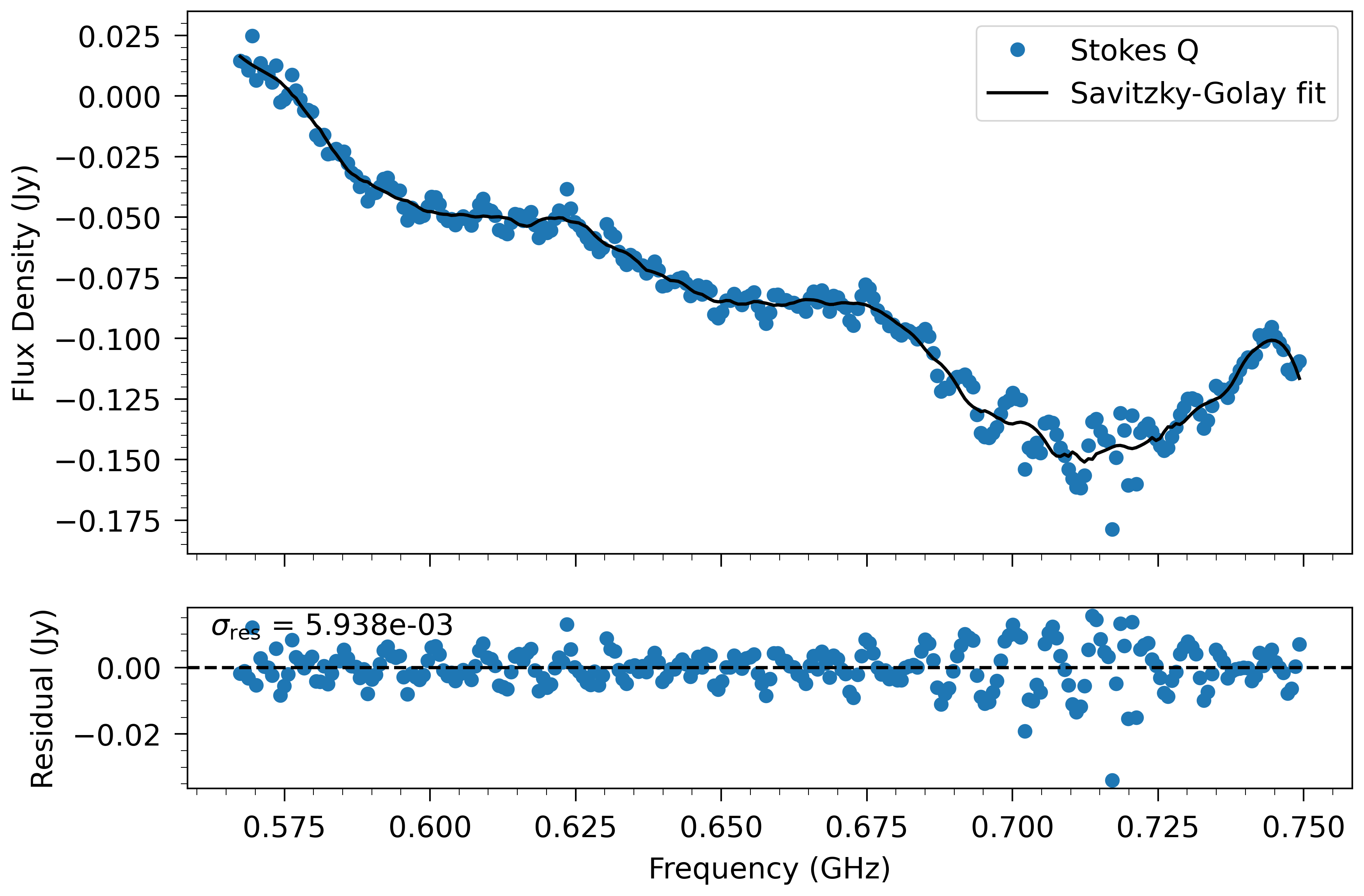}
    \includegraphics[width=0.24\textwidth]{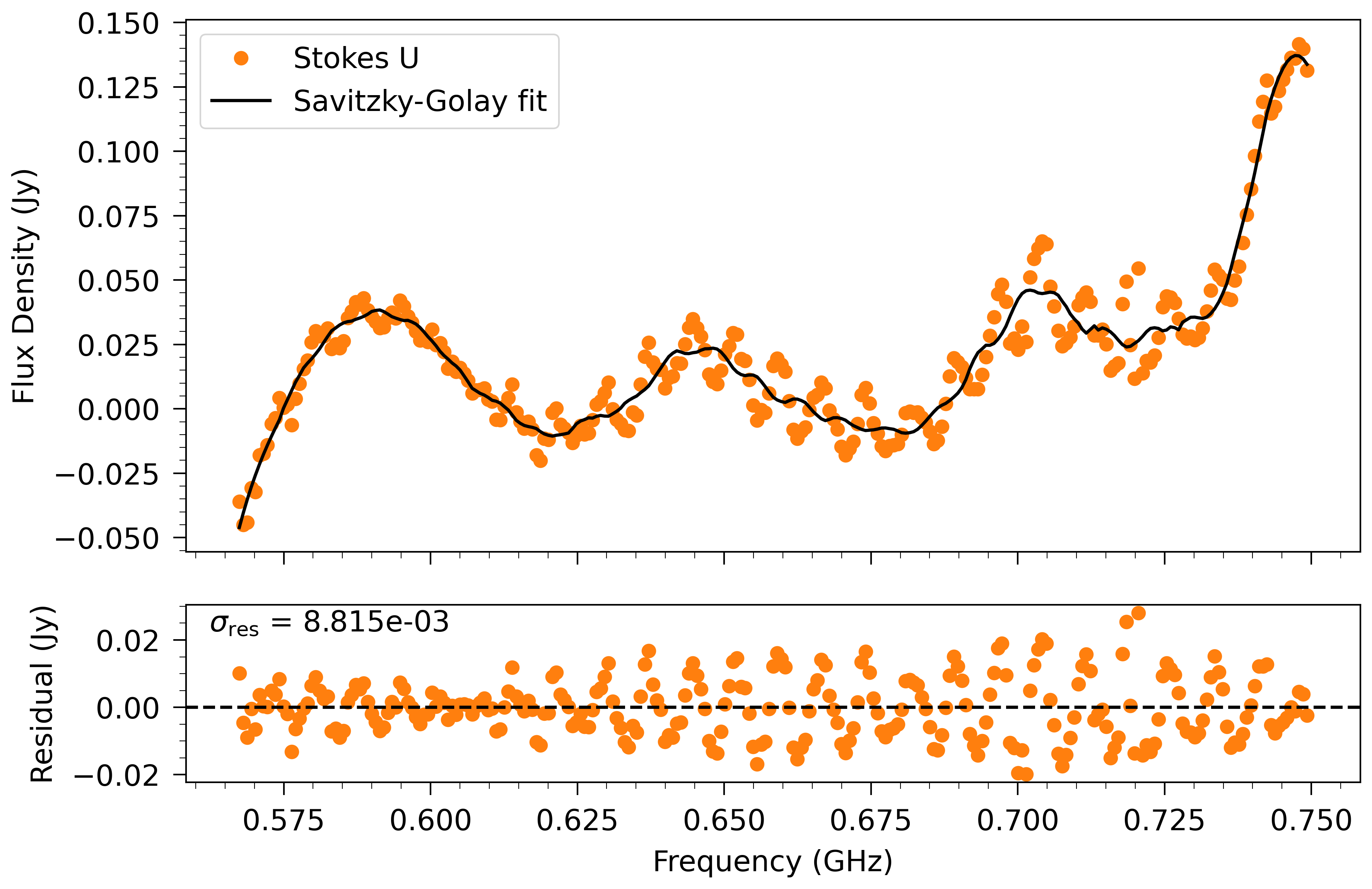}
    \includegraphics[width=0.24\textwidth]{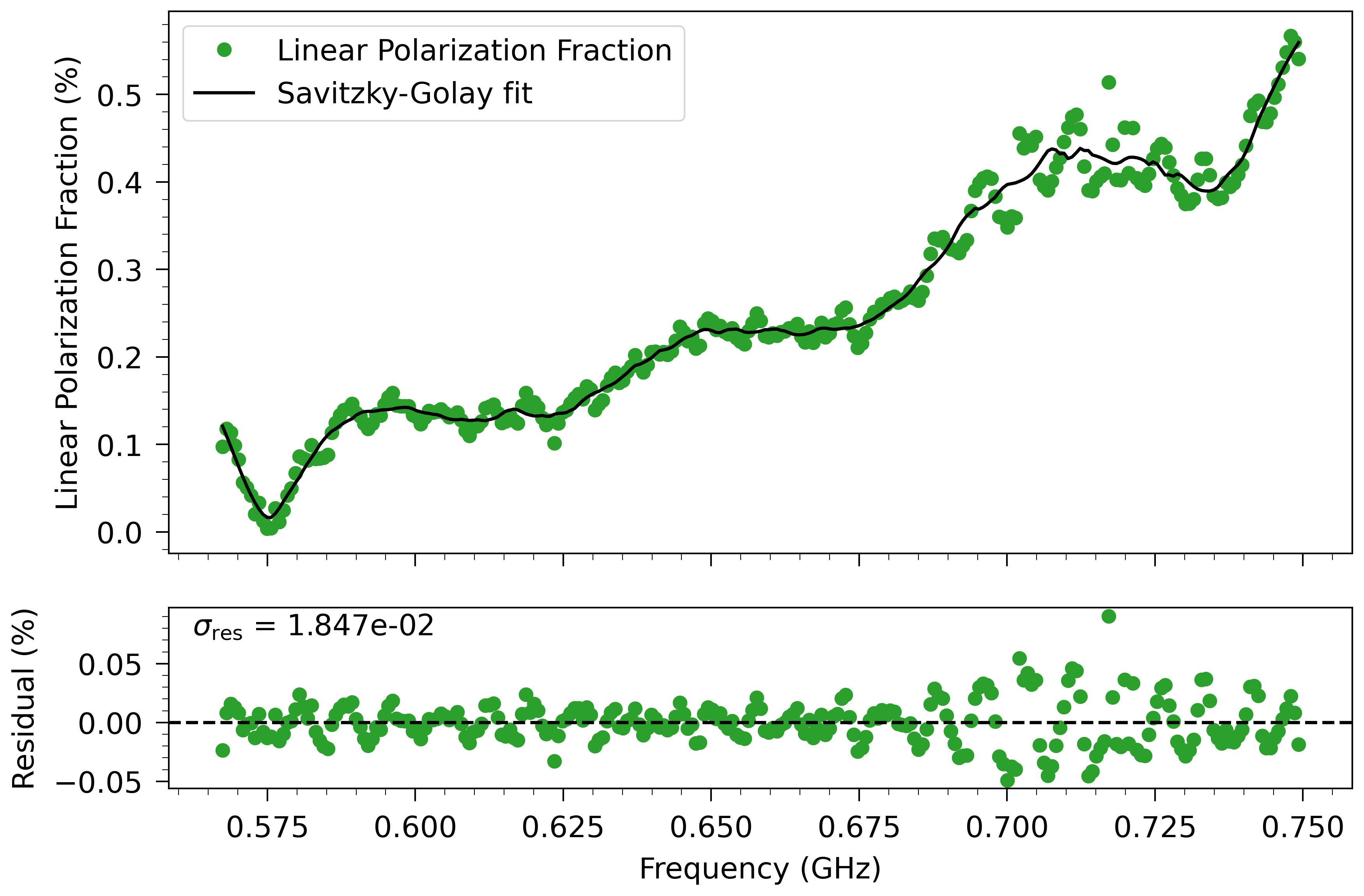}    
\caption{\small Same as Figure \ref{fig:3C138_IQUf_residuals} but for 3C147. As discussed in Section~\ref{sec:poln}, the Stokes parameter values were averaged over the shorter baselines.}
    \label{fig:3C147_IQUf_residuals}
\end{figure}

Figures \ref{fig:3C138_IQUf_residuals} \& \ref{fig:3C147_IQUf_residuals} present the Stokes I, Q, U, and FPOL spectra, which were fitted using a Savitzky–Golay (SG) smoothing function\footnote{It performs a local least-square polynomial fit within a moving window, preserving the underlying smooth spectral behaviour while suppressing the high-frequency fluctuations, allowing the residual spectra to provide a robust estimate of the remaining frequency-dependent instrumental variations.} \citep{Savitzky1964}.
For 3C138, the Stokes' values were averaged over the shorter baselines. Additionally, a few frequency intervals exhibiting large drops in the visibility amplitudes were excluded. The presence of these gaps prevented the SG filter from providing a reliable continuous fit across the intervening frequency ranges, in spite of the presence of well-behaved data outside the excluded frequency intervals. The residual spectrum was obtained by subtracting the smoothed spectrum from the measured data. The $\sigma_{\mathrm{res}}$ values presented in Table \ref{tab:propertiesfromplot} are estimated as the standard deviation of the residuals,
\begin{equation}
\sigma_{\mathrm{res}} =
\sqrt{\frac{1}{N}
\sum_{i=1}^{N}
\left(r_i-\bar{r}\right)^2},
\end{equation}

where,

\begin{equation}
r_i = S_i - S_{i,\mathrm{fit}},
\end{equation}


was adopted as a measure of the channel-to-channel spectral fluctuations. Here, $S_i$ is the measured value in the $i$th frequency channel, $S_{i,\mathrm{fit}}$ is the corresponding value from the Savitzky-Golay fit, $\bar{r}$ is the mean residual, and $N$ is the number of frequency channels used after excluding the edge channels. For 3C138, the Walsh ON data show significantly reduced residuals in Stokes Q and U, with $\sigma_{\mathrm{res}}$ values of 5 and 8 mJy~beam$^{-1}$, respectively, compared to 8 and 13 mJy~beam$^{-1}$ for the Walsh OFF data. Similarly, the $\sigma_{\mathrm{res}}$ in fractional polarization decreases from 0.067\% with Walsh OFF to 0.039\% with Walsh ON. For the unpolarized calibrator 3C147, the $\sigma_{\mathrm{res}}$ in Q, U, and FPOL are reduced by factors of $\sim$3, $\sim$2, and $\sim$2.4, respectively, compared to the Walsh OFF mode. The recovered fractional polarization remains unchanged within the uncertainties for both modes. The results obtained for 3C147 display the overall accuracy of the polarization calibration. The Walsh ON data exhibit reduced channel-to-channel spectral scatter and improved precision in the recovered polarization products. However, the presence of a broad frequency-dependent oscillatory structure in the Walsh ON spectra warrants further investigation before attributing all spectral differences solely to the Walsh scheme. The overall frequency dependence of the fractional polarization is consistent with the 3C138 model presented by \citet{Perley2026}. However, the uGMRT measurements systematically deviate from the MeerKAT model by $\approx1$\% in fractional polarization for both modes.

We note, for completeness, that noticeable differences were observed in the Stokes I, Q, U, and FPOL signals between the Walsh ON and OFF modes, as shown in Figures~\ref{fig:3C138_IQUf_residuals} and \ref{fig:3C147_IQUf_residuals} for 3C138 and 3C147, respectively. The Stokes Q and U signals for 3C138 and 3C147 show larger undulations with Walsh ON than with Walsh OFF. Therefore, the small value of $\sigma_{\mathrm{res}}$ in the Walsh ON case may arise because the smoothing filter has captured the large-scale oscillations. It is also possible that the Walsh scheme itself introduces spurious frequency-dependent structure into the Stokes Q and U signals. The non-simultaneity of the Walsh OFF and ON scans must also be taken into account, as both the polarizer response and the ambient RFI environment are likely to have changed between scans. These effects warrant closer scrutiny using additional data acquired with Walsh modulation.

\begin{table*}
\small
\centering
\caption{A comparison of the full-Stokes spectral properties of the calibrators 3C138 and 3C147 obtained with Walsh switching OFF and ON at 650~MHz (band-4). The Stokes data were fitted using a Savitzky-Golay smoothing function after excluding a few frequency intervals. The $\sigma_{\mathrm{res}}$ values represent the standard deviation of the residual data obtained over the frequency range. The improvement factor is estimated as the ratio of the values with Walsh OFF to those with Walsh ON, indicating that Walsh ON produces less scatter in the calibrated data. The I, Q, U, linear polarization intensity (PPOL), and FPOL values are from the fit at 650~MHz. The quoted uncertainties on the fitted I, Q, and U values are estimated from the standard deviation of the residuals.}
\label{tab:propertiesfromplot}

\begin{tabular}{lcccc}
\hline
Source & Parameter & Walsh OFF & Walsh ON & Improvement factor \\
\hline
\hline
3C138 
& $\sigma_{\mathrm{res}}$($I$) (Jy) 
& 0.023 & 0.019 & 1.2 \\
& $\sigma_{\mathrm{res}}$($Q$) (Jy) 
& 0.009 & 0.005 & 1.8 \\
& $\sigma_{\mathrm{res}}$($U$) (Jy) 
& 0.013 & 0.008 & 1.6 \\
& $\sigma_{\mathrm{res}}$(FPOL) (\%) 
& 0.067 & 0.039 & 1.7 \\
\hline

& $I$ (Jy) 
& $12.386 \pm 0.023$ 
& $12.906 \pm 0.019$ & -- \\

& $Q$ (Jy) 
& $0.494 \pm 0.009$ 
& $0.495 \pm 0.005$ & -- \\

& $U$ (Jy) 
& $-0.007 \pm 0.013$ 
& $-0.023 \pm 0.008$ & -- \\

& PPOL (Jy) 
& $0.494 \pm 0.009$ 
& $0.496 \pm 0.005$ & -- \\

& FPOL (\%) 
& $3.987 \pm 0.067$ 
& $3.844 \pm 0.039$ & -- \\
\hline

3C147
& $\sigma_{\mathrm{res}}$($I$) (Jy)
& 0.0372 & 0.0258 & 1.4 \\

& $\sigma_{\mathrm{res}}$($Q$) (Jy)
& 0.0172 & 0.00594 & 2.9 \\

& $\sigma_{\mathrm{res}}$($U$) (Jy)
& 0.0162 & 0.00882 & 1.8 \\

& $\sigma_{\mathrm{res}}$(FPOL) (\%)
& 0.0442 & 0.0185 & 2.4 \\

\hline

& $I$ (Jy)
& $37.889 \pm 0.037$
& $37.974 \pm 0.026$ & -- \\

& $Q$ (Jy)
& $-0.085 \pm 0.017$
& $-0.085 \pm 0.006$ & -- \\

& $U$ (Jy)
& $-0.017 \pm 0.016$
& $0.021 \pm 0.009$ & -- \\

& PPOL (Jy)
& $0.086 \pm 0.017$
& $0.087 \pm 0.006$ & -- \\

& FPOL (\%)
& $0.228 \pm 0.044$
& $0.230 \pm 0.019$ & -- \\
\hline
\end{tabular}
\end{table*}

\begin{table*}
\small
\centering
\caption{A comparison of the image properties for 3C138 with Walsh OFF and Walsh ON at 650 MHz. PPOL = Linear polarized intensity. FPOL = Linear fractional polarization.}
\label{tab:walsh_test}
\begin{tabular}{lcccccc}
\hline
 & \multicolumn{2}{c}{r.m.s. noise (Jy beam$^{-1}$)}
 & \multicolumn{2}{c}{Peak intensity (Jy beam$^{-1}$)}
 & \multicolumn{2}{c}{Flux density (Jy)} \\
Stokes Parameter
& Walsh OFF & Walsh ON
& Walsh OFF & Walsh ON
& Walsh OFF & Walsh ON \\\hline
$I$ & $8.0\times10^{-3}$ & $6.4\times10^{-3}$ & 11.0 & 11.9 & 12.8 & 13.3 \\
$Q$ & $6.2\times10^{-4}$ & $5.9\times10^{-4}$ & 0.42 & 0.46 & 0.48 & 0.51 \\
$U$ & $4.6\times10^{-4}$ & $5.1\times10^{-4}$ & $-0.034$ & $-0.023$ & $-0.018$ & $-0.029$ \\\hline
Mean PPOL (Jy) & & & $0.205 \pm 0.002$
     & $0.331 \pm 0.001$ & & \\
Mean FPOL (\%) & & & $3.72 \pm 0.05$
          & $3.899 \pm 0.007$ & & \\\hline
\end{tabular}
\end{table*}
\subsection{Imaging}\label{sec:imaging}
We imaged the polarized source 3C138 using the multi-term multi-frequency synthesis (MTMFS) deconvolver and Briggs' weighting with a {\tt Robust} parameter of 0.5, and self-calibrated the images for three phase-only cycles. Further self-calibration rounds (amplitude+phase) did not improve the image further. Although the self-calibration rounds reduced the r.m.s. noise drastically, and increased the source flux density, a lot of solutions failed for the image with the Walsh OFF mode. This resulted in greater flagging of antennas, which increased the beam size. Quantitatively, during the self-calibration rounds, there was additional flagging of 31.5\% ($1.5\times10^8$ out of $4.8\times10^8$) and 25.1\% ($1.4\times10^8$ out of $5.7\times10^8$) data in the Walsh OFF and Walsh ON modes, respectively. This is an empirical difference observed in this experiment and should not be interpreted as a general consequence of disabling Walsh modulation.

\begin{figure}
\centering
\includegraphics[width=7.5cm,trim=50 50 0 30]{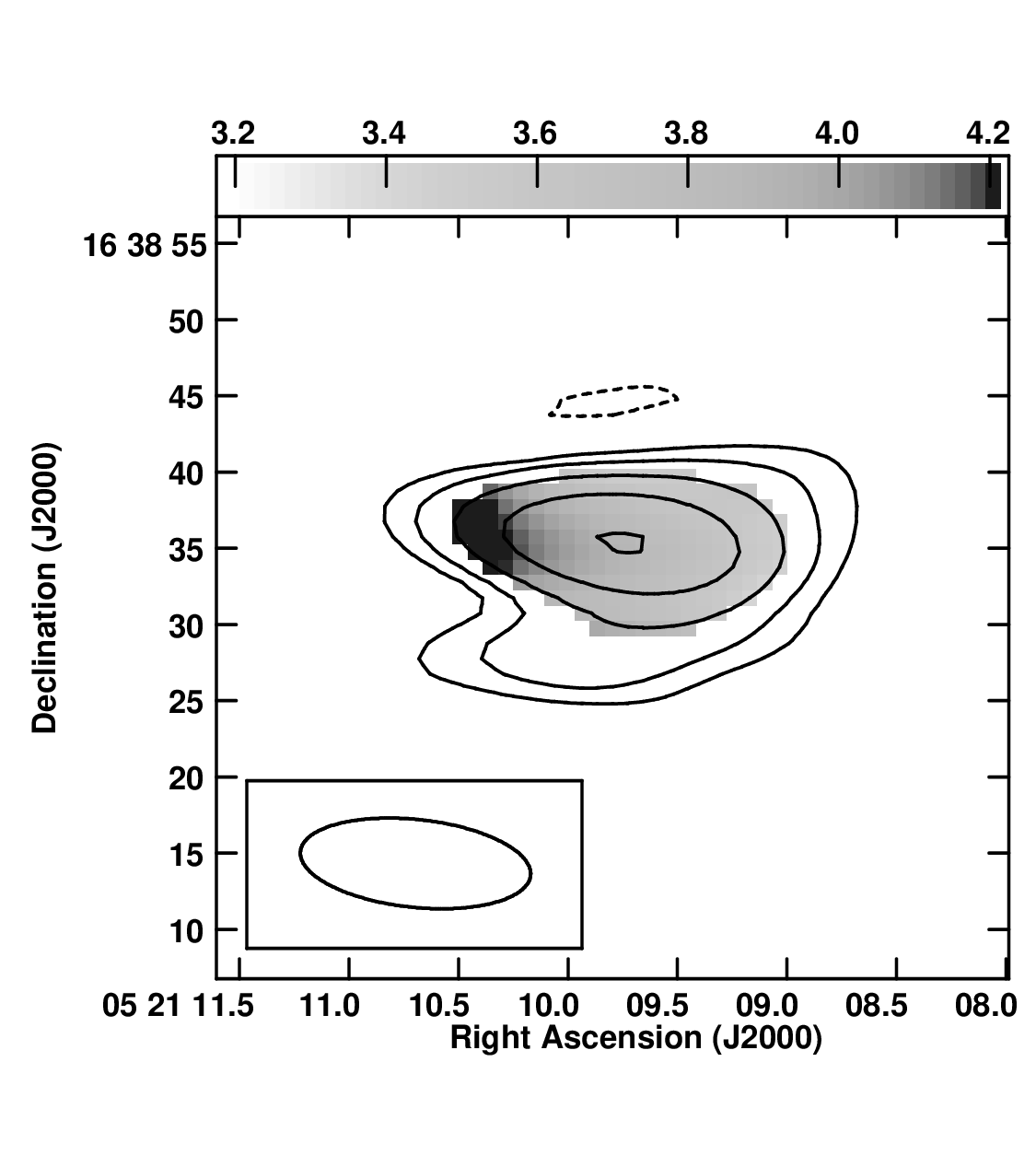}
\includegraphics[width=7cm,trim=0 50 40 30]{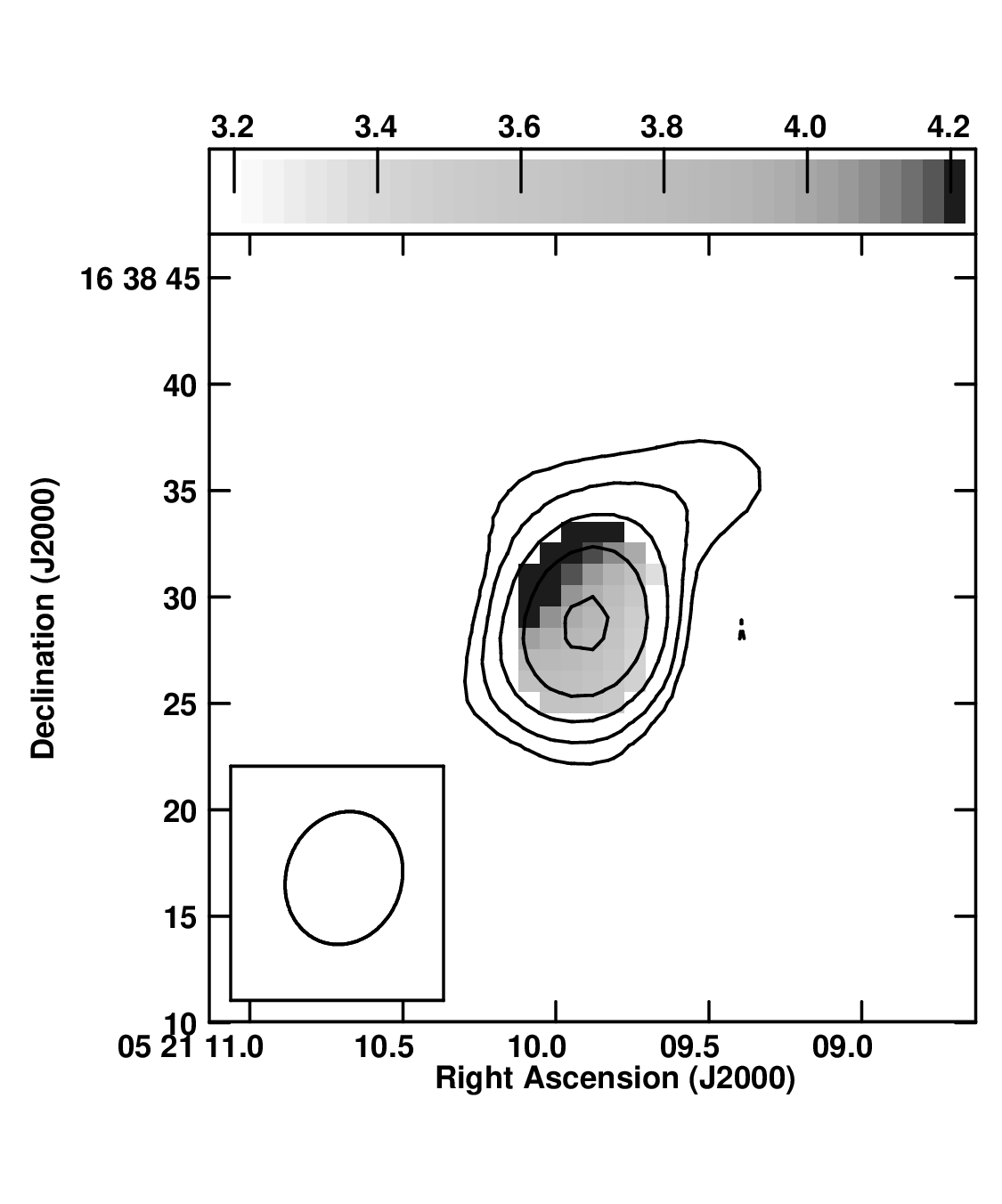}
\caption{\small The contour image of 3C138 produced after several rounds of self-calibration, with Walsh OFF (left) and Walsh ON (right). The fractional polarization has been shown in greyscale. The total intensity contours in black are at levels $0.12\times(-5.6, 5.6, 11.25, 22.5, 45, 90)$ Jy beam$^{-1}$. The fractional polarization values range from 3.2 to 4.2\% in both panels. The synthesized beam sizes are plotted on the bottom left corner with values $15.18''\times5.77''$, 84 degrees and $6.3''\times5.4''$, $-19.3$ degrees, for the left and the right panel, respectively.}
\label{fig:3C138_Walsh-ONOFF_images_p3}
\end{figure}

Both the 3C138 images show slight extensions in the north-south direction. The polarization fraction is higher in the north-east direction consistent with higher degrees of polarization observed in north-eastern jet in the 22 GHz VLA image of \citet{Akujor1993}. The final polarization fraction images with total intensity contours for Walsh OFF and Walsh ON are shown in Figure \ref{fig:3C138_Walsh-ONOFF_images_p3}. 
%The images evidently depict the difference between the Walsh OFF and Walsh ON mode, as a much better image, in terms of a calibrator source shape and near-source r.m.s. noise, is obtained with the Walsh ON mode. 
A better image is obtained with the Walsh ON mode in terms of the lower near-source r.m.s. noise and higher fidelity to the true source structure seen in the resolved VLA image. 
The values of Stokes I, Q, and U r.m.s. noise, peak intensity, flux density, PPOL, and FPOL values for 3C138 have been tabulated in Table~\ref{tab:walsh_test}. Overall, we found a $4.8\%$ improvement in the fractional polarization value with the Walsh ON scheme for this experiment, with a 7.4 times reduction in the polarization fraction uncertainty. 

\section{Acknowledgements}
We thank Prof. C. H. Ishwara-Chandra (NCRA) for suggesting various observational setups for carrying out experiments using the Walsh scheme. We thank Dr. Silpa Sasikumar (University of Concepcion, Chile; formerly NCRA) for initiating tests using the Walsh scheme.

{\small
\bibliographystyle{plainnat}
\bibliography{Walsh}
}

\end{document}